\documentclass[iop]{emulateapj}
\pdfoutput=1

\newif\ifpreprint
\preprinttrue
\preprintfalse

\usepackage{comment}
\usepackage[usenames,dvipsnames]{color}
\usepackage{graphics}
\usepackage{lscape}
\usepackage{xspace}
\usepackage{ulem}
\usepackage{enumitem}

\newcommand{\lya}        {Ly$\alpha$\xspace}
\newcommand{\llya}       {$L_{\rm Ly\alpha}$\xspace}
\newcommand{\unitcgssb}  {erg\,s$^{-1}$\,cm$^{-2}$\,arcsec$^{-2}$\xspace}
\newcommand{\unitcgsflux}{erg\,s$^{-1}$\,cm$^{-2}$\xspace}
\newcommand{\unitcgsflam}{erg\,s$^{-1}$\,cm$^{-2}$\,\AA$^{-1}$\xspace}
\newcommand{\unitcgslum} {erg\,s$^{-1}$\xspace}

\newcommand{\um}         {\micron\xspace}
\newcommand{\kms}        {\ifmmode{\rm \,km\,s^{-1}}\else\,km\,s$^{-1}$\xspace\fi} 
\newcommand{\bptx}       {$\log$([\ion{N}{2}]/H$\alpha$)\xspace}

\newcommand{\ha}         {H$\alpha$\xspace}
\newcommand{\halamb}     {H$\alpha$ $\lambda$6563\xspace}
\newcommand{\hb}         {H$\beta$\xspace}
\newcommand{\hblamb}     {H$\beta$ $\lambda$4861\xspace}
\newcommand{\heii}       {\ion{He}{2}\xspace}
\newcommand{\civ}        {\ion{C}{4}\xspace}
\newcommand{\oiii}       {[\ion{O}{3}]\xspace}
\newcommand{\oiiilamb}   {[\ion{O}{3}] $\lambda\lambda$4959,5007}
\newcommand{\oiiilamba}  {[\ion{O}{3}] $\lambda$4959\xspace}
\newcommand{\oiiilambb}  {[\ion{O}{3}] $\lambda$5007\xspace}
\newcommand{\oii}        {[\ion{O}{2}]\xspace}
\newcommand{\oiilamb}    {[\ion{O}{2}] $\lambda\lambda$3727,3729}
\newcommand{\nii}        {[\ion{N}{2}]\xspace}

\newcommand{\civlamb}    { \ion{C}{4}  $\lambda$1546}
\newcommand{\helamb}     {\ion{He}{2}  $\lambda$1640}
\newcommand{\dvlya}      {\ifmmode\Delta v_{\rm Ly\alpha}\else$\Delta v_{\rm Ly\alpha}$\xspace\fi}
\newcommand{\dvis}       {\ifmmode{\Delta v_{\rm IS}}\else$\Delta v_{\rm IS}$\xspace\fi}
\newcommand{\vexp}       {\ifmmode{v_{\rm exp}}\else$v_{\rm exp}$\xspace\fi}
\newcommand{\vabs}       {\ifmmode{v_{\rm abs}}\else$v_{\rm abs}$\xspace\fi}
\newcommand{\dvmax}      {\ifmmode{\Delta v_{\rm max}}\else$\Delta v_{\rm max}$\xspace\fi}
\newcommand{\NHI}        {$N_{\rm{H I}}$}
\newcommand{\HI}         {{\ion{H}{1}}\xspace}
\newcommand{\sig}        {$\sigma$\xspace}
\newcommand{\msun}       {$M_{\sun}$}
\newcommand\avg[1]       {\langle#1\rangle}
\newcommand\E[1]         {$\times$$10^{#1}$}
\newcommand\ff[1]        {\tablenotemark{#1}}

\slugcomment{Accepted for publication in ApJ.}

\shorttitle{Properties of \lya Nebulae: Gas Kinematics}
\shortauthors{Yang et al.}

\begin{document}

\title{The Properties of \lya Nebulae: 
       Gas Kinematics from Non-resonant Lines\altaffilmark{*}}

\author{
   Yujin Yang\altaffilmark{1,2},
   Ann Zabludoff\altaffilmark{3},
   Knud Jahnke\altaffilmark{2},
   Romeel Dav\'e\altaffilmark{3,4,5,6}
}
\altaffiltext{1}{Argelander Institut f\"ur Astronomie, Universit\"at Bonn, Auf dem H\"ugel 71, 53121 Bonn, Germany, yyang@astro.uni-bonn.de}
\altaffiltext{2}{Max-Planck-Institut f\"ur Astronomie, K\"onigstuhl 17, Heidelberg, Germany}
\altaffiltext{3}{Steward Observatory, University of Arizona, 933 North Cherry Avenue, Tucson AZ 85721}
\altaffiltext{4}{University of the Western Cape, Bellville, Cape Town 7535, South Africa}
\altaffiltext{5}{South African Astronomical Observatories, Observatory, Cape Town 7525, South Africa}
\altaffiltext{6}{African Institute for Mathematical Sciences, Muizenberg, Cape Town 7545, South Africa}
\altaffiltext{*}{Based on observations made with an ESO telescope at the 
                 Paranal Observatory, under the program ID 086.A-0804.}

\begin{abstract}

With VLT/X-shooter, we obtain optical and near-infrared spectra of six
\lya blobs at $z$ $\sim$ 2.3.  For a total sample of eight \lya blobs
(including two that we have previously studied), the majority (6/8)
have broadened \lya profiles with shapes ranging from a single peak to
symmetric or asymmetric double-peaked.  
The remaining two systems, in which the \lya profile is not significantly
broader than the \oiii or \ha emission lines, have the most spatially
compact \lya emission, the smallest offset between the \lya and the
\oiii or \ha line velocities, and the only detected \civ and \heii lines
in the sample, implying that a hard ionizing source, possibly an AGN,
is responsible for their lower optical depth.
Using three measures --- the velocity offset between the \lya line
and the non-resonant \oiii or \ha line (\dvlya), the offset of
stacked interstellar metal absorption lines, and a new indicator, the
spectrally-resolved \oiii line profile --- we study the kinematics of
gas along the line of sight to galaxies within each blob center.
These three indicators generally agree in velocity and direction, and are
consistent with a simple picture in which the gas is stationary or slowly
outflowing at a few hundred \kms from the embedded galaxies.  The absence
of stronger outflows is not a projection effect: the covering fraction for
our sample is limited to $<$\,1/8 (13\%).  The outflow velocities exclude
models in which star formation or AGN produce ``super'' or ``hyper''
winds of up to $\sim$1000\,\kms.  The \dvlya offsets here are smaller
than typical of Lyman break galaxies (LBGs), but similar to those of
compact \lya emitters.  The latter suggests a connection between blob
galaxies and \lya emitters and that outflow speed cannot be a dominant
factor in driving extended \lya emission.
For one \lya blob (CDFS-LAB14), whose \lya profile and metal absorption
line offsets suggest no significant bulk motion, we use a simple
radiative transfer model to make the first column density measurement
of gas in an embedded galaxy, finding it consistent with a damped \lya
absorption system.
Overall, the absence of clear inflow signatures suggests that the
channeling of gravitational cooling radiation into \lya is not significant
over the radii probed here.  However, one peculiar system (CDFS-LAB10)
has a blueshifted \lya component that is not obviously associated
with any galaxy, suggesting either displaced gas arising from tidal
interactions among blob galaxies or gas flowing into the blob center.
The former is expected in these overdense regions, where {\sl HST}
images resolve many galaxies. The latter might signify the predicted
but elusive cold gas accretion along filaments.

\end{abstract}


\section{Introduction}
\label{sec:intro}

Giant \lya nebulae, or ``blobs,'' are extended sources at $z$ $\sim$
2--6 with typical \lya sizes of $\gtrsim$\,5\arcsec\ ($\gtrsim$\,50\,kpc)
and line luminosities of $L_{\rm{Ly\alpha}}\gtrsim10^{43}$\,\unitcgslum\
\cite[e.g.,][]{Keel99, Steidel00, Francis01, Matsuda04, Matsuda11, Dey05,
Saito06, Smith&Jarvis07, Hennawi09, Ouchi09, Prescott09, Prescott12a,
Yang09, Yang10, Erb11}.  The low number counts, strong clustering,
multiple embedded sources, and location in over-dense environments of
the largest \lya blobs indicate that they lie in massive ($M_{\rm halo}$
$\sim$ $10^{13}$\msun) dark matter halos, which will evolve into those
typical of rich galaxy groups or clusters today \citep{Yang09, Yang10,
Prescott08, Prescott12b}.  Therefore, \lya blobs are unique tracers of
the formation of the most massive galaxies and their early interaction
with the surrounding intergalactic medium (IGM).

This interaction is probably tied on some scale to the source of the
blobs' extended \lya emission, but that mechanism is poorly understood.
Emission from \lya blobs could arise from several phenomena, which may
even operate together, including shock-heating by galactic superwinds
\citep{Taniguchi&Shioya00} or gas photoionized by active galactic
nuclei \citep{Haiman&Rees01,Geach09}.  Another possibility is smooth gas
accretion, which is likely to play an important role in the formation
of galaxies \cite[e.g.,][]{Keres05, Keres09} and which should channel
some of its gravitational cooling radiation into atomic emission lines
such as \lya \citep{Haiman00, Fardal01, Dijkstra&Loeb09, Goerdt10}.
Another scenario is the resonant scattering of \lya photons produced
by star formation or active galactic nuclei (AGN)  \citep{Steidel10,
Hayes11} in the embedded galaxies.

\begin{figure*}
\epsscale{1.0}
\ifpreprint\epsscale{0.90}\fi
\plotone{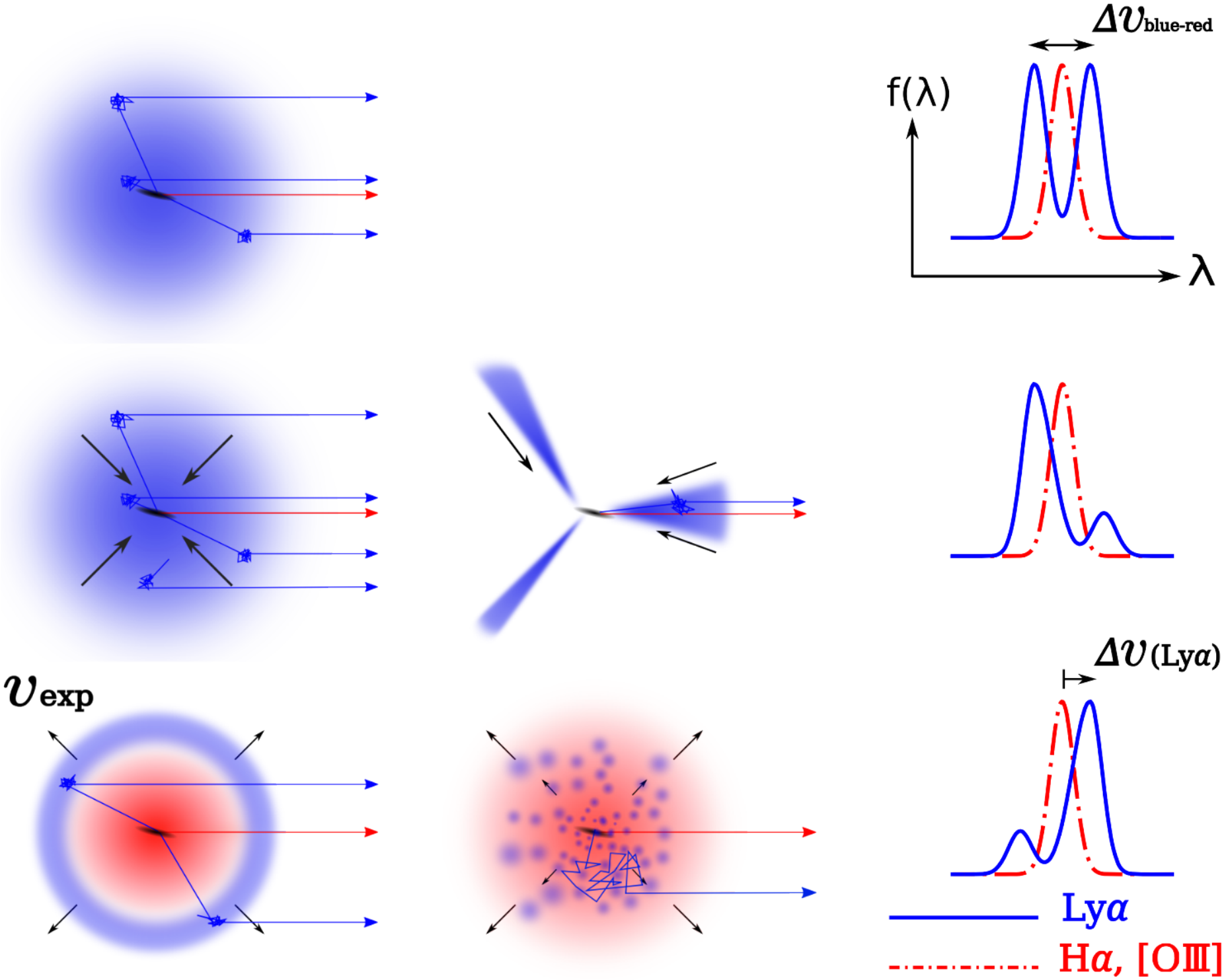}
\caption{
\ifpreprint\small\fi
Schematic diagrams of the expected \lya and non-resonant H$\alpha$
or \oiii profiles if the gas in the blobs is static, inflowing, or
outflowing. The solid (blue) and dashed (red) lines represent the \lya
and non-resonant (optically-thin to \HI) lines, respectively.
{\bf (Top: static cloud)} Because the surrounding gas cloud is optically
thick to the \lya line, \lya photons in the Doppler wings should
escape the cloud through random scatterings in the frequency domain.
Therefore, if the surrounding gas is static, the \lya line should have a
double-peaked profile, while the non-resonant line photons will escape
without any radiative transfer effects.
{\bf (Middle: spherical infall)} If the gas is infalling toward the
embedded galaxy, \lya photons within the cloud on the red side of
the double-peaked profile will see higher optical depth due to the
line-of-sight infalling gas, and the red peak will be depressed.
{\bf (Bottom: expanding shell)} In contrast, if the gas is outflowing,
the blue side of the profile will be more diminished.
In summary, if there is gas infall (or outflow), the \lya profiles will
be asymmetric and blue-shifted (or redshifted) against the non-resonant 
line due to radiative transfer in the optically thick medium, but the
corresponding non-resonant line profiles should be symmetric.  Note that
one cannot determine whether the \lya line is redshifted or blue-shifted
against the background velocity field unless there is a optically thin
reference line, i.e., H$\alpha$ or \oiii.  {\it Both \lya and non-resonant 
lines are therefore required to resolve the nature of the \lya blobs.}
{\bf(Middle column)}  However, the actual geometry of infall
and outflow is unknown. For example, gas infall may take place in
narrow streams \cite[e.g.,][]{Dekel09,Goerdt10}. Or, an outflow may be
bipolar/highly collimated \cite[e.g., M82,][]{Bland&Tully88} or ``clumpy''
\citep{Steidel10}.  A large statistic sample of blob gas kinematics,
like that presented in this paper, is required to constrain that geometry.
}
\label{fig:cartoon}
\end{figure*}


To resolve the debate about the nature of \lya blobs requires ---
at the very least --- that we discriminate between between outflowing
and inflowing models.  Because \lya is a resonant line and typically
optically thick in the surrounding intergalactic medium, studies
even of the same \lya blob's kinematics can disagree.  On one hand,
\citet{Wilman05} argue that their integral field unit (IFU) spectra of
a \lya blob are consistent with a simple model where the \lya emission
is absorbed by a foreground slab of neutral gas swept out by a galactic
scale outflow. On the other, \citet{Dijkstra06b} explain the same data
as arising from the infall of the surrounding intergalactic medium.
Worse, \citet{Verhamme06} claim that the same symmetric \lya profiles
are most consistent with static surrounding gas.

To distinguish among such possibilities requires a comparison of the
center of the \lya line profile with that of a non-resonant line
like \halamb or \oiiilambb.  These rest-frame optical nebular lines
are better measure of the \lya blob's systemic velocity, i.e., of
the precise redshift, because it is not seriously altered by radiative
transfer effects and is more concentrated about the galaxies in the \lya
blob's core.
We illustrate this line offset technique in Figure \ref{fig:cartoon}.
While gas accretion models predict different line profile shapes depending
on various assumptions, e.g., the location of ionizing sources, the
detailed geometry, and the velocity field, they all predict that the
overall \lya line profile, originating from the central source or the
surrounding gas, will be blue-shifted with respect to the center of
a non-resonant line such as H$\alpha$ that is optically thin to the
surrounding \HI gas \citep{Verhamme06, Dijkstra06a}.
This is because the red-side of the \lya profile will see higher
optical depth due to the infalling (approaching) gas.  In other words,
the H$\alpha$ kinematics represent the true underlying velocity field if
the \lya blob is accreting gas from the intergalactic medium.  The same
is true if the gas is outflowing, except that the \lya line will be
redshifted with respect to H$\alpha$.  Thus, if we measure the direction
of the \lya--H$\alpha$ and/or \lya--\oiii line offset (hereafter defined
as \dvlya), we can distinguish an inflow from an outflow.

The first such analysis for two \lya blobs shows that \lya is coincident
with or redshifted by $\sim$200\,\kms\ from the H$\alpha$ line center
\cite[][see also McLinden et al.~2013]{Yang11}.  These offsets are
much smaller than the $\sim$1000\kms expected from superwind models
\citep{Taniguchi&Shioya00} and even smaller than those typical of
Lyman break galaxies (LBGs), which are widely believed to have galactic
outflows \citep{Steidel04, Steidel10}.  Thus, if \dvlya\ is a proxy
for outflow velocities (\vexp), our initial results suggest that star
formation- or AGN-produced winds may not be required for powering \lya
blobs, making other interpretations of their emission more likely.

However, we do not yet know if these results are representative of {\it
all} \lya blobs or if we have failed to detect strong flows due to the
projected orientations of these two sources.
For example, the gas flow may not be isotropic.  As in bipolar outflows
in M82 \cite[e.g.,][]{Bland&Tully88}, a galactic-scale outflow may occur
in the direction of minimum pressure in the surrounding interstellar
medium (ISM), often perpendicular to the stellar disks.  Or, if gas accretion
is taking place in \lya blobs, numerical simulations suggest that the gas
infall may occur preferentially along filamentary streams \citep{Keres05,
Keres09, Dekel09}.
Thus, if the bulk motion of gas (either infalling or outflowing) happens
to be mis-aligned with our line of sight (LOS), then we may underestimate
or even fail to detect the relative velocity shifts.  Therefore,
it is critical to measure \dvlya for a larger sample to average over
any geometric effects and obtain better constraints on the incidence,
direction, speed, and isotropy of bulk gas motions in blobs.

In this paper, in order to overcome this geometry effects,
we present new X-shooter optical and near-infrared (NIR)
spectroscopy of the six more \lya blobs at $z\approx2.3$. Note
that the survey redshift of this \lya sample has been carefully
selected to allow all important rest-frame optical diagnostic lines
(e.g., [\ion{O}{2}]\,$\lambda$3727, [\ion{O}{3}]\,$\lambda$5007,
H$\beta$\,$\lambda$4868, H$\alpha$\,$\lambda$6563) to fall in NIR
windows and to avoid bright OH sky lines and atmospheric absorption
\citep{Yang09, Yang10}.  With the resulting large sample of  \dvlya
measurements (a total of eight), we determine the relative frequency of
gas infall versus outflow.

Benefiting from X-shooter's high spectral resolution and wide spectral
coverage (3000\AA\ -- 2.5\micron), we constrain the gas kinematics
in \lya blobs using three different tracers: (1) the offset of the
\lya profile with respect to a non-resonant nebular line, (2)
the offset of an interstellar metal absorption line in the rest-frame
UV with respect to the nebular emission line, and (3) a new indicator,
the profile of the spectrally-resolved \oiii emission line.


This paper is organized as follows.  In Section \ref{sec:observation},
we review our sample selection and describe the X-shooter observations
and data reduction.
In Section \ref{sec:result}, we present the results from the
X-shooter spectroscopy, confirming the \lya blobs' redshift (Section
\ref{sec:spec1d}).
We present 1--D and 2--D spectra in Section \ref{sec:spec1d} and Section
\ref{sec:spec2d}, respectively. We briefly summarize the properties of
individual systems in Section \ref{sec:individual}.
In Section \ref{sec:kinematics}, we constrain the gas kinematics using
the three different techniques noted above. In Section \ref{sec:shift},
we compare the \lya profiles with the H$\alpha$ or \oiii line centers
to discriminate between simple infall and outflow scenarios and present
the \dvlya\ statistics for the sample.
In Section \ref{sec:absorption}, we describe the interstellar absorption
lines detected in three galaxies.
In Section \ref{sec:O3profile}, we inspect the \oiii emission line
profiles in detail to look for possible signatures of warm outflows.
In Section \ref{sec:column_density}, we constrain the \HI column density
of a \lya blob by comparing its \lya profile with a simple radiative
transfer (RT) model.
In Section \ref{sec:cdfs-lab10}, we focus on a \lya blob with a
blue-shifted \lya component not directly associated with any detected
galaxy, a possible marker of gas inflow.
Section \ref{sec:conclusion} summarizes our conclusions. Throughout
this paper, we adopt the cosmological parameters: $H_0$ =
70\,\kms\,Mpc$^{-1}$, $\Omega_{\rm M}=0.3$, and $\Omega_{\Lambda}=0.7$.

\section{Observations and Data Reduction}
\label{sec:observation}

\subsection{Sample}

We observe six \lya blobs from the \citet{Yang10} sample. These targets
were chosen such that they are not X-ray detected [$L({\rm 2-32keV})$ $<$
(0.3 -- 4.2)\E{43} \unitcgslum; \cite{Lehmer05, Luo08}], and thus are
not obvious AGN, as our primary goal is to cleanly detect gas infall
or outflow.  These six blobs lie in the Extended Chandra Deep Field
South (ECDFS) and were discovered via deep narrowband imaging with the
CTIO-4m MOSAIC--II camera and a custom narrowband filter ({\sl NB}403).
This filter has a central wavelength of $\lambda_c \approx 4030$\AA,
designed for selecting \lya-emitting sources at $z\approx2.3$.
In Figure \ref{fig:image}, we show the images of these six \lya blobs
(CDFS-LAB06, 07, 10, 11, 12, 13, 14) at various wavelengths ({\sl UBK},
\lya, {\it Spitzer} IRAC 3.6\um and {\sl HST} F606W; \citealt{Gawiser06a},
\citealt{Yang10}, \citealt{Damen11}, \citealt{Rix04}).

With X-shooter, we are targeting intermediate luminosity (\llya $\sim$
10$^{43}$ \unitcgslum) \lya blobs:  the higher \lya blob ID indicates
the lower \lya luminosity in our sample.  Our sample was obtained from
a blind survey, so combined with the two brightest \lya blobs presented
in \citet{Yang11}, the full sample (a total of eight) spans a wide and
more representative range of \lya luminosity and size.  Furthermore,
the transition from compact \lya emitters (LAEs; isophotal area of a
few arcsec$^2$) to extended \lya blobs ($>$10\,arcsec$^2$) is continuous
\citep{Matsuda04, Yang10}, thus the gas kinematics for our faintest \lya
blobs might share the properties with those of bright LAEs.  We refer
readers to \citet{Yang10} for details of the sample selection and to
\citet{Yang11} for the first results of our spectroscopic campaign.

\begin{figure*}
\epsscale{0.8}
\ifpreprint\epsscale{0.9}\fi
\plotone{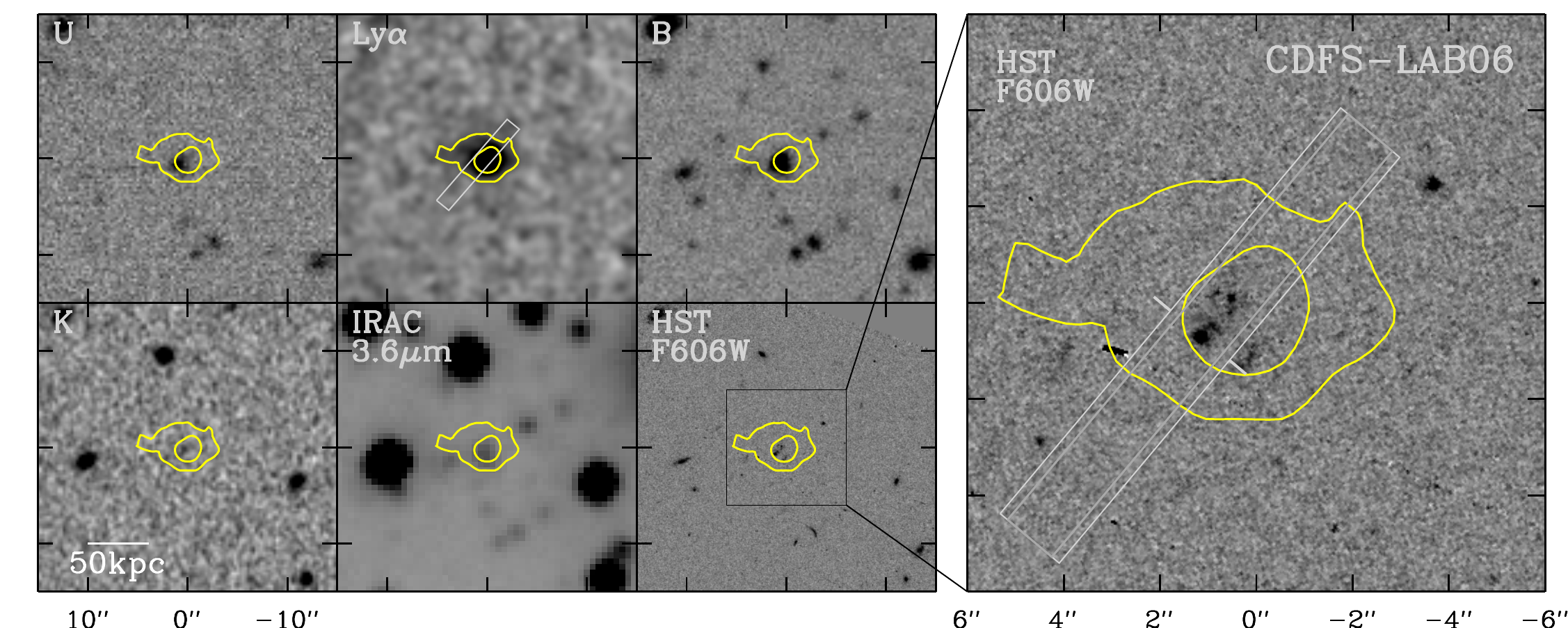} 
\plotone{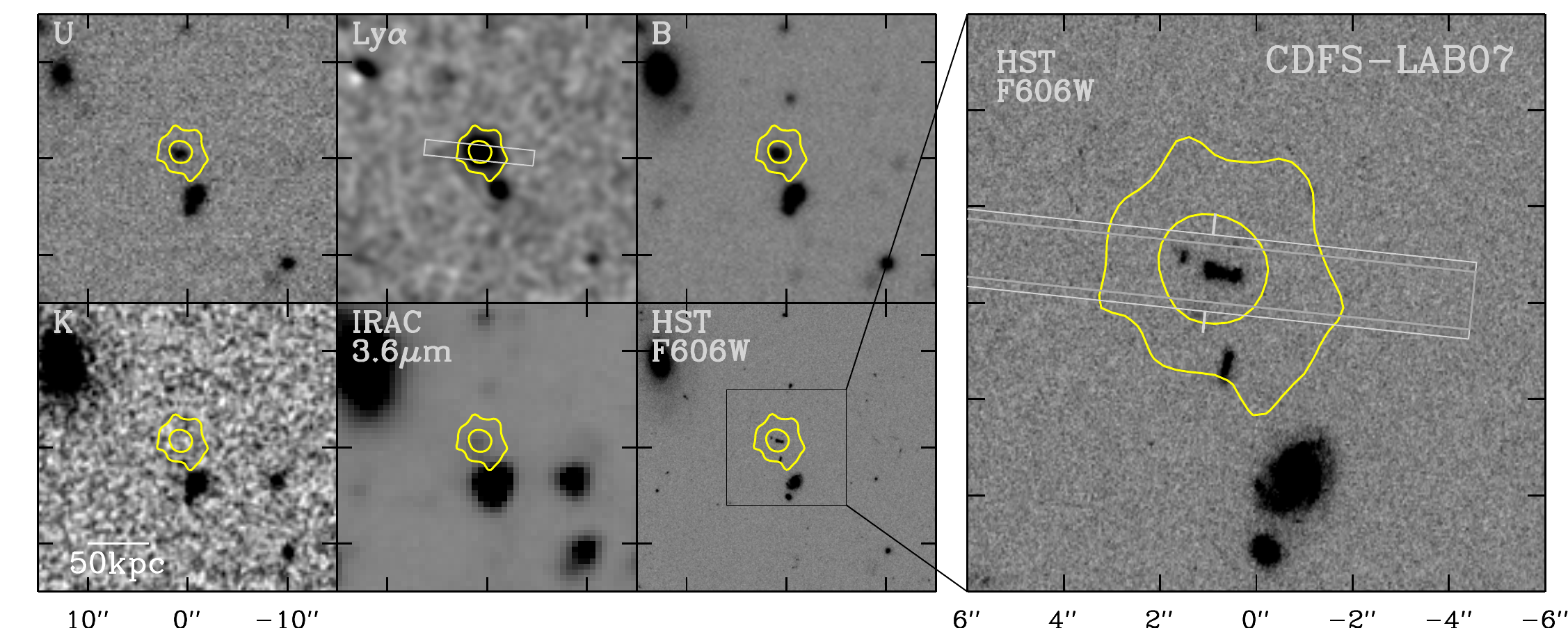} 
\plotone{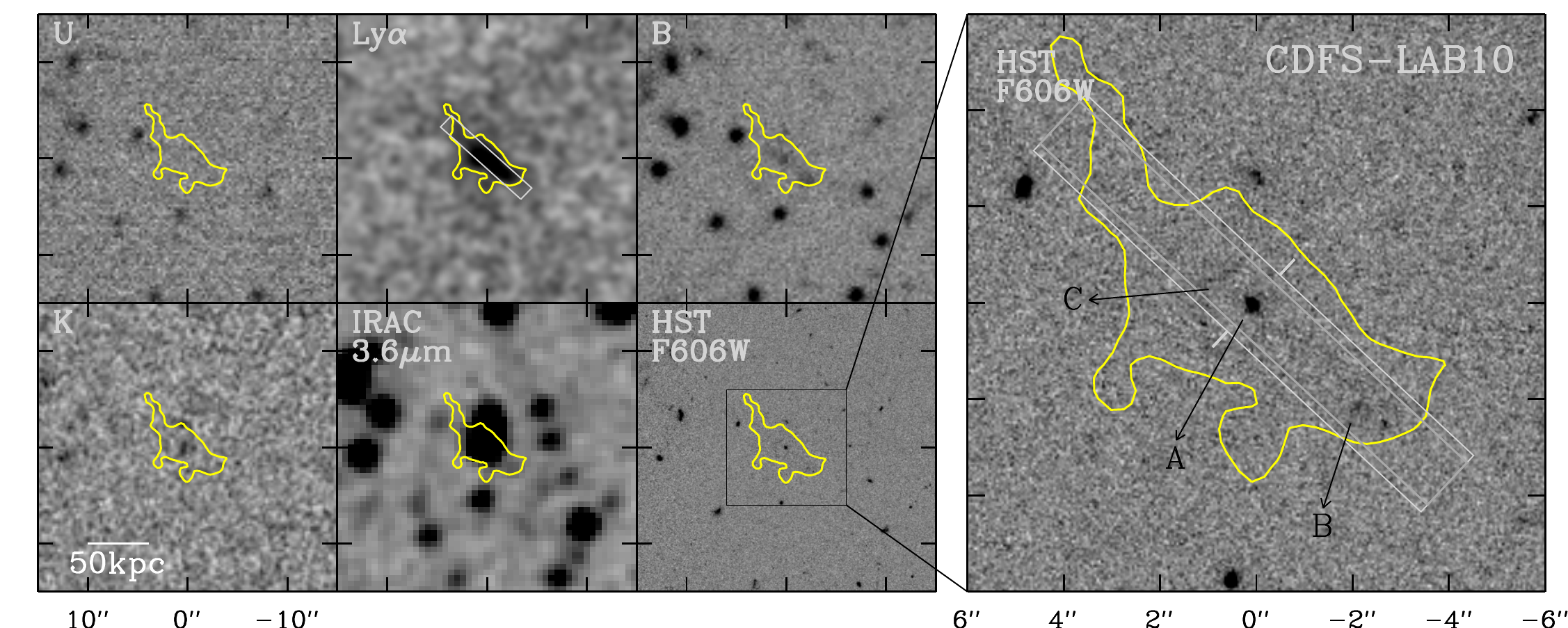} 
\caption{
Images of \lya blobs at various wavelengths: {\sl U}, continuum-subtracted
\lya line, {\sl B}, {\sl K}, {\it Spitzer} IRAC 3.6\micron, and {\sl
HST} F606W images.  Contours represent a surface brightness of 4 and
20$\times$10$^{-18}$ \unitcgssb\ from CTIO-4m narrowband imaging. Ticks
indicate 10\arcsec\ (82 physical kpc) intervals.  Two gray boxes indicate
the X-shooter slit positions of the UVB (1\farcs6) and NIR (1\farcs2)
arms (12\arcsec-longslit), respectively. Center of a slit is marked
with small outward ticks.  X-shooter provides the unique capability of
targeting the optical and NIR diagnostic lines simultaneously.
}
\label{fig:image}
\end{figure*}

\begin{figure*}
\addtocounter{figure}{-1}
\epsscale{0.8}
\ifpreprint\epsscale{0.9}\fi
\plotone{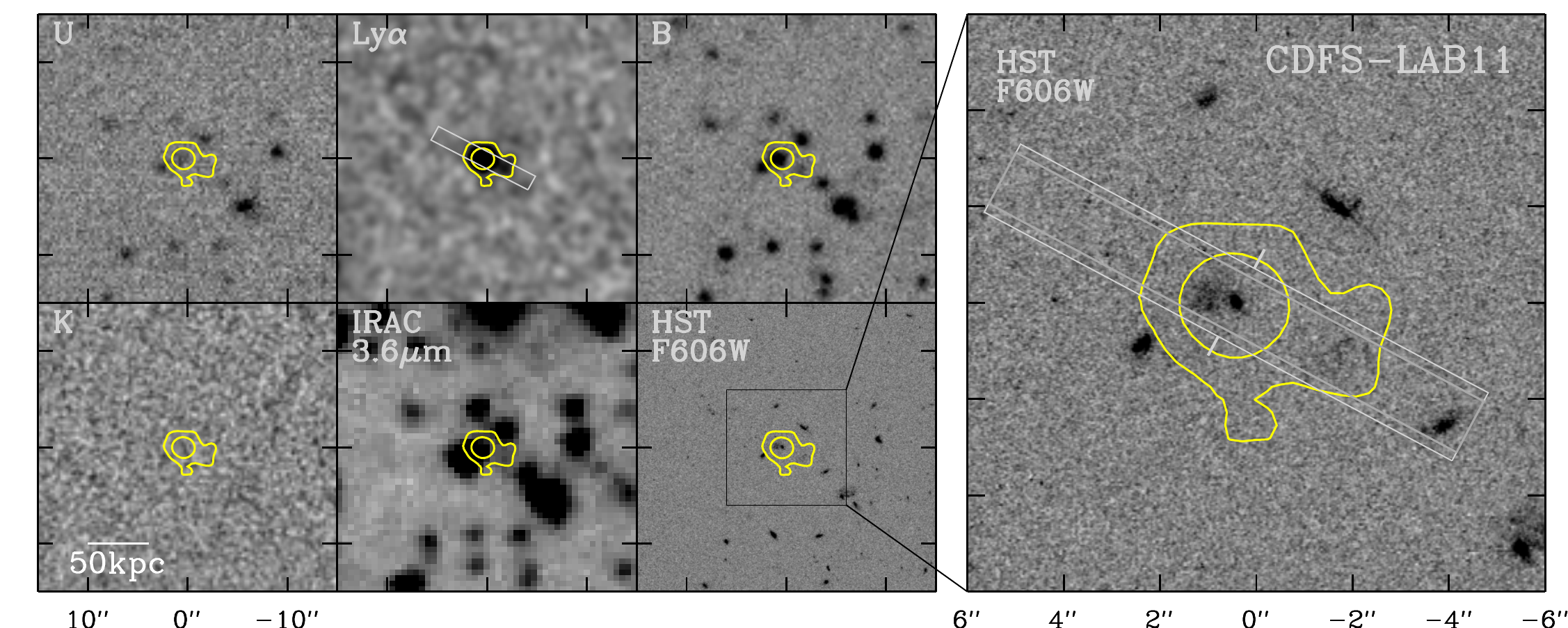} 
\plotone{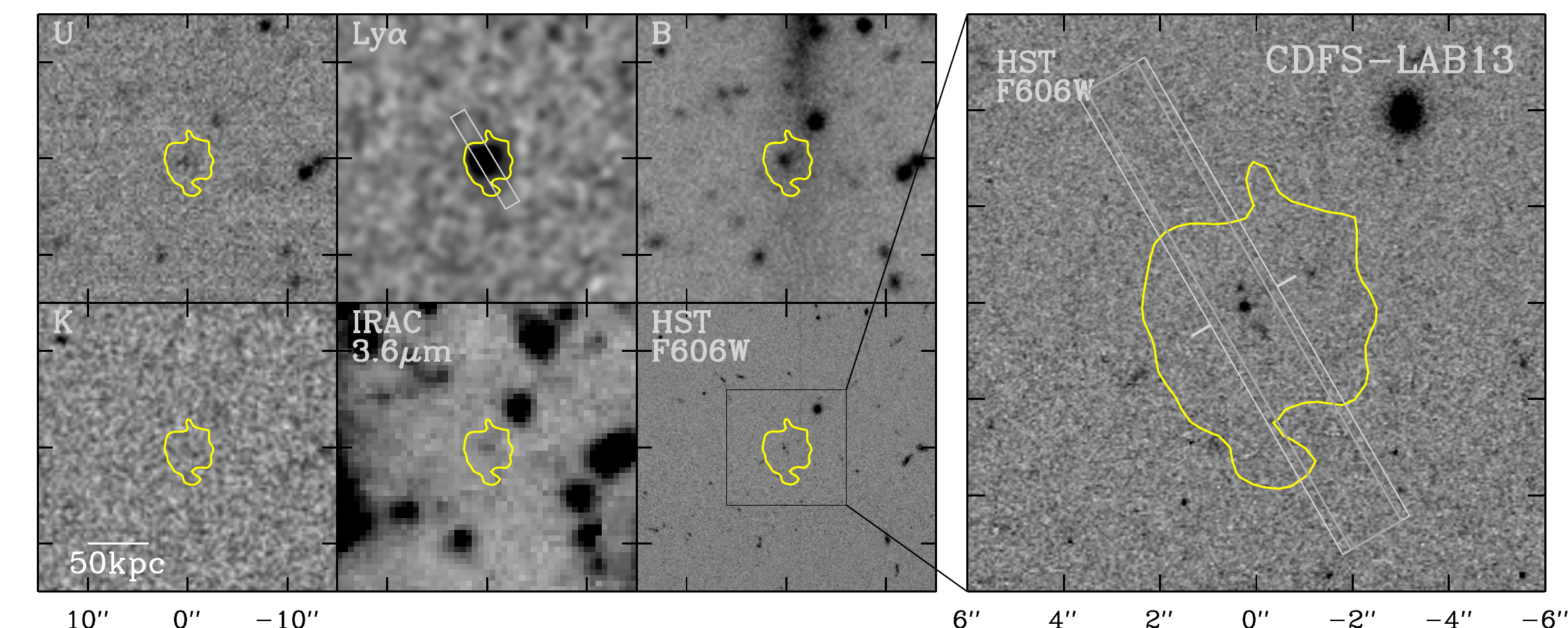} 
\plotone{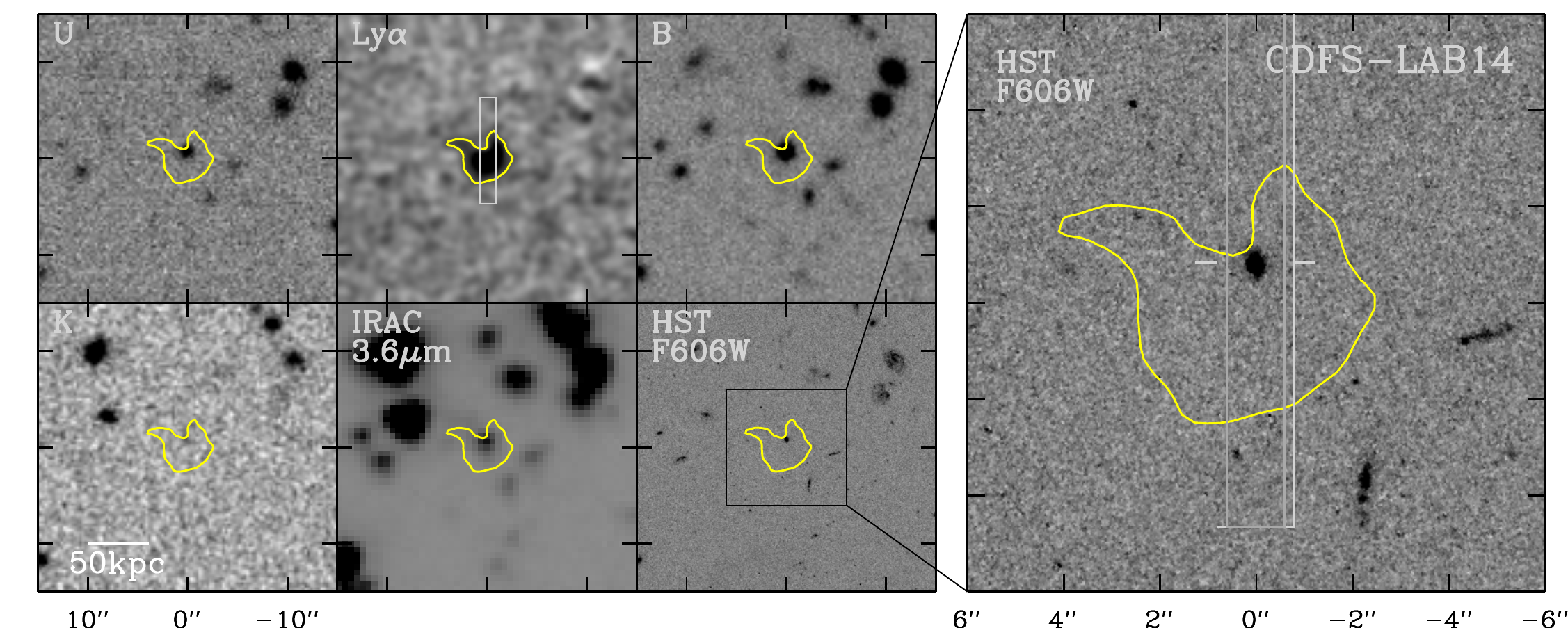} 
\caption{Continued.}
\end{figure*}


\subsection{UV-to-NIR Spectroscopy}

We obtained high resolution optical--NIR (3000\AA -- 2.5\um) spectra of
the six \lya blobs using X-shooter, a single object echelle spectrograph
\citep{Vernet11}, on the VLT UT2 telescope in service mode between 2010
November 6 and 2011 January 28.  
In Table \ref{tab:obslog}, we summarize the X-shooter observations.
In Figure \ref{fig:image}, we show the location of the spectrograph slit
on the sky, which was placed on UV-brightest galaxy or galaxies embedded
at or near the \lya blob center.  These galaxies or galaxy fragments
also lie in the region of brightest \lya emission.  Later, we assume
that their redshifts mark the systemic redshift of the \lya blob.

X-shooter consists of three arms (UVB, VIS, NIR), which covers the
spectral ranges of 3000\AA--5500\AA, 5500\AA--1\um, and 1\um--2.5\um,
respectively.  This enormous spectral coverage allows us to obtain
both \lya and \ha lines with a {\it single} exposure in contrast to
our previous approach \citep{Yang11} involving both optical and NIR
spectrograph.  Furthermore, X-shooter can detect at least one of the
nebular lines ([\ion{O}{2}], \hb, [\ion{O}{3}], H$\alpha$),
which will provide the systemic velocity of the embedded galaxies.
Because at the redshift of our targets ($z \simeq 2.3$) most of the
emission lines of interest (\lya, \civ, \heii, [\ion{O}{2}], [\ion{O}{3}],
H$\alpha$) are located in the UVB or NIR arms, we focus only on the UVB
($\lambda_{\rm rest}$ = 900\AA--1660\AA) and NIR ($\lambda_{\rm rest}$
= 3020\AA--7500\AA) part of spectra in this paper.

The observations were carried out over eight nights and 14 observing
blocks (OBs) of one hour duration each. The sky condition was either
clear or photometric, and the guide camera seeing ranged from 0\farcs6
to 1\farcs2 with a median of 0\farcs8 depending on the OB.
We adopted 1\farcs6 and 1\farcs2-wide slits for UVB and NIR, yielding a
spectral resolution of $R$ $\sim$ 3300 and $R\sim3900$, respectively. The
slit length is rather small (12\arcsec) compared to typical long-slits.
In each OB, we placed the slit on a target using a blind offset from
a nearby star.  Using acquisition images taken after the blind-offset,
we estimate that the telescope pointing and position angle of the slit
are accurate within 0.2\arcsec\ and 0.5\degr\ on average, respectively.
The individual exposure times were 680s and 240s for UVB and NIR,
respectively, and the telescope was nodded along the slit by $\pm$
2\arcsec\ while keeping the science targets always on the slit but at
different detector positions.  Total exposure times were 0.8 -- 3.2 hours
depending on the targets. In general, we were always able to detect both
\lya and at least one of the optical nebular lines within one OB.

For accurate wavelength calibration, we took ThAr lamp frames through
pinhole mask right before the science exposures and at the same telescope
pointing to compensate for the effect of instrument flexure.  For the UVB
arm, we obtained another ThAr arc frame at the end of each OB in order
to verify the wavelength solution where no bright sky lines are available.
Telluric standard stars (B--type) were taken after or before the science
targets with similar airmass to correct for atmospheric absorption
in the NIR.  Spectrophotometric standard stars were observed with
5\arcsec-wide slits once during the night as a part of the observatory's
baseline calibration plan.

\subsection{Data Reduction}
\label{sec:reduction}

We reduce the data using the ESO X-shooter pipeline (version 1.3.7).
In the UVB arm, the frames are overscan-corrected, bias-subtracted,
and flat-fielded with halogen and deuterium lamps. The sky background
is then subtracted in the ``stare''-mode of the pipeline by modeling
the sky in 2--D as described in \citet{Kelson03}.
In the NIR arm, dark current and sky background are removed from each
science frame by subtracting the dithered ``sky'' frame (``nodding''-mode
of the pipeline). Then, we flat-field the data and correct for cosmic-ray
hits and bad pixels.  In both arms, these flat-fielded, sky-subtracted
frames were corrected for the spatial distortions using multi-pinhole
arc frames.

Because we will compare the velocity centers of the \lya and H$\alpha$
lines, we carefully verify the wavelength calibration.  In the
NIR, we compare the wavelength solutions obtained from the OH sky
lines in the science frames to those from the daytime arc lamps and
flexure-compensation frames, i.e., the pipeline solutions. In the UVB,
we also compare the wavelength solutions obtained from the attached ThAr
arc frames with the pipeline solutions.  Our wavelength calibration is
accurate within $\sim$4\kms in both arms.  Furthermore, in the cases
where we visited sources multiple times, all spectra agree each other.
These frames are then rectified (resampled) and combined to create 2--D
spectra. 
We collapse the 2--D spectra in the wavelength direction to measure the
spatial extent of each emission line. Then we extract 1--D spectra from
[$-2\sigma$, $+2\sigma$] apertures, where the $\sigma$ is the Gaussian
width of the spatial profile. The aperture sizes are 2\arcsec -- 3\farcs5
in the UVB and 1\farcs5 -- 2\farcs5 in the NIR depending on the target.
Finally, the 1--D spectra are corrected to heliocentric velocities and
transformed to the vacuum wavelength.

\section{Results}
\label{sec:result}

\subsection{Systemic Redshift from {\rm\oiii} and {\rm\ha}}
\label{sec:redshift}

\begin{figure*}
\epsscale{1.15}
\ifpreprint\epsscale{0.95}\fi
\ifpreprint\vspace{-0.8cm}\fi
\begin{center}
\plotone{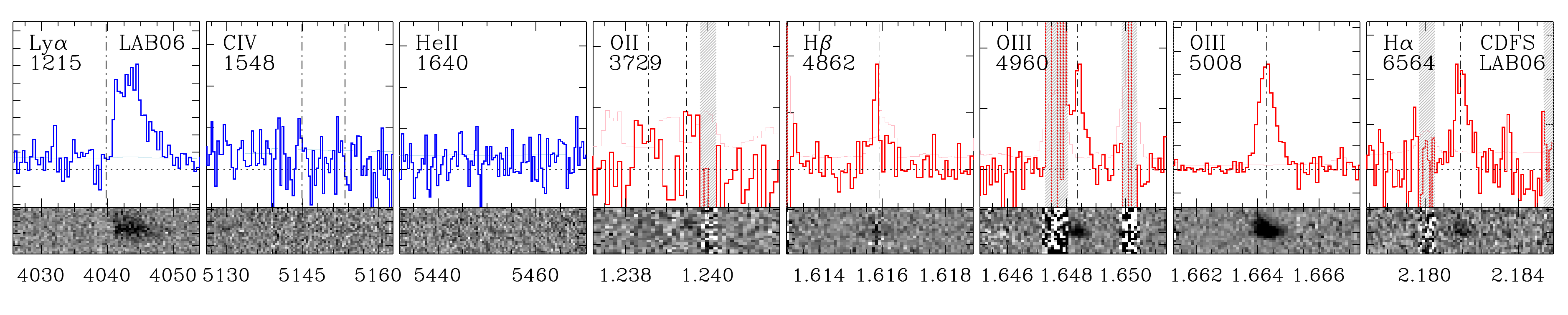}\\[-1.8ex]
\plotone{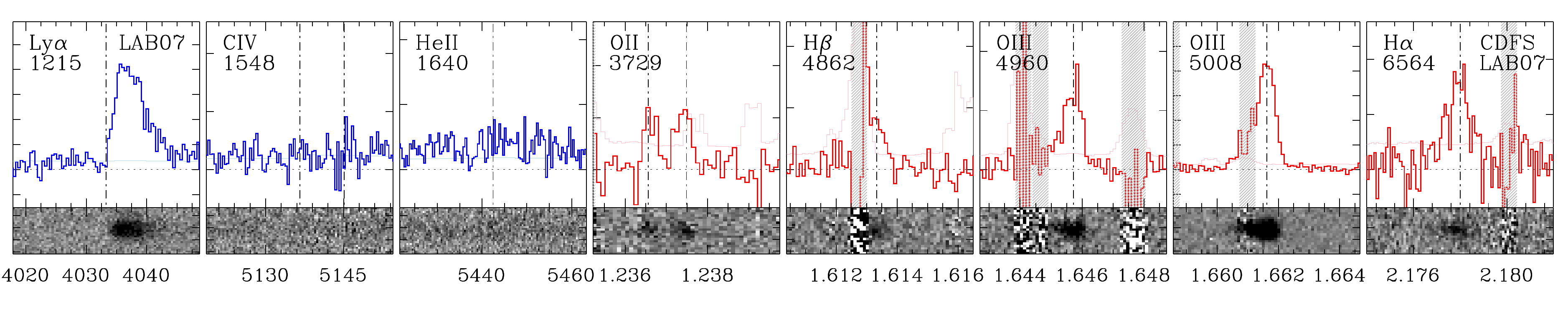}\\[-1.8ex]
\plotone{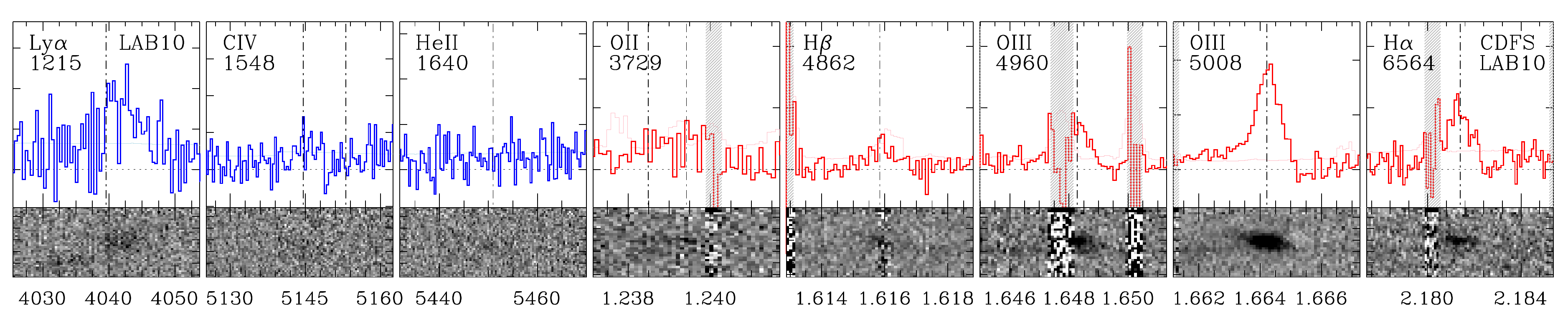}\\[-0.8ex]
\plotone{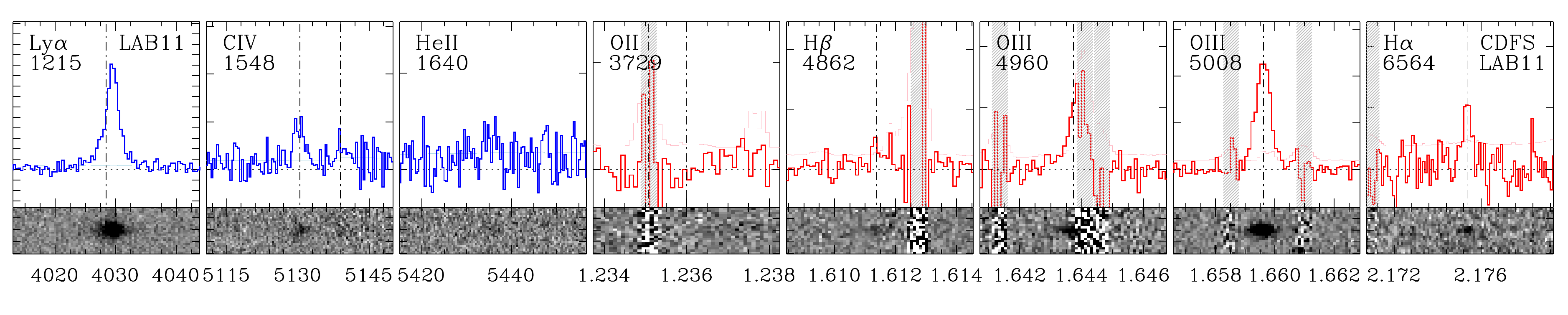}\\[-1.8ex]
\plotone{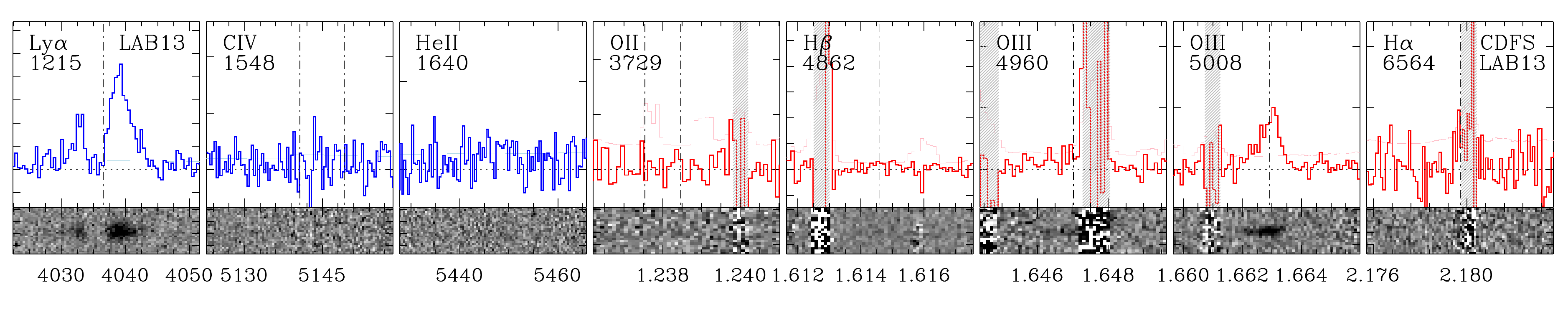}\\[-1.8ex]
\plotone{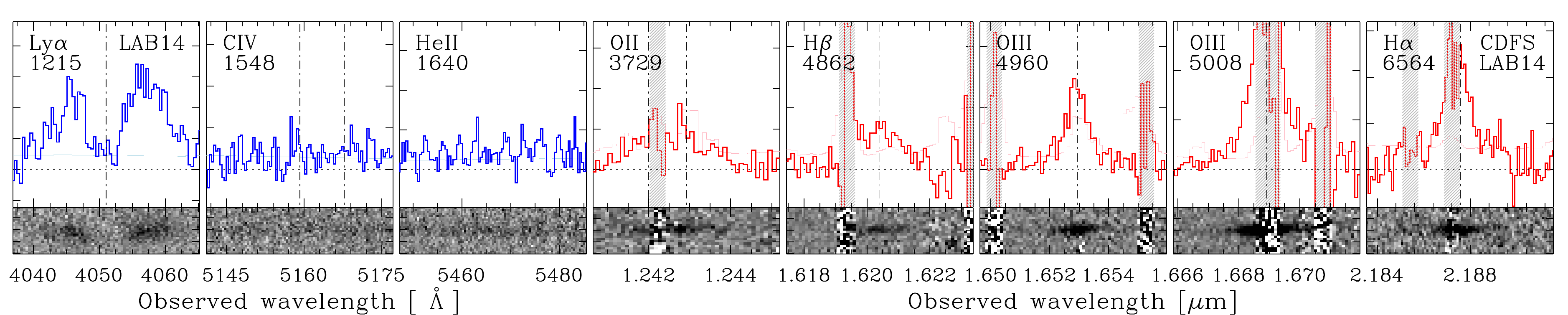}\\[-1.8ex]
\ifpreprint\vspace{-0.3cm}\fi
\end{center}
\caption{
\ifpreprint\small\fi
1--D integrated line profiles for the rest-frame UV (\lya, \civ,
\heii) and optical (\oii, \hb, \oiii, \ha) lines detected in the
sightline toward the central, embedded galaxies.  The velocity ranges
are $\pm$1000\,\kms and $\pm$550\,\kms for the UV and optical spectra,
respectively.  The $y$-axis ticks are spaced every 1\E{-17} and 0.5\E{-17}
\unitcgsflam for the rest-frame UV and optical spectra, respectively.
At the bottom of each spectrum, we show the 2-D spectra with 4\arcsec
or 6\arcsec width.  The vertical lines indicate the systemic velocity
determined from the \oiii\ line. The other \oii, \ha and \hb lines
are consistent with the \oiii\ center, suggesting that all provide a
good reference to the center of the \lya blob defined by the galaxy or
galaxies embedded in the blob core.  In all cases, the  \lya lines are
broader than the H$\alpha$ or \oiii lines, implying that the \lya is
resonantly scattered by an optically thick medium.  In CDFS-LAB11, the
most compact of the sources, which is also detected in \civ and \heii,
the \lya line is sharply peaked with broad wings.
}
\label{fig:spec1d}
\end{figure*}

Various emission lines in the UVB and NIR arms confirm that \lya blobs
lie at the survey redshift, $z\sim2.3$.  In addition to \lya and \ha,
we cover other UV emission lines and the non-resonant \oiilamb,
\oiiilamb, and \hblamb lines.  In Figure \ref{fig:spec1d}, we show 1--D
and 2--D spectra of the six \lya blobs.  The first three columns show
the rest-frame UV emission lines (\lya, \civ, \heii) from the X-shooter
UVB arm, and the remaining columns show the rest-frame optical nebular
emission lines (\oii, \hb, \oiii, \ha) from the NIR arm.

Among the rest-frame optical nebular lines,  the \oiii\ line is the
brightest and detected with highest signal-to-noise (S/N) ratio in
all cases, partly due to the low sky background and thermal instrument
background in {\sl H}-band.
We determine the systemic redshift using \oiii\ doublets. In one case
(CDFS-LAB14), the brighter \oiii\ line ($\lambda5007$) falls on top of
an OH sky line, thus making it impossible to determine the line center.
In this case we used the fainter \oiii\ line ($\lambda4959$).
The vertical dashed lines in Figure \ref{fig:spec1d} indicate the line
centers determined by \oiii\ lines, which are then overlayed on other
emission line profiles. As expected, we find that the line centers
of all non-resonant emission lines agree well each other, to within
$\sim$10\kms, showing that all of these lines are good indicators of
systemic velocity. Thus, the brightest \oiii\ line can serve as best
emission line to target for this survey redshift and instrument.

\subsection{1--D Ly$\alpha$ Profiles}
\label{sec:spec1d}

The Ly$\alpha$ profiles are significantly broad compared to the non-resonant 
lines (\ha and/or \oiii) in six cases out of the sample of eight
\lya blobs, including the two from our previous work \citep{Yang11}.
These integrated 1--D profile shapes range from an asymmetric
single-peaked profile (CDFS-LAB06, 07, 10), to a double-peaked profile
with a stronger red peak (CDFS-LAB02, 13), to a double-peaked profile
with two similar intensity peaks (CDFS-LAB14).
The \lya profiles of even this small sample show extremely diverse
morphologies consistent with simple radiative model predictions
\citep{Verhamme06,Verhamme08,Dijkstra06a} with varying geometry and
outflow velocities \cite[see also][]{Matsuda06,Saito08,Weijmans10}.

The remaining two \lya profiles are narrower {\it relative to} the
\oiii lines and show slightly extended wings (CDFS-LAB01 and 11). Note
that the \lya line width of CDFS-LAB01 is one of the largest among our
sample, but the blue side of its \lya profile agrees well with its \ha
profile \citep{Yang11}. While there is an underlying broad component
in CDFS-LAB11, the width of the dominant narrow component is small,
comparable to that of the \oiii line (see \S\ref{sec:type2}).
These are the two \lya blobs where \heii and \civ emission lines are
also detected indicating that they contain a hard ionizing source,
possibly an AGN.  If photo-ionization by an AGN is indeed responsible
for the \heii and \civ emission, and possibly the extended \lya emission
as well, the discovery of narrow \lya profiles suggests that the \lya
blob gas is highly-ionized, i.e.,  that the resonant scattering of \lya
is not effective enough to alter the profile significantly.  We will
further investigate the details of these \heii, \civ emission lines and
the implications for AGN in a future paper (Y.\,Yang in preparation).

The fraction of double-peaked profiles is significant: $\sim$38\% (3/8),
which is roughly consistent with the findings for LBGs and LAEs at $z$
= 2--3. Among LBGs, \citet{Kulas12} find that $\sim$30\% of LBGs with
\lya emission show multiple-peaked profiles.  \citet{Yamada12} also find
that $\sim$50\% of LAE's profiles have multiple peaks.

\begin{figure*}
\epsscale{0.95}
\ifpreprint\epsscale{0.71}\fi
\ifpreprint\vspace{-0.8cm}\fi
\plotone{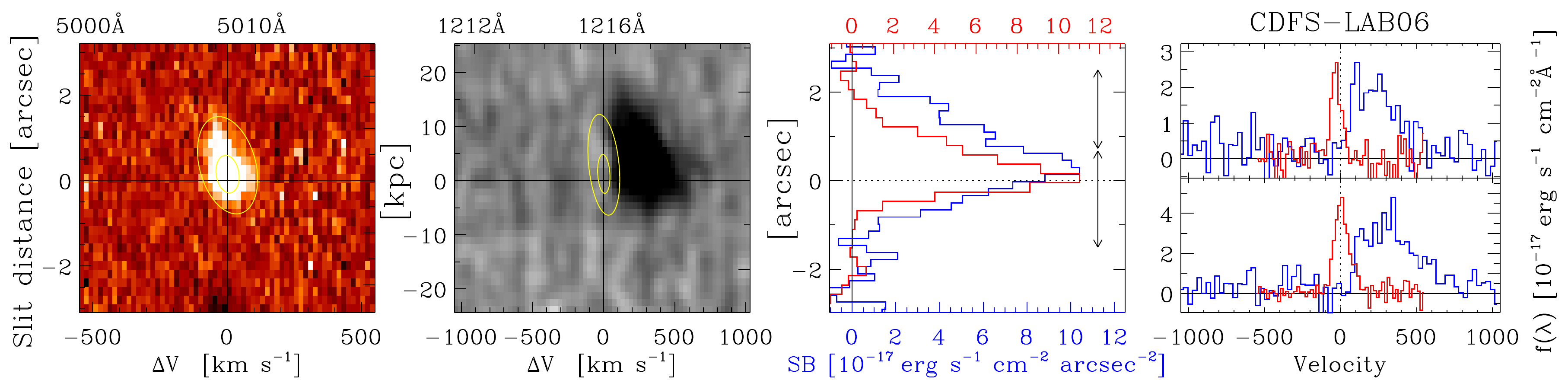}\\[-0.2ex]
\plotone{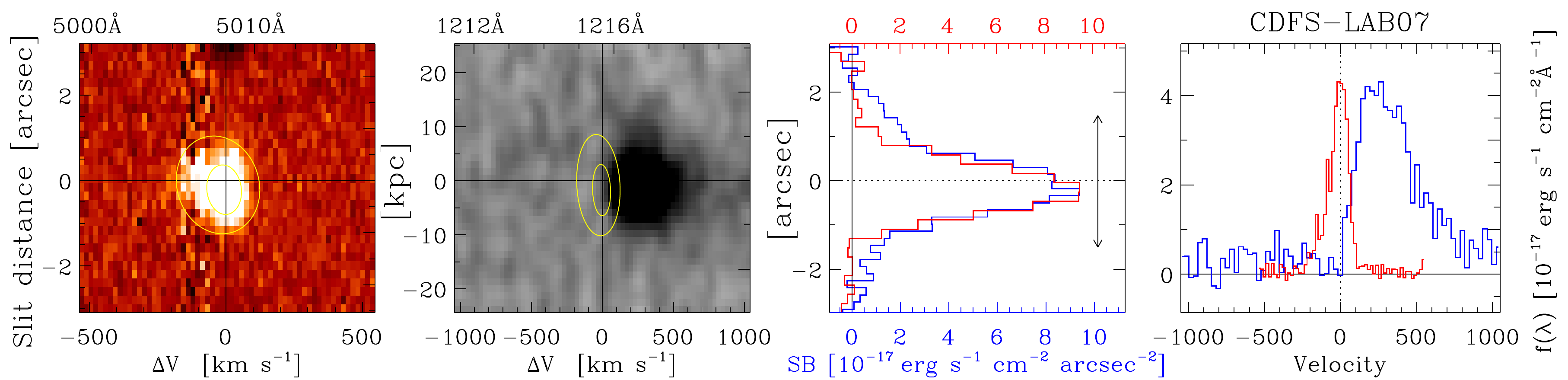}\\[-0.2ex]
\plotone{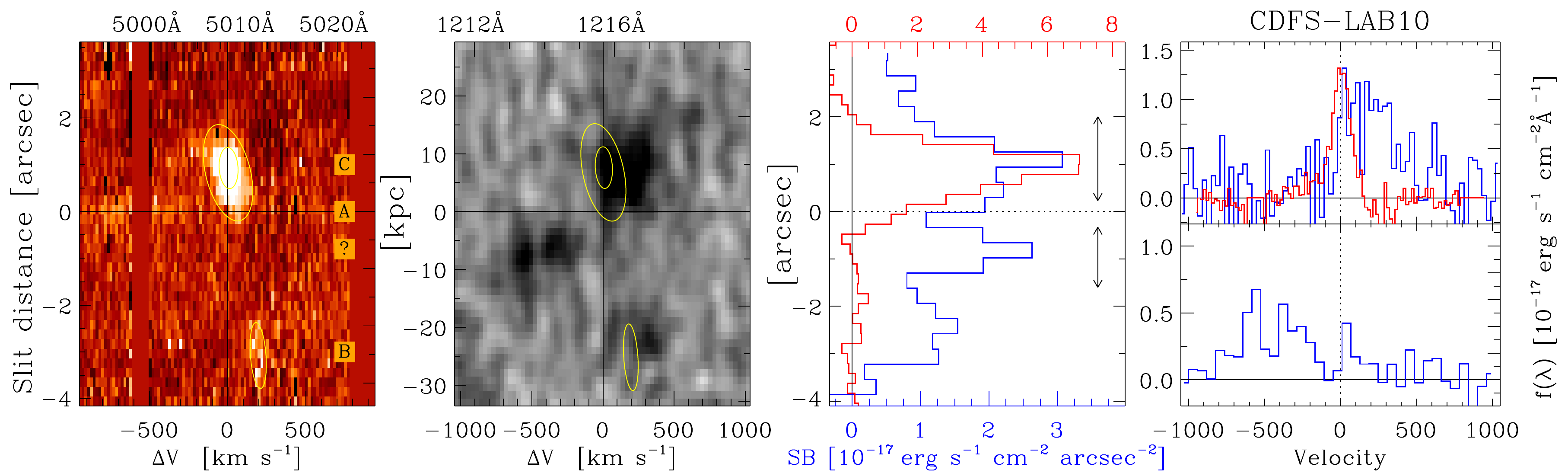}\\[-0.2ex]
\plotone{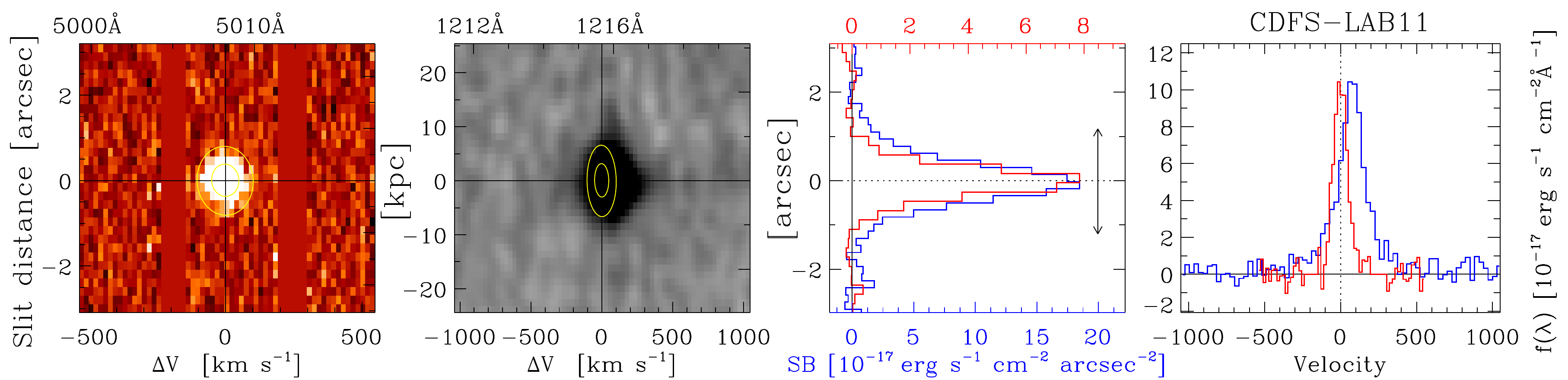}\\[-0.2ex]
\plotone{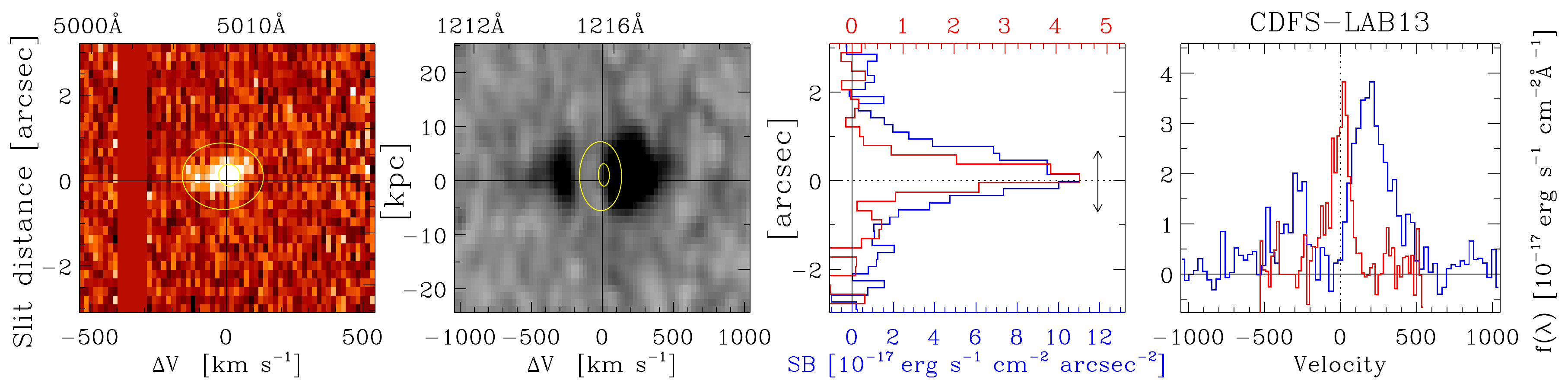}\\[-0.2ex]
\plotone{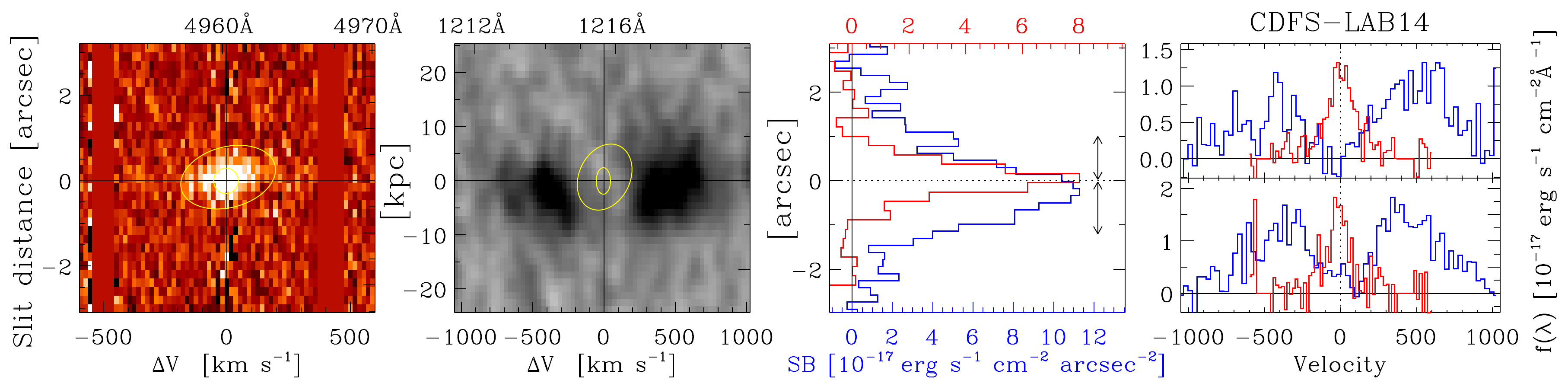}
\caption{
\ifpreprint\small\fi
2--D \oiii (first column) and \lya (second column) spectra for \lya
blobs at $z\sim2.31$.  The \lya spectra were smoothed to enhance low
surface brightness features.  Contours bounding the \oiii spectra are
overlayed on the \lya panels to aid comparison between the two lines.
The third column shows the spatial profile along the slits (collapsed
in the spectral direction). The slit profiles are normalized for easy
comparison, so the plots show different surface brightness scales on the
top and bottom axes as indicated by different colors. The last column
shows the extracted 1--D spectra from different regions defined by the
arrows in third column.  The \lya profiles are significantly broader
than the \oiii lines due to resonant scattering.  Spectra toward three
\lya blobs show spatially extended \lya spectra (CDFS-LAB06, 10, 14). In
CDFS-LAB06, both \oiii\ and \lya are extended due to a neighboring galaxy
to the north.  In the other two \lya blobs, \lya is more extended than
the compact \oiii line, implying that \lya is originating from the gas
outside the galaxies.
}
\label{fig:spec2d}
\end{figure*}


\subsection{2--D Ly$\alpha$ and {\rm[\ion{O}{3}]} Profiles}
\label{sec:spec2d}

Using 2--D spectra, we detect the extended \lya lines and identify the
exact locations from which the \lya or \oiii line originates.  In Figure
\ref{fig:spec2d}, we show close-ups of the 2--D \oiii and \lya profiles
in the first and second columns, respectively.  To aid the comparison
between the \oiii and \lya profiles, we overlay the rough boundary of the
\oiii spectrum on each \lya panel with ellipses.  The third column shows
the spatial profiles along the slit, i.e., collapsed in the wavelength
direction.  The last column shows the 1--D \lya and \oiii profiles that
are extracted from the different parts of the slit, as indicated with
vertical arrows in the third column.

In four cases (CDFS-LAB06, 10, 13, 14), the Ly$\alpha$ spectrum is
spatially extended relative to the \oiii line, confirming the narrowband
imaging result that the \lya-emitting gas extends beyond the embedded
galaxies that are probably responsible for the \oiii emission.  Note that
the X-shooter spectroscopy reaches much shallower \lya surface brightness
limit ($\sim$1.5--3\E{-17}\,\unitcgssb) than the narrowband imaging
($\sim$5\E{-18}\,\unitcgssb; 3\sig limit).

In two \lya blobs (CDFS-LAB06 and 10), the \oiii lines are spatially
resolved.  However,  there is no evidence that the \oiii or \ha lines
are extended beyond the UV continuum emission arising from stars in the
embedded galaxies seen in Figure 2.
Deeper NIR spectroscopic, preferentially IFU, observations are required
to better define the spatial extent of these lines and to measure the
spatially-resolved gas kinematics.  Throughout the paper, we assume that
the \oiii and \ha lines originate from the central embedded galaxies,
not from the extended \lya-emitting gas, and thus that their line centers
represent the systemic velocity of the \lya blob.  In the next section,
we briefly describe individual systems in detail.

\subsection{Notes for Individual Objects}
\label{sec:individual}

\subsubsection{CDFS-LAB06}

CDFS-LAB06 has two rest-frame UV sources with small separation (0\farcs8;
6.5\,kpc) in the {\sl HST} image (Fig.~\ref{fig:image}). Both components
(or clumps) were placed in the X-shooter slit and are detected in \oiii\
and \lya (Fig.~\ref{fig:spec2d}). These two sources are separated by only
$\sim$50\kms in velocity space, thus it is not clear whether they belong
to one galaxy or are interacting with each other.  We adopt the
redshift of the brighter UV source as the systemic velocity of CDFS-LAB06.
The \lya emission detected in the spectrum (between $\Delta\theta$ = $-1$ and
$+2$\arcsec) is spatially extended due to the other galaxy, so it
does not represent the IGM or circum-galactic medium (CGM).

\subsubsection{CDFS-LAB07}

The galaxy within CDFS-LAB07 has a bar-like morphology in the {\sl HST}
image with which we align the slit. This galaxy or galaxy fragments
were marginally resolved in the \oiii emission line.  All optical
nebular emission lines (\oii, \hb, \oiii, and \ha) are detected. Faint
UV continuum emission is marginally detected, allowing us to study the
gas kinematics with metal absorption lines (\S\ref{sec:absorption}).
The \oiii and \ha lines show asymmetric profiles extending toward
the blue. This profile can be fitted with a narrow Gaussian component
at the velocity center superposed on the blueshifted broad component
(\S\ref{sec:O3profile}).

\subsubsection{CDFS-LAB10}

CDFS-LAB10 is the most puzzling and complex source in the X-shooter
sample.  The \lya emission in the narrow-band image is elongated over
10\arcsec\ ($\sim$80\,kpc).  In the NIR spectrum, three sources are
detected: two with \oiii emission lines (galaxies B and C), and the other
(galaxy A) with very faint continuum (Fig.~\ref{fig:spec2d}).  This NIR
continuum source (galaxy A) is located at the slit center, while the
strongest \oiii-emitting source (galaxy C) is offset by $\sim$1\arcsec\
toward north-east from the center and barely detected in the {\sl HST}
image. The 2--D \oiii spectrum of galaxy C shows a velocity shear
indicative of rotating disk. In the 2--D \lya spectrum, there are also
three distinct components. Although the 1--D \lya spectrum of the entire
blob looks single-peaked with a broad linewidth, it is in fact composed
of these three components.  We will investigate these various emission
line and continuum sources in \S\ref{sec:cdfs-lab10}.

\subsubsection{CDFS-LAB11}

There are one compact UV source at the slit center and a diffuse emission
toward northeast in the {\sl HST} image.  It appears that the \oiii
emission originates from the central compact source.  Unlike commonly
observed broad \lya profiles found in \lya blobs and emitters, CDFS-LAB11
has a peculiar \lya profile that is almost symmetric and narrow.
As will be discussed in \S\ref{sec:type2}, both narrow and underlying
broad components are required to explain the \lya profile.  The \lya is
redshifted against \ha by small amount: \dvlya = 84$\pm$6\kms. Both \oiii
and \lya are spatially compact (Fig.~\ref{fig:spec2d}).  Both \civlamb\
and \helamb\ emission lines are also detected (Fig.~\ref{fig:spec1d}),
implying the presence of hard ionizing source.  A total of two \lya blobs
from our ECDFS sample (CDFS-LAB01 and 11; 2/8) show these emission lines.
Note that \civ and {\it narrow} \heii line emission are commonly detected
in bright \lya blobs \citep{Dey05, Scarlata09, Prescott09, Yang11},
while the nature of the hard-ionizing source is unknown.

\begin{figure}
\epsscale{1.10}
\ifpreprint\epsscale{0.65}\fi
\plotone{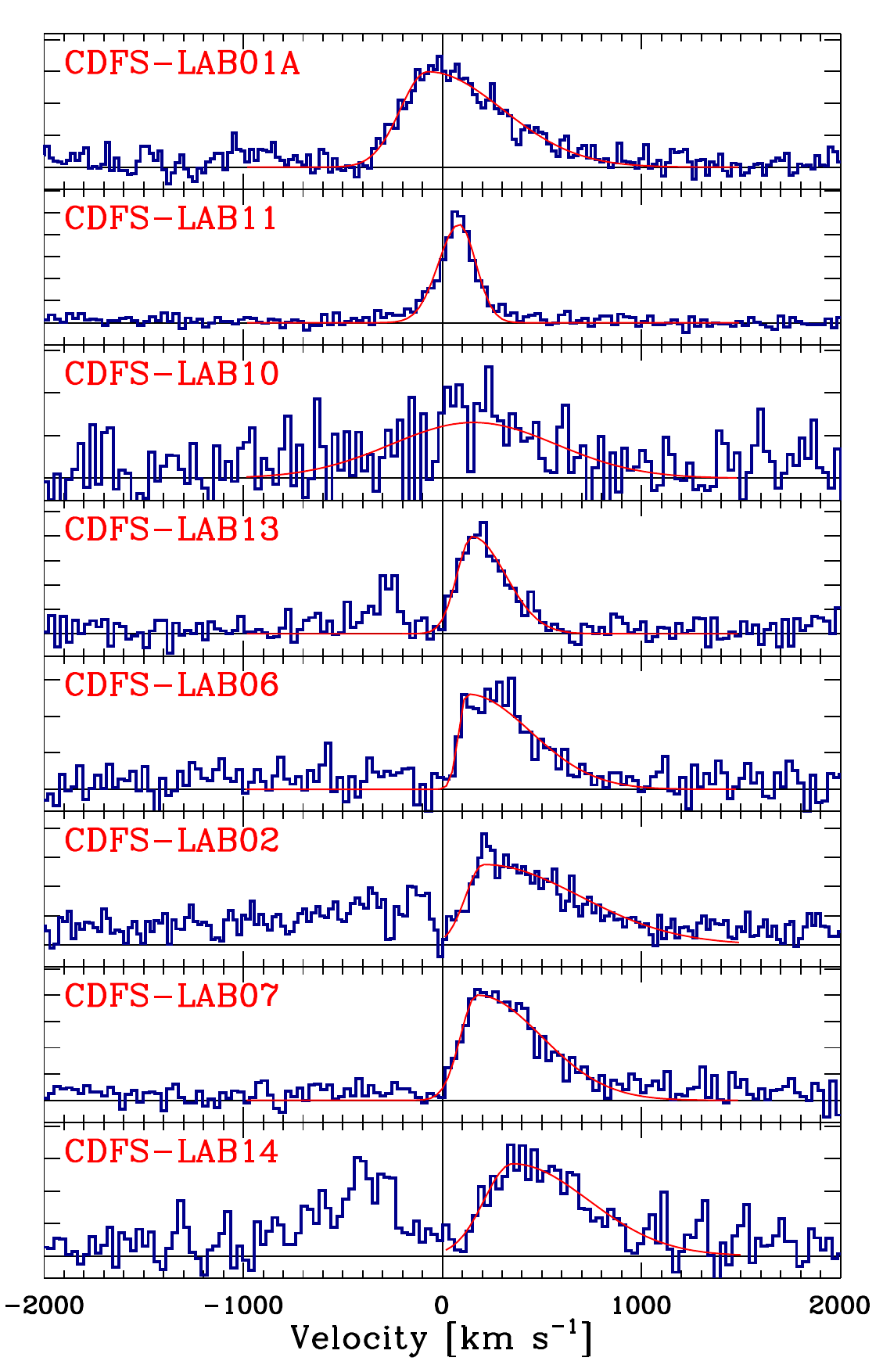}
\caption{
Comparison of \lya line profiles (blue histogram), centered on the \ha
or \oiii line arising from the embedded galaxy.  The \lya profiles are
broad and diverse, ranging from symmetrically double-peaked (CDFS-LAB14),
to double-peaked, but dominated by the red peak (CDFS-LAB02 and 13),
to a single red peak (CDFS-LAB06, 07, 10), to almost symmetric profiles
(CDFS-LAB01 and 11).  The profiles are consistent with a single family
of objects explained by a simple RT model in which the gas along the
line-of-sight to the blob center is static (double-peaked with similar
intensity), outflowing in a spherical shell from the center (red peak
dominated), or outflowing and relatively optically thin (symmetric). The
thin red lines indicate the fits with asymmetric Gaussian profiles used
to determine the velocity offset between \lya and either \ha or \oiii,
\dvlya (see \S\ref{sec:shift}).
}
\label{fig:line_comparison}
\end{figure}


\subsubsection{CDFS-LAB13}

CDFS-LAB13 has a double-peaked profile with a stronger red peak, and
asymmetric \oiii profile (\S\ref{sec:O3profile}). \lya in CDFS-LAB13
is likely more extended, but the low S/N of the \oiii line makes the
comparison difficult. There are multiple galaxy fragments in the {\sl
HST} UV continuum image, but they were not spatially resolved in the
X-shooter observations.

\subsubsection{CDFS-LAB14}

In the {\sl HST} and narrowband images, a UV source is located at the
upper boundary of the \lya emission contours.  In the X-shooter 2--D
spectra (Figure \ref{fig:spec2d}), the \oiii line is centered on this UV
continuum source and \lya is more extended toward the south in agreement
with narrowband imaging.  CDFS-LAB14 is one of two cases where extended
\lya emission is well detected in the spectroscopy.
The \lya profile has two peaks with similar intensities.  The peak
separation narrows as the slit distance from the central galaxy
increases.  Faint UV continuum is also detected at the location of the
UV source allowing us to measure outflow speed from metal absorption
lines (\S\ref{sec:absorption}).  Combining the \lya and the absorption
profiles, we will put constraints on the gas kinematics and the neutral
column density of this system (\S\ref{sec:column_density}).  \oiii has
a broad wing on top of the narrow component (\S\ref{sec:O3profile})

\subsection{Gas Kinematics}
\label{sec:kinematics}

In this section, we constrain the gas kinematics in the \lya blobs using
three different techniques and compare those results.
First, for the six blobs in the X-shooter survey, we compare the optically
thick \lya and non-resonant \oiii line (either $\lambda5007$
or $\lambda4959$ in the case of CDFS-LAB14) to measure \lya velocity
offsets from the systemic velocity of the \lya blobs.  For the two \lya
blobs from our previous work (CDFS-LAB01A and 02; \cite{Yang11}), where
an \oiii line is unavailable, we use \ha.  As mentioned previously,
the line centers of both \oiii lines and \ha are all consistent.
Second, we constrain the outflow speed from the interstellar metal
absorption lines in three galaxies where we are able to detect the
rest-frame UV continuum in the spectrum.
Lastly, we present a new tracer of kinematics in four \lya blobs:
characterizing the breadth and asymmetry of the \oiii line profile.
Note that while we detect an asymmetric wing in some \oiii line profiles,
it does not affect the line centroid, which is essential in determining
the systemic velocity in the first technique above.

\subsubsection{\lya\ -- {\rm \oiii} offset}
\label{sec:shift}

We compare the peak of each \lya profile with the center of a non-resonant
nebular emission line, particularly \oiii.  Throughout the paper,
the \lya--\ha and \lya--\oiii offsets are used interchangeably to
represent the \lya offset from the systemic velocity: \dvlya.  Figure
\ref{fig:line_comparison} shows eight \lya profiles from this work and
from \citet{Yang11}, plotted with increasing \dvlya.

Because \lya spectra are somewhat noisy because of the small spectral
dispersion (20--30\kms\ per pixel), we measure \dvlya\ using the
following two methods.  First, we measure the velocity of the \lya peak
after smoothing the spectra with a Gaussian filter with FWHM = 90\kms,
corresponding to our velocity resolution.  Second, we fit the red peaks
with an asymmetric Gaussian function, which consists of two Gaussian
functions with different FWHMs being joined at the center.  The two
measurements agree to within $\sim$50\kms, except for CDFS-LAB06 where
the second method gives \dvlya\ = 120\kms compared to 320\kms from simple
smoothing due to the very sharp edge at the blue side of \lya profile.
We adopt the second measurements in this paper and show these fits in
Figure \ref{fig:line_comparison} and Table \ref{tab:Lyman}.  None of
the conclusions in this paper are affected by this choice.

\begin{figure}
\epsscale{1.18}
\ifpreprint\epsscale{0.90}\fi
\plotone{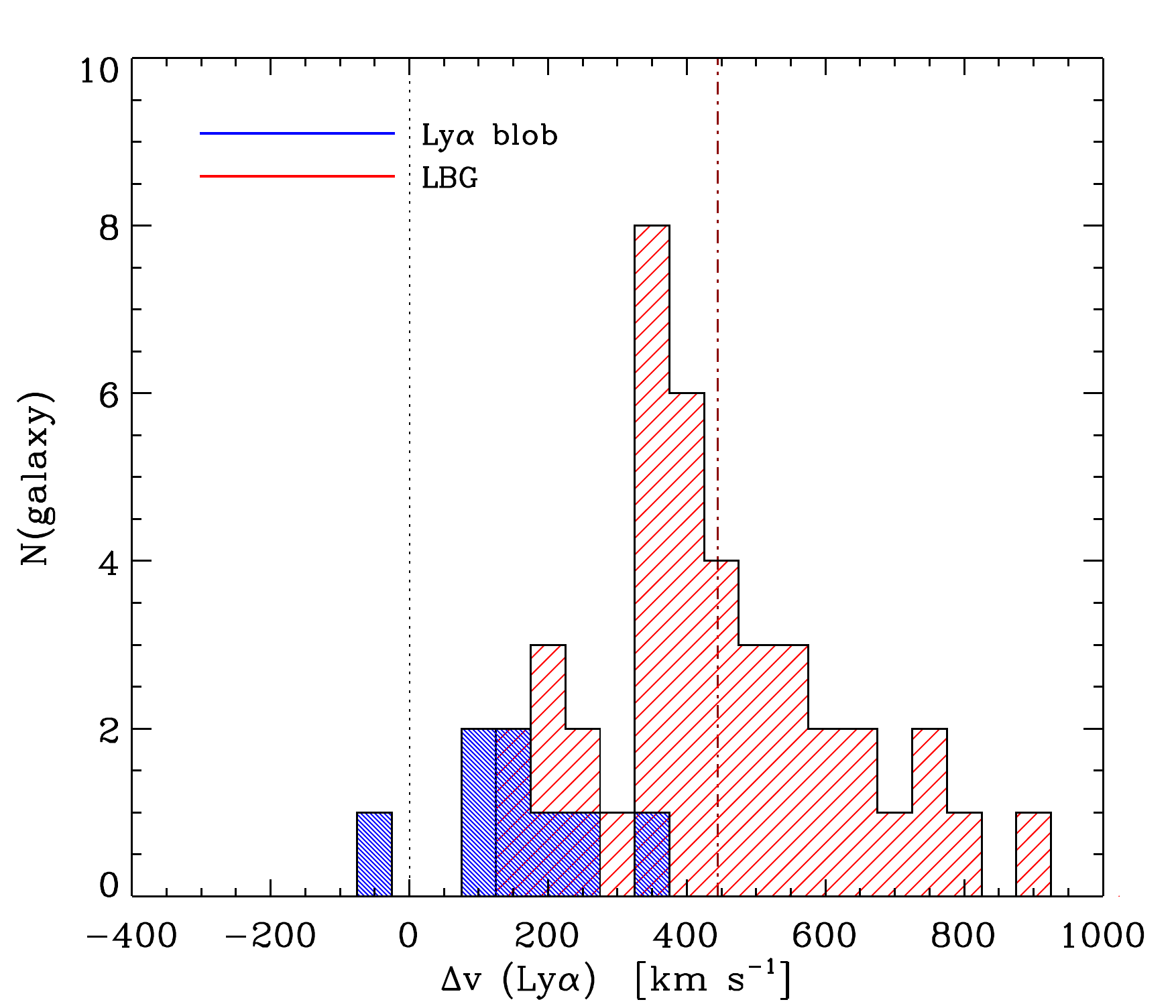}
\caption{
Distribution of \dvlya\ for eight \lya blob galaxies in comparison
to those of 41 LBGs \citep{Steidel10}.  The vertical dot-dashed line
indicates the average of LBG sample, $\avg{\dvlya}$ = 455\kms.  The \dvlya
offsets for blob galaxies are generally smaller than those of LBGs.
If \dvlya\ is a proxy for the outflow velocity of neutral material, the
outflow speeds in \lya blobs are smaller than for typical star-forming
galaxies at $z=2-3$.
}
\label{fig:voffset}
\end{figure}


Figure \ref{fig:voffset} shows the distribution of \dvlya\ of the eight
\lya blob galaxies in comparison with 41 LBGs that have both H$\alpha$
and \lya spectra \citep{Steidel10}.  Because the spectral resolutions of
our \lya spectra obtained from VLT/X-shooter and Magellan/MagE are higher
than those of the \lya profiles of the LBGs (FWHM $\simeq$ 370\kms),
we test if the different spectral resolutions affect the \lya blob--LBG
comparison.  We repeat the measurement of \dvlya\ after convolving our
\lya profiles to the spectral resolution of LBG sample.
The \dvlya\ values change by a negligible amount (only $\pm$50\kms)
except for CDFS-LAB02, where \dvlya\ increases by +100\kms because of its
sharp red peak.  Therefore, we conclude that the \dvlya\ distributions
of \lya blobs and LBGs can be compared directly.

The  \dvlya\ distribution for galaxies within our \lya blobs reveals
smaller velocity offsets than typical of LBGs, confirming previous
claims \cite[][see also McLinden et al.~2013]{Yang11}.  Galaxies within
blobs have \dvlya = $-$60 $\rightarrow$ +400\kms\ with an average of
$\langle\dvlya\rangle$ = 160\kms, while LBGs at similar redshifts have
\dvlya = 250 -- 900\kms\ with $\langle\dvlya\rangle$ = 445\kms.

The remaining question is how to interpret these small \dvlya\ values. Two
possibilities are:
(1) \dvlya\ is a proxy for the gas outflow velocity (\vexp), suggesting that 
the outflows here are weaker than in other
star-forming galaxies at $z$ = 2--3 \citep{Verhamme06}, or
(2) there is less neutral gas close to the systemic velocity of the
embedded galaxies \citep{Steidel10}, and \dvlya\ is independent of the outflow speed.
We discussed the caveats associated with interpreting \dvlya\ in
\citet{Yang11} and revisit this issue in \S\ref{sec:dvlya}.

In the direction of the embedded galaxies, we do not find any \lya profile
that is blue-peak dominated, i.e., \dvlya $<$ 0\,\kms.  Therefore, there
is no evidence along these lines-of-sight for infalling gas (but see
\S\ref{sec:cdfs-lab10}).  While the statistics are still small, we place
an upper limit on the covering factor of infalling gas (if any) detectable
with the \dvlya\ technique.  Here the covering factor of any inflowing
streams like those predicted by cold-mode accretion \citep{Keres05,
Keres09, Dekel09} must be less than $\sim$13\% (1/8). We further discuss
the covering factor of cold streams in \S\ref{sec:covering_factor}.
\vspace{1cm}

\subsubsection{Interstellar Metal Absorption Lines}
\label{sec:absorption}

\begin{figure*}
\epsscale{1.00}
\begin{center}
\includegraphics[ width=0.80\textwidth]{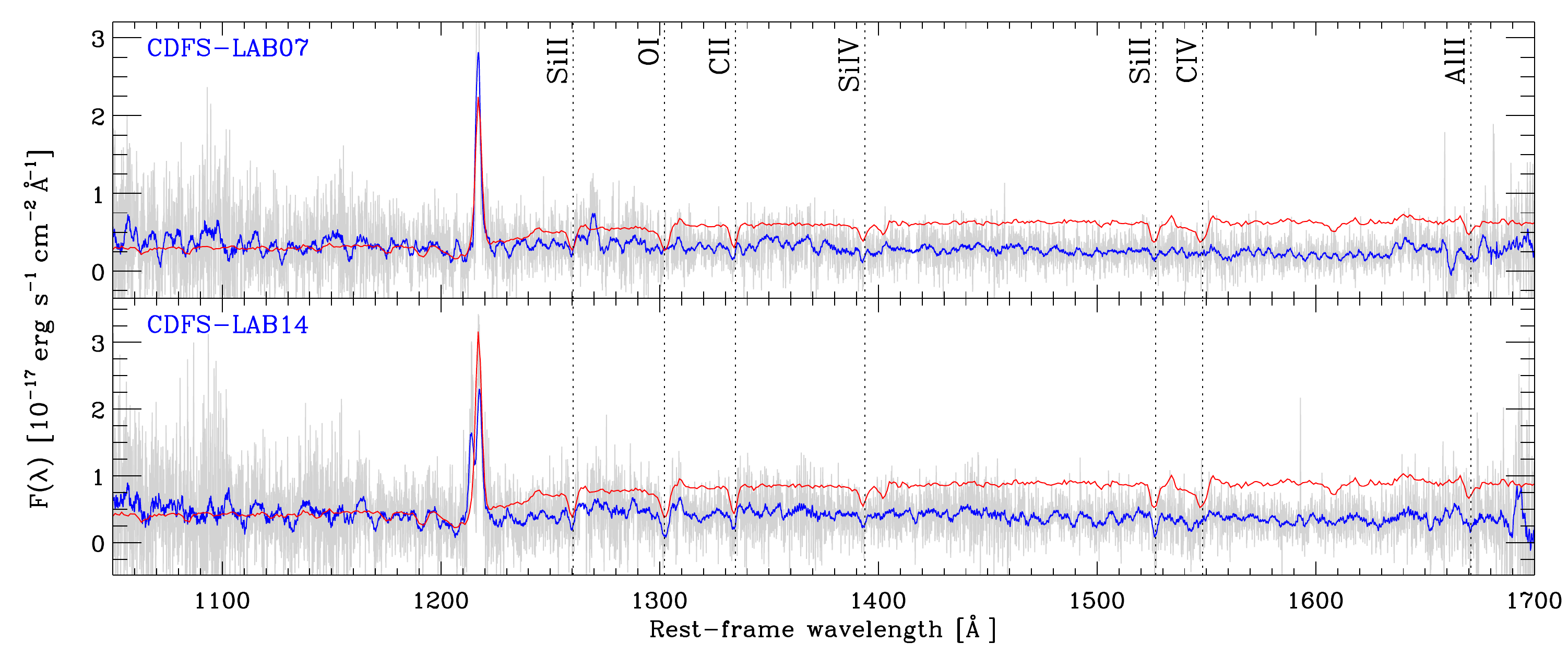}
\includegraphics[height=0.28\textwidth,trim=00 0 0 0,clip=true]{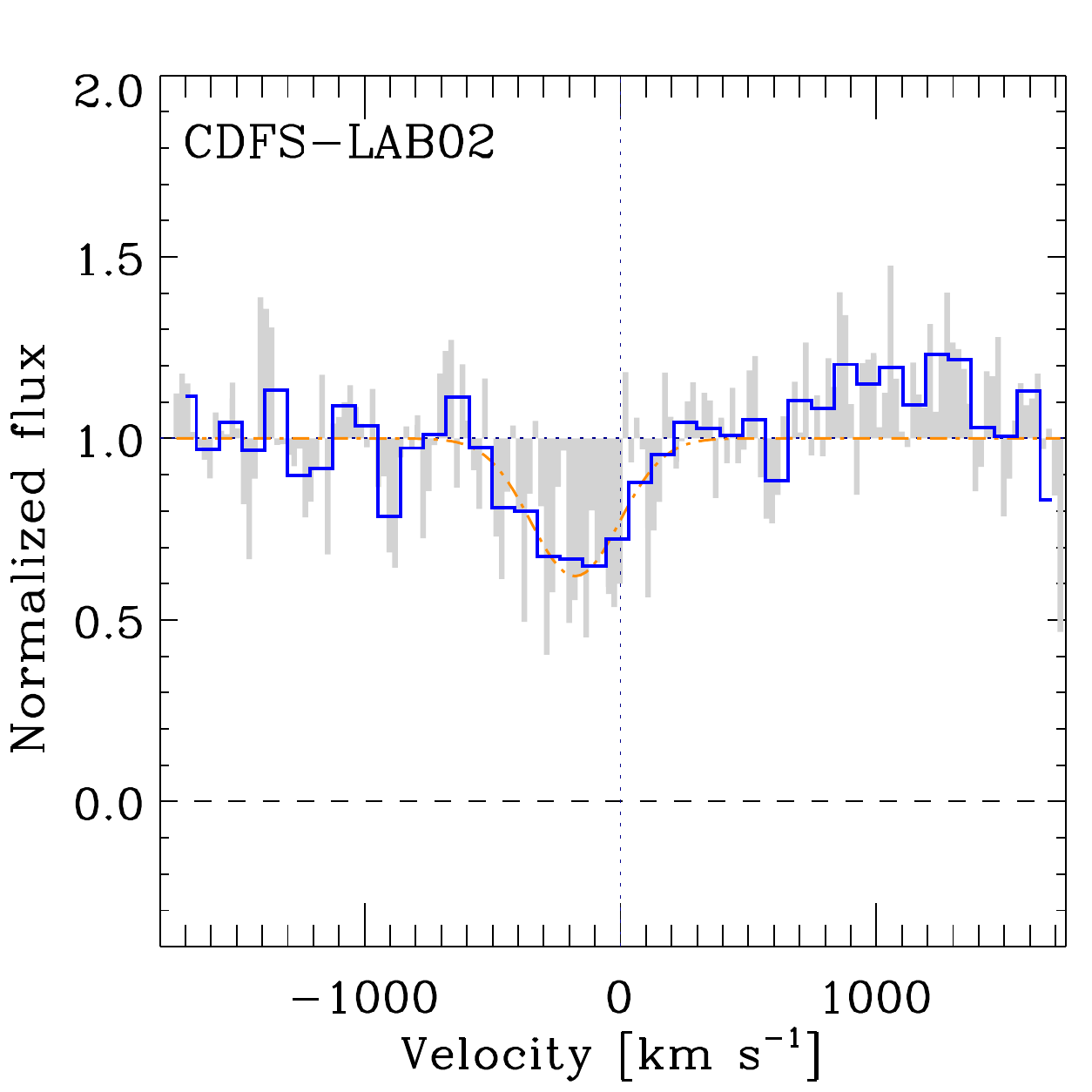}
\includegraphics[height=0.28\textwidth,trim=20 0 0 0,clip=true]{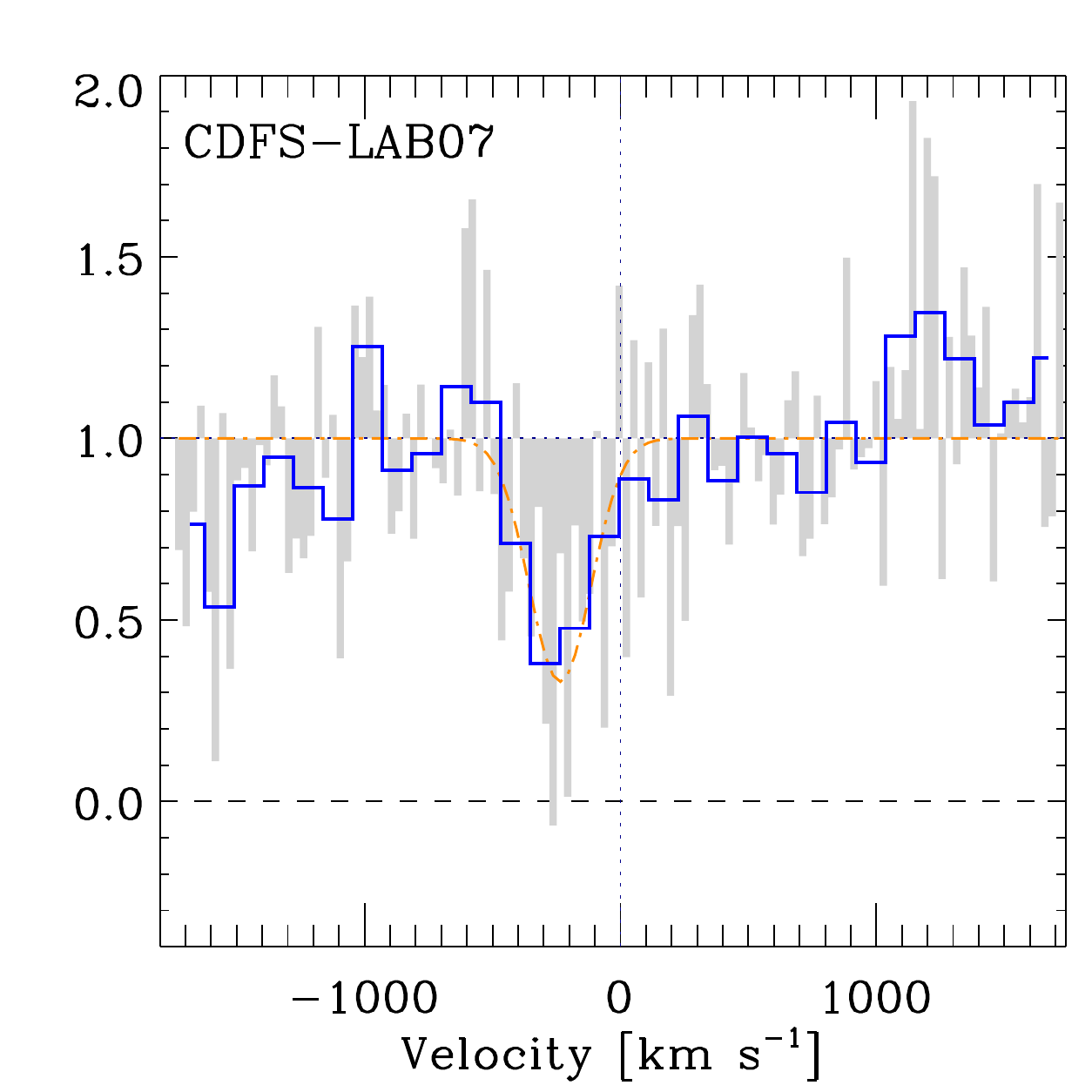}
\includegraphics[height=0.28\textwidth,trim=20 0 0 0,clip=true]{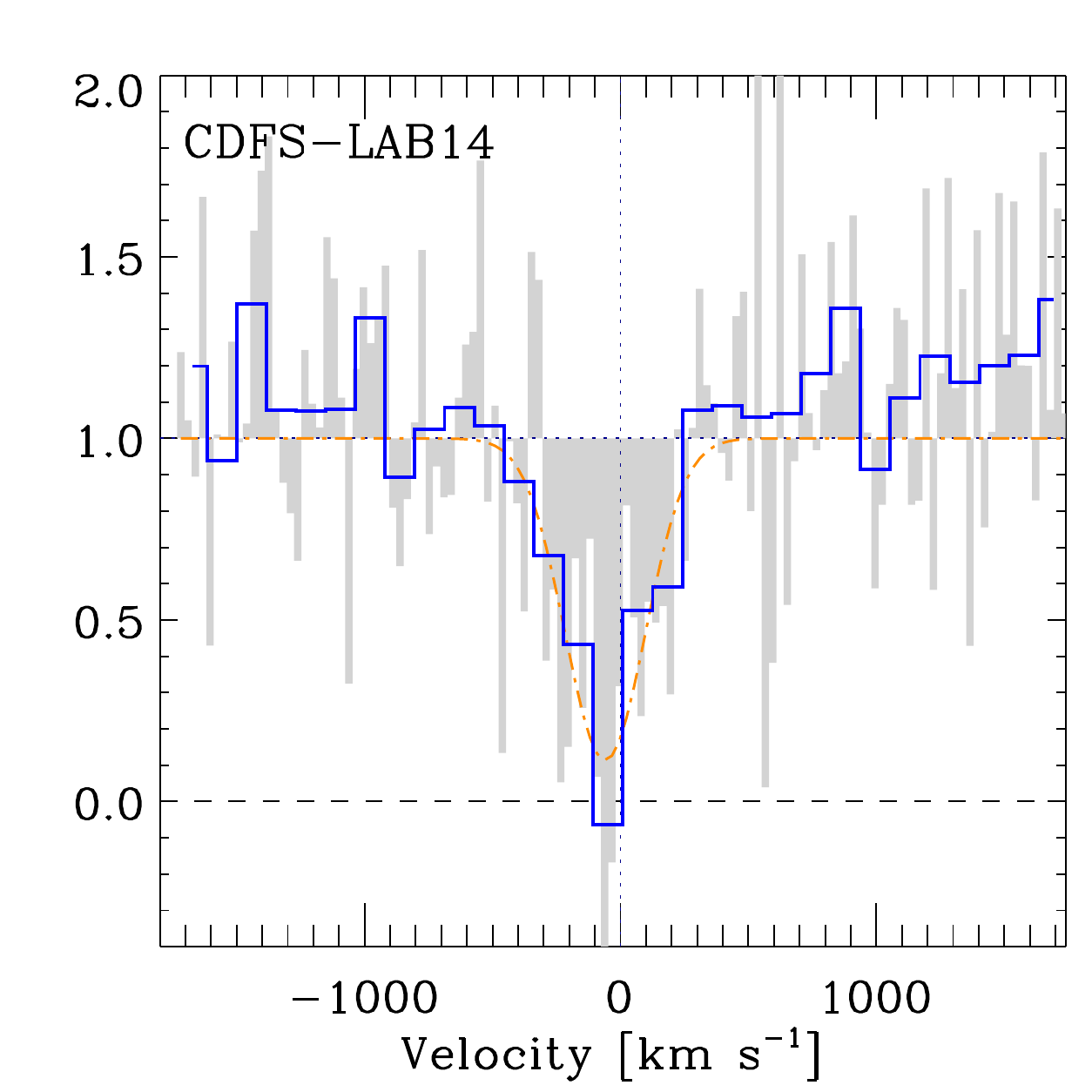}
\end{center}
\caption{
({\it Top})
Rest-frame UV spectra of two embedded blob galaxies with S/N $\gtrsim$ 1.5
pixel$^{-1}$ (CDFS-LAB07 and 14).  The noisy gray lines show the unbinned
spectra, and the thick (blue) solid lines represent the boxcar-smoothed
spectra with $\sim$500\,\kms.  The thin (red) lines indicate the LBG
composite spectrum \citep{Shapley03}, showing that several absorption
lines are detected in these two galaxies.
({\it Bottom})
The stacked absorption profiles for three galaxies including CDFS-LAB02,
which we have previously studied \citep{Yang11}, and the two above. The
velocity is relative to the systemic velocities determined from the \oiii
or \ha lines.  For each galaxy, we stack five low-ionization lines (two
\ion{Si}{2}, \ion{O} {1}, \ion{C} {2}, \ion{Al}{2}) to boost the S/N.
The gray and blue lines are the unbinned spectra and the spectra binned
over 4 pixels, respectively. The orange dot-dashed lines are the Gaussian
fits. 
Two galaxies (CDFS-LAB02 and 07) have small outflow velocities around
\dvis = $-$177\,$\pm$\,31\kms and $-$236\,$\pm$\,31\kms, respectively.
The profiles may extend up to $-$500\kms, but there is no sign of
strong outflows ($\sim$1000\kms). The absorption profile of another
galaxy (CDFS-LAB14) is symmetric around $v \sim 0$\kms (\dvis =
$-$59\,$\pm$\,32\kms), suggesting no significant bulk motion.
Therefore, at least for three galaxies that have \lya, \oiii or \ha, and
ISM absorption profiles, there is no fast ($\sim$1000\kms) outflowing
material along the line-of-sight, which excludes the super/hyperwind
model \citep{Taniguchi&Shioya00} for extended \lya emission in \lya
blobs. The outflows here are similar to or weaker than in LBGs.
}
\label{fig:spec_abs}
\end{figure*}


While strong emission lines such as \lya and \oiii\ are relatively easy
to detect, interstellar metal absorption lines provide a less ambiguous
way to measure the outflow velocity of neutral gas lying {\it in front
of} the galaxies targeted with our spectroscopic slit.
However, due to the faint UV continuum ($B$ = 23.8--26.5 mag) of our
targets, even 8-m telescope struggles to detect their continuum for
absorption line studies.  Therefore, we obtain higher-S/N absorption line
profiles by stacking several lines in each galaxy.  Our aim is to test
the mild outflow interpretation of our \dvlya\ results by comparing the
velocity offset of the stacked ISM absorption profile to the systemic
velocity of the galaxy:  \dvis.

Among the six \lya blobs, only the UV continua of the galaxies in
CDFS-LAB07 and CDFS-LAB14 have S/N higher than 1.5 per spectral pixel
($\sim$20--30\kms), allowing us to marginally extract the absorption
profiles. Figure \ref{fig:spec_abs}(top) shows the rest-frame UV
spectra of these two galaxies in comparison with the LBG composite
\citep{Shapley03}.  For comparison, we also show the boxcar-smoothed
spectra over 20 pixels, corresponding to a velocity width of 450--600\kms.
While it is difficult to identify the absorption lines in the un-binned
spectra, the smoothed spectra show a good match to the composite spectrum,
revealing several low- and high-ionization lines.

As suggested above, it is still difficult to measure individual absorption
line profiles, so we stack the five low-ionization lines:
\ion{Si}{2}  $\lambda$1260,
\ion{O} {1}  $\lambda$1302,
\ion{C} {2}  $\lambda$1334,
\ion{Si}{2}  $\lambda$1526,
\ion{Al}{2}  $\lambda$1670 to increase the S/N.
We do not include the \ion{C}{4} and \ion{Si}{4} lines in the stacking,
as these high ionization lines are contaminated with broader absorption
features arising from stellar winds from massive stars \citep{Shapley03},
which are hard to remove in our low S/N spectra. Furthermore, it is
possible that these high ionization lines trace different state of gas,
while the low ionization lines show similar profiles \citep{Steidel10}.
In Figure \ref{fig:spec_abs}(bottom), we show the stacked absorption
profiles for three galaxies, including the re-analyzed CDFS-LAB02 spectrum
that we obtained earlier with Magellan/MagE \citep{Yang11}.

The stacked absorption profiles here are consistent with weaker outflows
than required by the super/hyperwind hypothesis for \lya\ blob emission
\citep{Taniguchi&Shioya00}.  These profiles are also consistent with
the interpretation of small \dvlya\ indicating small outflow speed,
as discussed in \S\ref{sec:shift}.
In two galaxies (within CDFS-LAB02 and 07), the absorption profiles have
minima around $-$200\kms and might extend up to $-$500\kms, although the
exact end of the profile is uncertain due to the low S/N.  The shapes of
these two absorption profiles are similar to those of typical star-forming
galaxies (i.e., LBGs) at the same epoch \citep{Steidel10}, and the
implied outflow velocities are comparable to or slower than in the LBGs.
By fitting a Gaussian profile, we obtain \dvis = $-$177\,$\pm$\,31\kms
and $-$236\,$\pm$\,31\kms for CDFS-LAB02 and CDFS-LAB07, respectively.
We do not find any redshifted absorption component, which is
consistent with the absence of any blueshifted \lya emission line
associated with these galaxies.

In contrast, the profile of CDFS-LAB14's galaxy has a minimum at
$v$ $\simeq$ 0\,\kms\ (\dvis = $-$59\,$\pm$\,32 \kms) without any
significant blueshifted (outflowing) component.  While the other two
profiles terminate roughly at $v$ $\simeq$ 0\,\kms, CDFS-LAB14's profile
extends to $v$ $>$ 0\,\kms.  Thus, its ISM absorption and \lya\ profiles
are both consistent with the simple RT model expectation that a symmetric,
double-peaked \lya profile should emerge from static gas or the absence of
of bulk motions.  Furthermore, the absorption profile is almost saturated,
indicating that the column density in this \lya blob could be higher than
for the other two systems.  We will place a constraint on the \ion{H}{1}
column density of its \lya-emitting gas in \S\ref{sec:column_density}.

For CDFS-LAB14, the location of the dip between the blue and red \lya
velocity peaks changes little for the two extraction apertures shown in
Figure \ref{fig:spec2d}:  one along the line of sight toward the embedded
galaxy (the \oiii source) and the other encompassing the spatially
extended gas around $\Delta \theta$ = $-$0.5\arcsec. Note that such a
trough between \lya peaks is often interpreted as arising from absorption
by the neutral media between the \lya source and observers.  For example,
\citet{Wilman05} claim that their IFU spectra of a \lya blob (SSA22-LAB02;
Steidel blob 2) suggest that the \lya emission is absorbed by a foreground
slab of neutral gas swept out by a galactic scale outflow.  More recently,
\citet{Martin14b} show instead that the absorption troughs in the \lya
emission are actually located at the systemic velocity determined by
the \oiii emission line, i.e., there are negligible velocity offsets
between any foreground screen and the systemic velocity.
CDFS-LAB14 also demonstrates that a coherent velocity trough, at least
over a $\sim$10\,kpc scale, can arise entirely from complicated radiative
transfer effects even if there are no significant bulk motions in the
\lya-emitting gas.

\subsubsection{Broadening and Asymmetry in {\rm \oiii} Profile}
\label{sec:O3profile}

\begin{figure*}
\epsscale{0.80}
\ifpreprint\epsscale{0.80}\fi
\plotone{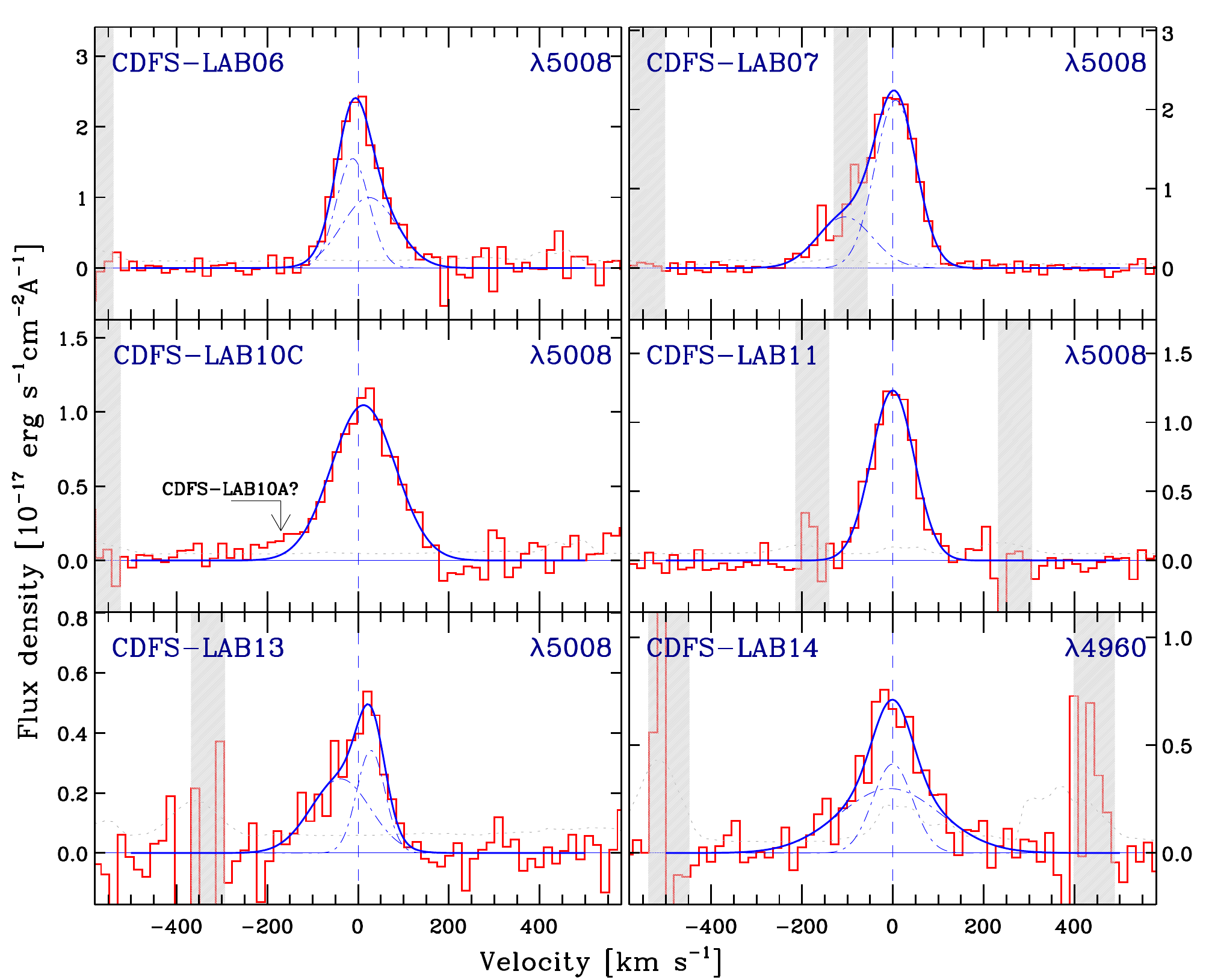}
\caption{
Close up of the \oiii profiles of six galaxies embedded in \lya blobs.
The dot-dashed and solid lines indicate the individual Gaussian components
and the combined profiles, respectively.  Four galaxies (CDFS-LAB06,
07, 13, 14) show a broad component in addition to a narrow component,
suggesting a warm, ionized outflow. These broad components are slightly
blue or redshifted against the brighter narrow components in three
galaxies (CDFS-LAB06, 07 and 13), making the \oiii lines asymmetric.
CDFS-LAB11 shows an almost symmetric single component only.  CDFS-LAB10's
profile is contaminated by a neighbor.  The maximum blueshifted velocity,
which ranges from \dvmax = 50 to 220\,\kms, is smaller than those found in
$z\sim2$ star-forming galaxies (380 -- 1000\,\kms; Genzel et al.~2011),
and consistent with our outflow speed estimates from \dvlya and the ISM
absorption lines.
}
\label{fig:O3profile}
\end{figure*}


The central galaxies in four \lya blobs (CDFS-LAB06, 07, 13, 14) show a
broad and/or shifted underlying component to the \oiii profile, which
provides additional constraints on the kinematics of warm ionized
gas in the vicinity of the galaxies. To the authors' knowledge,
this is the first detection of broadened, asymmetric \oiii profiles
from narrowband-selected \lya-emitting galaxies at high redshifts,
demonstrating that high spectral resolution is required to fully exploit
the NIR spectroscopy of \lya galaxies.

We emphasize again that these broad underlying components, even when
shifted in velocity, do not affect our measurements of \oiii line
centers, because the \oiii flux density at the core is dominated by the
narrower component, which is likely to arise from nebular regions in the
embedded galaxy.  As discussed previously, the line centers determined
from other non-resonant emission lines (e.g., \ha and \hb) agree to
within $\sim$10\kms. Therefore, our measurements of systemic velocity,
critical for determining \dvlya\  and \dvis, remain unchanged.

In Figure \ref{fig:O3profile}, we show close-ups of the \oiii profiles.
In CDFS-LAB11, which has the narrowest and the most symmetric \lya
profile, the \oiii line is also symmetric.  CDFS-LAB10 is excluded
from the following analysis because its \oiii line is elongated
in position-velocity space, suggesting some velocity shear, and the
neighboring galaxy (CDFS-LAB10A) makes it difficult to reliably extract
its profile (see \S\ref{sec:cdfs-lab10} for more details of this system).
In the remaining four \lya blobs, the \oiii profile either has an
asymmetric wing (CDFS-LAB06, 07, 13) or a symmetric broad component
or components (CDFS-LAB14).  Note that the CDFS-LAB06 profile in
Figure \ref{fig:O3profile} also includes the light from a faint
neighbor or clump to the northwest that is slightly blueshifted
($\sim$50\kms). Therefore, the somewhat redshifted wing of CDFS-LAB06
is not due to this contamination.

To extract the underlying broad components, we fit the line profiles
with two Gaussian functions (the two dot-dashed lines in Figure
\ref{fig:O3profile}) and list the results in Table \ref{tab:O3profile}. 
In the Appendix, we describe our fitting procedures in detail.
We adopt a two-component fit for simplicity and for comparison with
previous studies (although we cannot rule out the possibility that the
velocity wings consist of multiple small narrow components).
The line center of the broad component ($v_{\rm broad}$) agree with that
of the narrow component in CDFS-LAB14, is blueshifted in CDFS-LAB07,
and in CDFS-LAB13, and is marginally redshifted in CDFS-LAB06.
The widths of the broad components, corrected for the instrumental
resolution, are relatively small: $\sigma_{\rm broad}$ = 45 -- 120\kms
(FWHM = 100 -- 280\kms), which would not be detected with lower resolution
or lower S/N spectra.

To quantify the contribution of the broad components, we measure the
``broad-to-narrow ratio'' ($F_{\rm broad}/F_{\rm narrow}$), which
is defined as the ratio of the flux in the broad \oiii emission line
to the flux in the narrow component.  Because the fluxes in the two
Gaussian components are anti-correlated with each other, this ratio
has large uncertainties. We also list the ``broad flux fraction''
($f_{\rm broad}$), which is the ratio of the flux in broad emission
to the total \oiii flux ($F_{\rm broad}$/$F_{\rm total}$).  The broad
emission component is significant with $F_{\rm broad}/F_{\rm narrow}$
= 0.4 -- 0.8 constituting 30\%--45\% of the total \oiii line flux
(although the uncertainties are fairly large).  The measurements of
the underlying components are highly dependent on the S/N of spectra.
Due to the low S/N in the $K$-band, we are not able to reliably measure
the broad line components from permitted lines such as \ha or \hb, but
similar broad wings are also present in the \ha profiles of at least
two galaxies (CDFS-LAB07 and 14; see Figure \ref{fig:spec1d}).

What is the mechanism responsible for the broad component in the \oiii
profiles?  A broad component with a larger line-width ($\sigma_v$)
of a few hundred \kms is generally interpreted as a signature of
starburst-driven galactic winds, and often observed in local dwarf
starbursts \cite[e.g.,][]{Westmoquette07} and in local ultraluminous
infrared galaxies (ULIRGs) \cite[e.g.,][]{Soto12}.  While broad components
in forbidden lines such as \oiii are not related to the broad line
region (BLR) of Type 1 AGN, they might still arise from the shocked
narrow line region (NLR) in AGN.  However, the composite spectrum of all
of our X-shooter sample has a line ratio \bptx $<$ $-0.88$, excluding
any significant contribution from AGN (Y.~Yang et al.~in preparation).
Furthermore, in CDFS-LAB11, whose \heii and \civ emission lines hint at
the presence of an AGN, no broad component is detected.

At high redshift ($z\gtrsim2$), broadened emission lines were first
reported by \citet{Shapiro09}.  They found that the stacked spectrum of
$z\sim2$ star-forming galaxies (SFGs) shows broad (FWHM $\sim$ 550\,\kms)
emission underneath the H$\alpha$+\nii line complex.  More recently,
stacked spectra of higher S/N data have revealed that the broad emission
is spatially extended over a half-light radius \citep{Newman12b}.
Broad emission lines are now detected from individual SFGs and even from
giant star-forming clumps within them \citep{Genzel11}.  \citet{Genzel11}
show that $z\sim2$ SFGs show broad wings of \ha emission with FWHM $\sim$
300 -- 1000\kms ($\sigma_{\rm broad}$ $\sim$ 125 -- 425) and employ the
maximum blueshifted velocity, \dvmax = $| \langle v_{\rm broad} \rangle$
$-$ $2\,\sigma_{\rm broad} |$ as a proxy of the outflow speed, finding
that SFGs have \dvmax = 380 -- 1000\,\kms.  Thus, the broad component
in these galaxies is attributed to powerful galactic outflows.
For compact \lya emitters, although there are increasing number of
detections of \oiii and \ha \cite[e.g.,][]{McLinden11, Finkelstein11,
Nakajima12, Hashimoto13}, broad emission in \oiii has not been reported
so far, perhaps due to the lower spectral resolution or lower S/N of
these studies.  Therefore, at the moment, it is not clear whether the
broad component and sometime line asymmetry that we observe in \lya
blobs is a general property of \lya-selected galaxies.

As in local ULIRGs and in high redshift SFGs, the detection of a broad
component in the \oiii profile here suggests warm ionized outflows from
the galaxies within \lya blobs, presumably driven by supernovae and
stellar winds.  However, while the flux fraction of our broad emission
is comparable to those of SFGs, the broad \oiii wings are narrower
($\sigma_{\rm broad}$ = 45 -- 120\kms) and the inferred velocities
much smaller \dvmax = 150 -- 260\,\kms.  Therefore, at face value, the
warm ionized outflows from the \lya blob galaxies are not as strong
as those in SFGs at similar redshift. Furthermore, these estimates for
outflow velocity are roughly consistent with the values obtained from
our two other kinematic measures, \dvlya and \dvis (\S\ref{sec:shift}
and \S\ref{sec:absorption}), independently discounting shock-heating
via super/hyperwinds as a viable powering mechanism.

\subsection{Constraint on {\rm \ion{H}{1}} Column Density}
\label{sec:column_density}

Constraining the physical state of the \lya-emitting gas in a \lya blob
is a critical step to understand its emission mechanism and to directly
compare the observations with the numerical simulations \cite[e.g.,][]
{Faucher-Giguere10, Rosdahl&Blaizot12, Cen&Zheng13, Latif11}, which
predict the \lya emissivity maps from the gas density, temperature,
and UV radiation fields. In this section, as a first step, we estimate
the \ion{H}{1} column density of a \lya blob, CDFS-LAB14.

The \dvlya and \dvis analyses of CDFS-LAB14 are consistent with a simple
RT model in which the surrounding gas is static or without bulk motions.
The third kinematic indicator, the \oiii profile, which suggests a mild
outflow, either contradicts the other indicators or is sensitive to
gas in a different state (e.g., warm, ionized instead of cold, neutral)
or distribution (e.g., around galaxy instead of in galaxy).  Here we use
the \lya emission and ISM absorption line profiles to place a constraint
on the amount of neutral hydrogen in this \lya blob.  Ultimately, by
comparing the \lya profile with detailed RT models, one could extract a
wealth of information, including outflow speed, optical depth, and column
density \cite[e.g.,][]{Verhamme08}.  We defer such detailed analysis to
the future papers and consider the most simplistic case in this section.

\lya line transfer in an extremely thick medium of neutral gas has
been studied over many decades, and the analytic solutions for simple
geometries (like a static homogeneous slab or uniform sphere) are
known \citep{Harrington73, Neufeld90, Dijkstra06a}.  We consider a
simple geometry in which a \lya source is located at the center of a
static homogeneous slab with optical depth of $\tau_0$ from the center
to the edge.  Because \lya photons generated at the center escape the
system through random scattering in the frequency space, the emergent
line spectrum is double-peaked profile with its maxima at
$\Delta x_{\rm peak} \equiv (\nu-\nu_0)/\nu_D = \pm 1.173 (a \tau_0)^{1/3}$, 
where $x$ represents the line frequency in units of Doppler width $\nu_D$.
$a$ and $\tau_0$ are the Voigt parameter and  \lya optical depth at the
line center, respectively \citep{Dijkstra06a}\footnote{ Note that this
coefficient is slightly different from the traditional Neufeld solution
(1.066).}.
In terms of velocity, the separation between blue and red peaks with
same the intensity is
\begin{equation}
\Delta v_{\rm blue-red} = 424\kms \cdot \left(\frac{b}{12.85\kms}
             \frac{N_{\rm HI}}{10^{20}{\rm cm^{-2}}}\right)^{1/3},
\end{equation}
where $b$ and $N_{\rm HI}$ represent the Doppler parameter
($\sqrt{v^2_{\rm th} + v^2_{\rm turb}}$) and the column density of neutral
hydrogen, respectively.  If we adopt a uniform sphere geometry instead of
a slab, the coefficient of the above equation will decrease to 336\,\kms,
and the product ($b$ $N_{\rm HI}$) will increase by a factor of two for
a fixed $\Delta v_{\rm blue-red}$ value.  Note that the blue-to-red peak
separation is degenerate between two parameters, $b$ and $N_{\rm HI}$.

From the \lya profile of CDFS-LAB14 (Figure \ref{fig:spec1d} and
\ref{fig:spec2d}), we measure the separation between the blue and red
peaks, $\Delta v_{\rm blue-red}$ = 790\,$\pm$\,39\kms.  
In the analytic solution, the peaks on the blue and red sides of the
systemic velocity are each symmetric, whereas for CDFS-LAB14 they are
slightly asymmetric.  As a result, the RT in this system may require
ultimately a more complicated model than the assumed simple geometry.

To constrain $N_{\rm HI}$, we consider two extreme cases for the Doppler
parameter $b$:
First, as a lower limit, $b$ $>$ 12.85\kms, obtained by assuming a
temperature $T$ = 10$^4$\,K and ignoring the turbulence term.
Second, $b$ $<$ $\sigma_{\rm abs}$, the width of the metal absorption
lines.  In other words, we assume that the observed velocity width of
absorbing material is purely due to the turbulence or random motions
inside the slab. The width of the absorption line is $\sigma_{\rm abs}$
= 152 $\pm$ 35\kms from a single Gaussian fit (\S\ref{sec:absorption} and
Figure \ref{fig:spec_abs}) and after being corrected for the instrumental
line width.
For this range of $b$, we obtain 19.7 $<$ $\log N_{\rm HI}$ $<$ 20.8.

Because the \lya emission is spatially resolved in CDFS-LAB14, we
further apply this technique to the extended \lya-emitting gas in a
direction other than toward the embedded galaxy.  As shown in Figure
\ref{fig:spec2d}, the \lya profile remains double-peaked as we move away
from the sight-line directly toward the embedded galaxy.  The separation
between  the blue and red peaks decreases, indicating that $b$ $N_{\rm
HI}$ also decreases.  For $\Delta v_{\rm blue-red}$ $\simeq$ 650 \kms,
we obtain slightly smaller estimates for the column density: 19.5 $<$
$\log N_{\rm HI}$ $<$ 20.6.

Using the accurate measurement of systemic velocity, the metal absorption
lines, and the symmetric double-peaked \lya profile, we are able to
place constraints on the $N_{\rm HI}$ toward the galaxy and the extended
\lya-emitting gas, albeit with large uncertainties.
It is intriguing that this rough estimate of the \ion{H}{1} column density
is similar to those of damped \lya absorption systems (DLA; $N_{\rm HI}$
$>$ 2\E{20}\,cm$^{-2}$).  While we have identified this \lya blob by
searching for extended \lya emission, it would be interpreted as a
DLA if there were a background QSO whose continuum spectrum showed
absorption at the blob redshift.  There are similar systems where
extended \lya emission is identified from DLAs very close to the QSO
redshift \cite[e.g., ][]{Moller98, Fynbo99, Hennawi09}.

\subsection{Blueshifted \lya without Broadband Counterpart}
\label{sec:cdfs-lab10}

\begin{figure}
\epsscale{1.00}
\ifpreprint\epsscale{0.52}\fi
\ifpreprint\vspace{-0.6cm}\fi
\plotone{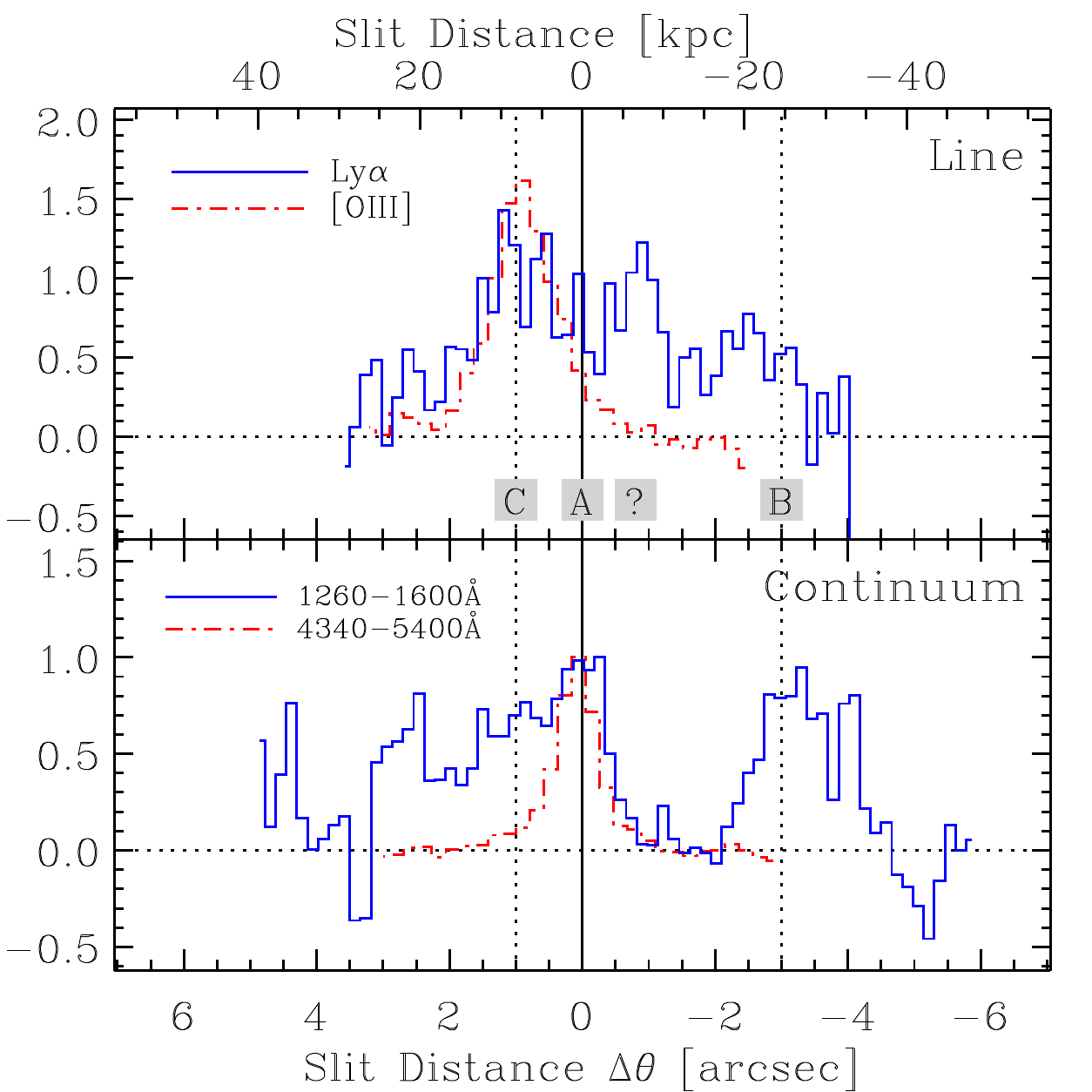}\\
\plotone{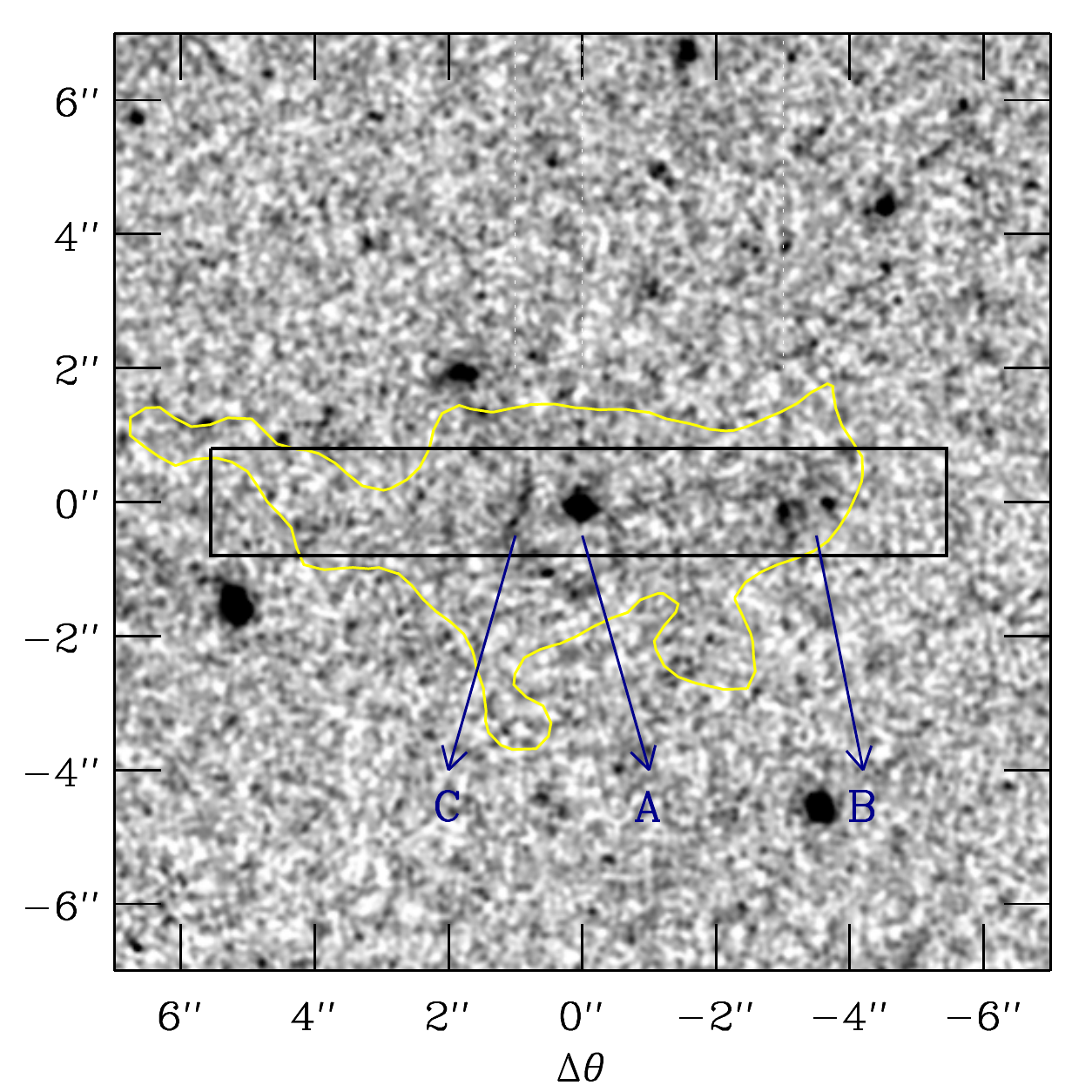}
\caption{
\ifpreprint\small\fi
({\it Top}) Spatial profiles of CDFS-LAB10 collapsed in the wavelength
direction for the emission lines (\lya and \oiii) and the continuum
(rest-frame UV and optical), respectively.  The upper panel is the
same as in Figure \ref{fig:spec2d}. The lower panel shows the spatial
profiles for the continuum around the \lya and \oiii lines.  The solid
(blue) and dot-dashed (red) lines represent the profiles obtained from
the UVB and NIR arms, respectively.
({\it Bottom}) {\sl HST} F606W image of CDFS-LAB10. The yellow contour
represents the \lya.
The detected \oiii emission line originates from a very faint UV source
(CDFS-LAB10C; $m_{\rm F606W}$ $\approx$ 28\,mag), indicating that
extremely deep and high spatial resolution imaging is necessary for
identifying possible energy sources for the extended \lya emission.
There is no counterpart of the \lya emission at $\Delta \theta$ =
$-1$\arcsec\ and $\Delta v$ $\sim$ $-$500\,\kms in either the {\sl
HST} image or in the X-shooter continuum spectrum (marked by a ``?").
Therefore, this isolated \lya emission probably arises from the extended
gas and represents the first unambiguous detection of a blueshifted \lya
line with respect to the systemic velocity of a \lya blob.
}
\label{fig:slitprof}
\end{figure}


Up to this point, we have examined only \lya emission along the
line-of-sight to galaxies embedded in our blobs.  In the 2--D spectrum
of CDFS-LAB10, however, we identify a region of blueshifted \lya emission
that does not spatially coincide with any of the blob galaxies identified
with {\it HST}.  This is the first case in our sample where we detect the
\lya emission from the extended gas itself.  Given that a blueshifted
\lya line is a long-sought signature of gas inflows as described in
\S\ref{sec:intro} and Figure \ref{fig:cartoon}, we present a detailed
analysis of CDFS-LAB10 here.

In Figure \ref{fig:slitprof}\,(bottom), we show the {\sl HST} F606W image
of CDFS-LAB10 now rotated and smoothed with a Gaussian kernel to increase
the contrast of faint sources. There are three broadband sources labeled
CDFS-LAB10A, B, and C within the \lya contour, which is elongated over
10\arcsec\ ($\sim$82\,kpc).  Figure \ref{fig:slitprof}(top) shows the
spatial profiles of the stellar continuum and the emission lines (\lya and
\oiii).  The latter are the same as in Figure \ref{fig:spec2d}.  To obtain
the spatial profiles for the continuum, we collapse the 2--D spectra
for the rest-frame wavelength range, [1260\AA, 1600\AA] and [4340\AA,
5400\AA], i.e., redward of \lya and both sides of the \oiii lines,
respectively.  Because of different data reduction modes, the spatial
profiles from the NIR arms (\oiii) cover only the central $\sim$6\arcsec,
while the profiles from the UVB arm (\lya) cover $\sim$14\arcsec.

The brightest UV source in the {\sl HST} image, CDFS-LAB10A at the slit
center ($\Delta \theta$ = 0\arcsec), is detected in both the rest-frame
UV and optical continua, but not in \lya or any strong emission lines
(\oii, \oiii, \hb, \ha).  It is still possible that there is a faint
\oiiilambb line under a sky line at $\Delta v$ $\sim$ $-550$\kms (see
Figure \ref{fig:spec2d}).  Therefore, it is not clear whether galaxy
A is at the same redshift as the \lya-emitting gas or is a fore- or
background galaxy.
The galaxy B at $\Delta \theta$ $\sim$\,$-$3\arcsec\ is detected in \lya,
\oiii, and UV continuum and is thus a member of the CDFS-LAB10 blob.
Galaxy B's \lya emission is slightly offset from its UV continuum.
CDFS-LAB10C, which has a filamentary or elongated morphology in the
{\sl HST} image, is the faintest ($m_{\rm F606W}$ = $27.9\pm0.2$) of
the three sources, but the brightest in \lya and \oiii.   Thus, galaxy
C is the dominant source of \lya emission from the known galaxies in
this \lya blob.  It appears to be a classic example of an object heavily
extincted in the UV whose \lya photons can still escape.  Its 2--D \oiii
spectrum shows a velocity shear, suggesting a disk.

There is no obvious counterpart of the \lya emission at $\Delta \theta$ =
$-1$\arcsec\ and $\Delta v$ $\sim$ $-$500\,\kms in either the {\sl HST}
image or in the X-shooter continuum spectrum.  Therefore,
this isolated \lya emission probably arises from the extended gas.
Its line profile is broad, double-peaked, and blueshifted from galaxy B
and galaxy C (Figure \ref{fig:spec2d}).  It is not possible to distinguish
whether the bulk motion is inflowing or outflowing (at $\sim$500\,\kms)
with respect to galaxy C, the brightest \oiii source and marker of
systemic velocity (Figure \ref{fig:spec2d}), because we do not know
whether the gas lies in front of or behind galaxy C.
This is the first unambiguous detection of a blueshifted \lya line with
respect to galaxies embedded within \lya blobs.  It is likely that the
blueshift is relative to the systemic velocity of the \lya blob as well.

It is not clear what this blueshifted \lya emission represents. We
consider three possibilities.
First, although unlikely, this \lya component could be associated with
a heavily-extincted galaxy that lies below the detection limit of the
{\it HST} image ($m_{\rm F606W}$ $\gg$ 28 mag) or with galaxy A, whose
redshift is unknown.
Second, this gas might be tidally-stripped material arising from a
galaxy-galaxy interaction:  the two galaxies (B and C) are separated by
$\sim$30\,kpc in projected distance and $\sim$200\,\kms in velocity space.
Lastly, but most interestingly, this blueshifted \lya emission could
be the long-sought, but elusive cold gas accretion along filamentary
streams \citep{Keres05, Keres09, Dekel09}.

Note that similarly blueshifted \lya emission has been reported in a
faint \lya emitter at $z$ = 3.344, which was discovered in an extremely
deep, blind spectroscopic search \citep{Rauch11}.  The spectrum of this
peculiar system also has very complex structure (e.g., diffuse fan-like
blueshifted \lya emission and a DLA system). \citet{Rauch11} suggest that
this blueshifted \lya emission can be explained if the gas is inflowing
along a filament behind the galaxy and emits fluorescent \lya photons
induced by the ionizing flux escaping from the galaxy.  Discriminating
among the above possibilities will require us to further constrain the
\lya line profile of this blueshifted component and to more fully survey
possible member galaxies (or energy sources) within CDFS-LAB10.

Still, we were able to successfully link the \lya and \oiii sources with
the embedded galaxies in this complex system and to find blueshifted
\lya emission that might point to gas inflow.  This example raises
concerns about how to identify the sources of \lya emission and to
interpret the gas morphologies and kinematics solely from \lya lines.
Elongated or filamentary \lya morphologies may be a sign of bipolar
outflows \citep{Matsuda04} or related to filamentary cold streams.
In CDFS-LAB10, the \lya spectrum shows two kinematically distinct
components around the brightest galaxy A: the upper ($\Delta\theta$ $>$
0\arcsec) and lower parts are blue- and redshifted, respectively.  In the
absence of \oiii spectroscopy, the \lya data appear to be consistent with
either of the above outflow or infalling stream scenarios.  However,
our detailed analysis including the \oiii line clearly shows that one
of the \lya components arises from the faint galaxy C, which would be
difficult to detect without {\sl HST} imaging and is likely the dominant
source of \lya emission.  Therefore, we stress that interpreting the
\lya morphology and spectra requires deep high-resolution imaging and
the determination of the systemic velocity through NIR spectroscopy.

\section{Discussion}
\label{sec:discussion}

\subsection{Covering Factor of Inflows and Outflows}
\label{sec:covering_factor}

Under the assumption that inflowing gas streams (if any) are randomly
distributed, and from the non-detection of any blueshifted \lya--\ha
or \lya--\oiii offset in the direction of the eight embedded galaxies 
tested here (although see Section \ref{sec:cdfs-lab10}), we constrain
the covering fraction of inflows to be $<$\,1/8 (13\%).
Likewise, if all of our \dvlya offsets are different projections of
the same collimated outflow from the galaxies and are a proxy for
outflow speed, the covering fraction of {\it strong} outflows, i.e.,
super/hyper-winds, is less than $\sim$13\%.

The covering fraction of gas flows to which we are referring here
has a different meaning than often discussed in the literature
\cite[e.g.,][]{Faucher-Giguere10&Keres11, Kimm11, Fumagalli11}.
In these theoretical papers, the covering factor is defined as how
many sight-lines will be detected in metal or \HI\ {\it absorption}
when bright background sources close to the galaxies are targeted.
What we measure in this paper is how often the inflowing gas is aligned
with observer's sight-lines so that the column of neutral gas becomes
optically thick and blue-shifts \lya lines against systemic velocity.
Nonetheless, \citet{Faucher-Giguere10&Keres11} predict that the covering
factor within the virial radius ($R_{\rm vir}$ $\simeq$ 75\,kpc) from
their simulated galaxies with a halo mass of $M_h$ = 3\E{11}\msun\ at
$z=2$ is relatively small: $\sim$3\% and 10\% for DLAs and Lyman limit
systems (LLS), respectively.  Within 0.5$\times$$R_{\rm vir}$, this factor
increases to $\sim$10\% and 30\%, respectively, and presumably will be
much higher directly towards the galaxies (i.e., for a pencil beam or
looking down-the-barrel).  Note that  \citet{Faucher-Giguere10&Keres11}
ignore galactic winds in order to isolate the inflowing streams, so
we expect that there will be enough dense material arising from  gas
accretion to affect the transfer of \lya photons in our experiments.

What is uncertain is that how the velocity field of this dense
material near the galaxy will shift the emerging \lya profiles and
the statistics of \dvlya, because answering this requires full \lya
radiative transfer treatment.  Unlike RT models with simple geometry
\cite[e.g.,][]{Dijkstra06a, Verhamme06}, cosmological simulations
\citep{Faucher-Giguere10} predict that the gas inflow can only slightly
enhance the blue \lya peak, implying that the overall profile will be
still red-peak dominated if the effects of outflows from the galaxies
and IGM absorption are fully considered.  Therefore, according to these
simulations, our non-detection of infall signatures does not contradict
the cold stream model.  The \lya profiles calculated from cold-stream
models are in general the integrated profiles of \lya-emitting gas around
the galaxies, which are not detectable by our study.  What we need are
the predictions for how \lya profiles are modified and/or shifted against
optically-thin lines along the line-of-sight to the galaxies embedded
in the extended gas. In this way, the distribution of predicted \dvlya
can be compared directly with the observations.

It is also critical to take into account the effect of simultaneous
outflows and inflows.  For example, the H$\alpha$ and \oiii detections
suggest that the galaxies embedded in the blobs are forming stars,
which could generate mechanical feedback into the surrounding gas cloud.
Thus, one might have expected that the innermost part of the gas cloud,
close to the galaxies, has galactic scale outflows similar to those
of other star-forming galaxies at $z=2-3$ \cite[e.g.,][]{Steidel10}.
Gas infall (if any) may dominate at larger radii (up to $\sim$50 kpc,
the typical blob size).  In this case, the emerging \lya profile will
be more sensitive to the core of the \lya blob, presumably the densest
part of the CGM, than to the gas infall.  As a result, it might be
difficult to detect the infalling gas by measuring \dvlya.  This kind
of more realistic, infall+outflow model has not been considered yet in
RT calculations, so its spectral signatures are unknown.

\subsection{Outflow Speed vs. \dvlya}
\label{sec:dvlya}

In our sample of eight \lya blobs, the embedded galaxies have smaller
velocity offsets ($\langle\dvlya\rangle$ = 160\kms) than those of LBGs
(445\kms).  The remaining question is how to interpret these small
\dvlya values.  We consider two possibilities here.

First, if we assume a simple geometry where the outflowing material forms
a spherical shell that consists of continuous media, \dvlya originates
from the resonant scattering of the \lya photons at the shell as discussed
in \citet{Verhamme08}.
Note that the emerging \lya profile is independent of the physical
size of the shell as long as the shell have the same outflow velocity.
Therefore, this model is applicable to \lya blobs. In this shell model,
a central monochromatic point source is surrounded by an expanding
shell of neutral gas with varying column density ($N_{\rm H I}$) and
Doppler $b$ parameter \cite[e.g.,][]{Verhamme06,Verhamme08}.  A generic
prediction is that the \lya emission is asymmetric, with the details of
the line shape depending on the shell velocity, Doppler parameter $b$,
and optical depth of \HI\ column in the shell.
In this simple geometry, the \dvlya\ values can be used as a
proxy for outflow velocity of expanding shell, i.e., $v_{\rm exp}$
$\simeq$ 0.5\,\dvlya.  If this is the case, outflow velocities from
blob galaxies are much smaller than the values expected from models
of strong galactic winds \cite[$\sim$ 1000\kms;][]{Taniguchi&Shioya00,
Taniguchi01, Ohyama03, Wilman05}.  Furthermore, these offsets are even
smaller than the typical \lya--H$\alpha$ offsets (250--900\kms) of LBGs
\citep{Steidel04, Steidel10}.
In particular, while CDFS-LAB14 has the largest \dvlya\ for its red peak,
its blue and red peaks have similar intensity, which is a characteristic
of a static medium or no bulk motions \citep{Verhamme06, Kollmeier10}.

Second, in an ``expanding bullet'' or ``clumpy CGM'' model \citep{Steidel10},
the outflowing material, which consists of small individual clumps, has a
wide velocity range rather than a single value, and the \lya profiles are
determined by the Doppler shift that photons acquire when they are last
scattered by the clumps just before escaping the system.  In this case,
the \lya--\ha\ offsets are primarily modulated by the amount of gas that
has $v = 0$ component (though it is not clear where this material is
spatially located); thus \dvlya\ is not directly correlated with outflow
velocity.  If this is the case, the small \dvlya\ value of the galaxies
in our \lya blobs simply indicates that there is less neutral gas at the
galaxies' systemic velocity (presumably near the galaxy) compared to LBGs.

Testing these two hypotheses for \lya blobs or LBGs first requires
a detailed comparison between high resolution \lya profiles and RT
predictions. This test is beyond the scope of this paper and will be
discussed in the future.  As emphasized by \citet{Steidel10}, one also
needs to check the consistency of the \lya profiles with ISM metal
absorption profiles, which depends on high S/N continuum spectra that
are not available here. Nonetheless, we attempted such an analysis
by stacking many low-ionization absorption lines in three galaxies
(section \ref{sec:absorption}).

For those three embedded galaxies with measured \dvlya\ and \dvis, the
outflow velocity estimates from both methods roughly agree within the
uncertainties arising from low S/N and from RT complications.  However,
closer inspection reveals a discrepancy between the observations and
the simplest RT models in that we obtain \dvis $\sim$ $-$200\,\kms and
also \dvlya $\sim$ 200\,\kms while the RT model with an expanding shell
geometry predicts \dvis = $-$\vexp = $-0.5$ \dvlya.  This discrepancy has
also appeared in LBGs \citep{Steidel10}, and \citet{Kulas12} show that
the stacked \lya profile of LBGs with double-peaked profiles cannot be
reproduced accurately by the shell model. For a wide range of parameters
(\vexp, $b$, \NHI), \citet{Kulas12} cannot reproduce the location of
the red peak, the width of the \lya\ profile, and the metal absorption
profile at the same time.

We attribute this discrepancy to the very simple nature of the shell
model. By construction or by definition of the expanding ``shell'', the
internal velocity dispersion of the media that constitute the shell should
be much smaller than the expansion velocity of the shell itself, i.e.,
\vexp/$b$ $\gg$ 1. Therefore, one expects that the metal absorption lines
arising from this thin shell have very narrow line widths $\sigma_{\rm
abs}$ similar to $b$, typically $\sim$tens of \kms.  However, such narrow
absorption lines are not observed in either LBGs or \lya blobs. Clearly,
RT calculations with more realistic geometries and allowing a more
thorough comparison with the observed \lya profiles are required.

\subsection{\dvlya in Context of LAEs and LBGs}
\label{sec:LAE}

Given that the galaxies within our \lya blobs were selected, by
definition, as \lya emitters with high equivalent width (EW), the
comparison with LAEs suggests that small \dvlya\ values are a general
characteristic of all high EW \lya-selected populations, be they compact
or extended.
In addition to this work and \citet{Yang11}, there are recent studies
that measure \dvlya\ for bright compact \lya-emitters \citep{McLinden11,
Finkelstein11,Hashimoto13,Guaita13} and \lya blobs \citep{McLinden13},
which also find small \lya--\oiii offsets for LAEs ranging from 35 --
340\kms.
A Kolmogorov--Smirnov test fails to distinguish the \dvlya distribution
of our eight \lya blobs from that of ten compact LAEs compiled from the four
studies mentioned above.
Thus, we now have a fairly large sample of high--EW \lya emitters (compact
or extended) with \dvlya\ = $-$60 $\rightarrow$ +400\kms, which is clearly
different from the distribution for LBGs (see Section \ref{sec:shift}).

The smaller \dvlya of LAEs and LABs might be related to their higher \lya\
equivalent widths. An anti-correlation between EW(\lya) and \dvlya in a
compilation of LAE and LBG samples has been suggested \citep{Hashimoto13}.
Even within the LBG population itself there is an indication that
\dvlya decreases as EW(\lya) increases.  \citet{Shapley03} find
an anti-correlation between the EW(\lya) and the velocity offsets
between interstellar absorption and \lya emission lines, $\Delta v_{\rm
Ly\alpha-abs}$.\footnote{The notation, $\Delta v_{\rm em-abs}$ is used in
\citet{Shapley03}.} From the stacked spectra of LBGs binned at different
EW(\lya), they show that with increasing \lya line strength from $-$15\AA\
to 53\AA, $\Delta v_{\rm Ly\alpha-abs}$ decreases from 800\kms to 480\kms.
Because $\Delta v_{\rm Ly\alpha-abs}$ = \dvlya + $|\dvis|$, we can
infer that \dvlya is likely to decrease as EW(\lya) increases unless
the observed anti-correlation is entirely due to $|\dvis|$.

The origin of the apparent relationship between larger EW(\lya) and
smaller \dvlya\ is not understood.  It is possible that other physical
properties drive this anti-correlation:  for example, LBGs tend to
have brighter continuum magnitudes, thus likely higher star formation
rates and stellar masses, than are of typical LAEs.  \citet{Hashimoto13}
propose that low \HI column density, and thus a small number of resonant
scatterings of \lya photons, might be responsible for the strong \lya
emission and small \dvlya of \lya emitters.

What we do know from our \dvlya\ measurements is that LAEs and \lya\
blobs are kinematically similar.  Therefore, the mystery remains as
to what powers \lya\ nebulae, or, in other words, why certain galaxies
have more spatially-extended \lya\ gas (i.e., blobs) than others (i.e.,
compact LAEs) with similar  \dvlya\ and EW(\lya).
While the answer may be related to photo-ionization from (buried) AGN
(e.g., see Section \ref{sec:type2} or \citealt{Yang14}), an extended
proto-intracluster medium that can scatter or transport the \lya
photons out to larger distances, or less dust to destroy \lya\ photons
\cite[cf.][]{Hayes13}, discriminating among these possibilities will
require a multi-wavelength analysis of a large sample of \lya blobs. For
now, the similarity here of LABs and LAEs suggests that differences in
gas kinematics are not responsible for the extended \lya halos.

\subsection{CDFS-LAB11: Photo-ionization by AGN?}
\label{sec:type2}

\begin{figure}
\epsscale{1.15}
\ifpreprint\epsscale{0.70}\fi
\plotone{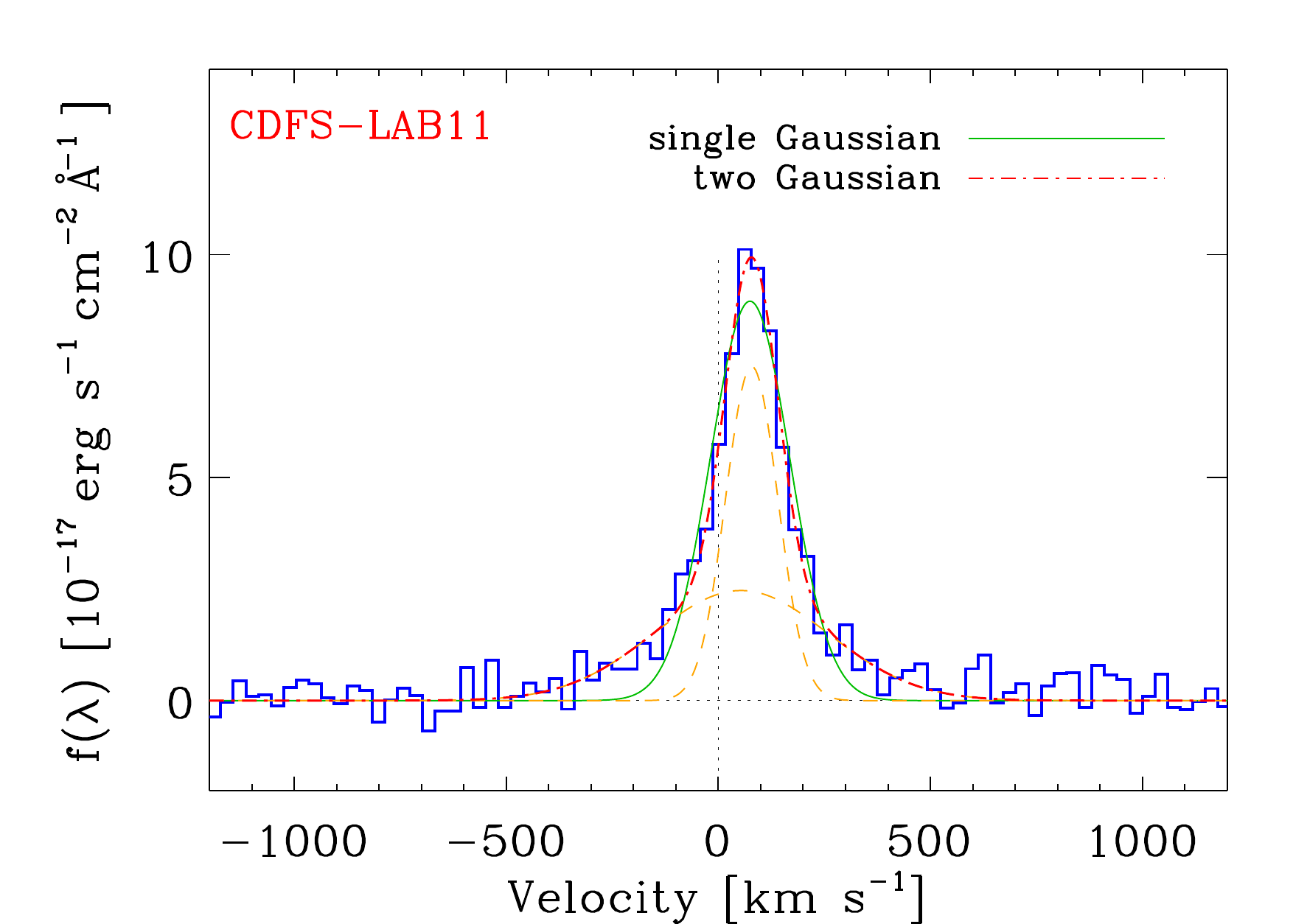}
\caption{
\lya profile of CDFS-LAB11. The solid and dot-dashed lines represent the
fits with one and two Gaussian components, respectively. The two dashed
lines are the narrow and broad components.  The narrow component may arise
from an AGN that ionizes the gas surrounding the embedded galaxies,
making \lya relatively optically thin and preventing the resonant
scattering of \lya photons from dominating the shape of the profile.
}
\label{fig:cdfs-lab11}
\end{figure}


From the nearly symmetric \lya profile (Fig.~\ref{fig:cdfs-lab11}),
\civ and \heii detection, and rest-frame optical line ratios, we
hypothesize that an AGN in CDFS-LAB11 ionizes the gas surrounding the
galaxies, making \lya\ {\it relatively} optically thin and preventing
the resonant scattering of \lya photons from dominating the shape of
the profile.  While fitting the profile requires an underlying broad
component (FWHM $\simeq$ 480\,\kms), the narrow component (redshifted by
$\simeq$ 80\,\kms) has a small velocity width (FWHM $\simeq$ 110\,\kms),
comparable to that of the \oiii line ($\simeq$80\,\kms).  Therefore,
if we assume that \lya and \oiii photons originate from the same region,
resonant scattering by the IGM/CGM does not broaden the intrinsic profile
significantly. In other words, at least along our LOS, the optical depth
of CDFS-LAB11's IGM/CGM is smaller than for the other \lya blobs in our
sample and for typical \lya emitters at high redshifts.

A nearly symmetric and narrow \lya profile such as CDFS-LAB11's is rarely
observed among high-$z$ \lya-emitting galaxies.  However, \lya radiative
transfer calculations show that such \lya profiles can be generated in the
presence of a central ionizing source.  For example, \citet{Dijkstra06a}
investigate the emerging \lya profiles from an collapsing/expanding sphere
centered on an ionizing source (an AGN) for different luminosities (see
their Figures 11 and 12).  They find that if the ionizing source is strong
and the \lya is {\it relatively} optically thin (line-center optical depth
$\tau_0$ $<$ $10^3$), the separation of the characteristic double-peaks
become smaller ($\Delta v_{\rm blue-red}$ $\ll$ 100\kms) as the ionizing
source gets stronger.  The overall profile becomes symmetric, and is blue
or redshifted depending on the flow direction, but again by small amount
(\dvlya \,$\ll$ 100\kms).  These authors were skeptical that the sign of
the \lya shift and the small \lya offset (\dvlya $\sim$ a few tens \kms)
from the systemic velocity could be measured.  However, with the right
strategy (i.e., high spectral resolution and a careful choice of survey
redshift), we are able to reliably measure the predicted small offset.
Clearly, more detailed comparisons with the RT models are required.

A broad, asymmetric \lya profile with a sharp blue edge (e.g., CDFS-LAB06
and 07) is characteristic of high-$z$ \lya-emitters.  In spectroscopic
follow-up observations of \lya-emitter candidates, where spectral
coverage is limited, these characteristics alone are often used to
discriminate high-$z$ \lya emitters from possible low-$z$ interlopers.
The example of CDFS-LAB11 demonstrates that caution is required because
a narrow and symmetric \lya line can also arise when the ISM or CGM
of a candidate galaxy is significantly photo-ionized, e.g., by AGN.
On the bright side, our \dvlya\ velocity offset technique could be used
for studying gas infall or outflow in high-$z$ QSOs instead of relying
on only the \lya line profile \cite[e.g.,][]{Weidinger04}.

\section{Conclusions}
\label{sec:conclusion}

Exploring the origin of \lya nebulae (``blobs") at high redshift requires
measurements of their gas kinematics that are difficult with only the
resonant, optically-thick \lya line.  To define gas motions relative
to the systemic velocity of the nebula, the \lya line must be compared
with non-resonant lines, which are not much altered by radiative
transfer effects.  We made first comparison of non-resonant \halamb\
to extended \lya emission for two bright \lya blobs in \citet{Yang11},
concluding that, within the context of a simple radiative transfer
model, the gas was static or mildly outflowing at $\lesssim$250\,\kms.
However, it was unclear if these two \lya blobs, which are the brightest
in the sample, are representative of the general \lya blob population.
Furthermore, geometric effects --- infall along filaments or bi-polar
outflows --- might hide bulk motions of the gas when only viewed from
the two directions toward these two \lya blobs.  With VLT X-shooter, we
obtain optical and near-infrared spectra of six additional \lya blobs
from the \citet{Yang10} sample.  With a total of eight \lya blobs, we
investigate the gas kinematics within \lya blobs using three techniques:
the \lya offset from the systemic velocity (\dvlya), the shape and shift
(\dvis) of the ISM metal absorption line profiles, and the breadth of
the \oiii line profile.

Our findings are:

\smallskip

\begin{enumerate} 

\item
Both \lya and non-resonant lines confirm that these blobs lie at the
survey redshift ($z\sim2.3$).  We also detect the \oiilamb, \oiiilamba,
\oiiilambb, and \hblamb lines.  All non-resonant line velocities
are consistent with each other and with arising from the galaxy or
galaxies embedded in the \lya blob.  \oiii, which is observed at high
signal-to-noise in all cases and whose profile is an RT-independent
constraint on the gas kinematics, is a particularly good diagnostic line
for this redshift and instrument.

\item
The majority of the blobs (6/8) have broadened \lya profiles indicating
radiative transfer effects.  These \lya profiles are consistent with being
in the same family of objects as predicted by RT, with profile shapes
ranging from symmetric double-peaked, to asymmetric red peak dominated, to
a single red peak.  The fraction of double-peaked profiles is $\sim$38\%
(3/8).

\item
The narrow \lya profile systems (CDFS-LAB01A, CDFS-LAB11), whose \lya
profile is not significantly broader than the \oiii or \ha lines, have
the smallest \dvlya offsets, the most spatially compact \lya emission,
and the only \civ and \heii lines detected, implying that a hard ionizing
source, possibly an AGN, is responsible for the lower optical depth
toward the central embedded galaxies.

\item
With a combination of \dvlya, the interstellar metal absorption line
profile, and a new indicator, the spectrally-resolved \oiii line profile,
we detect gas moving along the line of sight to galaxies embedded in the
\lya blob center.  Although not all three indicators are available for all
\lya blobs, the implied speeds and direction are roughly consistent for
the sample, suggesting a simple picture in which the gas is stationary
or slowly outflowing at a few hundred \kms from the embedded galaxies.
These outflow speeds are similar to those of LAEs, suggesting that
outflow speed is not the dominant driver of extended \lya emission.
Furthermore, these outflow speeds exclude models in which star formation
or AGN produce ``super'' or ``hyper'' winds of up to $\sim$1000\,\kms
\citep{Taniguchi&Shioya00}.

More specifically:


\begin{enumerate}[leftmargin=+0.2cm]

\item[$\bullet$]
We compare the non-resonant emission lines \oiii and \ha to the \lya
profile to obtain the velocity offset \dvlya.  The galaxies embedded
within our \lya blobs have smaller \dvlya than those of LBGs, confirming
the previous claims \citep{Yang11}.  The galaxies within \lya blobs have
\dvlya = $-$60 $\rightarrow$ +400\kms\ with an average of $\langle\dvlya\rangle$ =
160\kms, while LBGs at similar redshifts have \dvlya = 250 -- 900\kms.
The small \dvlya in the \lya blobs are consistent with those measured
for compact LAEs.  

\item[$\bullet$]
By stacking low-ionization metal absorption lines, we measure the outflow
velocity of neutral gas in front of the galaxies in the \lya blobs.
Galaxies in two \lya blobs show an outflow speed of $\sim$250\kms, while
another has an almost symmetric absorption line profile centered at \dvis
= 0\,\kms, consistent with no significant bulk motion.  The ISM absorption
line profiles here have low S/N, but are very roughly consistent with
those of some LBGs (at the level of several hundred \kms outflows).

\item[$\bullet$]
The high spectral resolution of our data reveals broad wings in the
\oiii\ profiles of four \lya blobs.  This new kinematic diagnostic
suggests warm ionized outflows driven by supernovae and stellar winds.
These broad line components are narrower ($\sigma_{\rm broad}$ = 45
-- 120\kms) and have a maximum blueshifted velocity (\dvmax = 150 --
260\,\kms) smaller than those of $z\sim2$ star-forming galaxies (SFGs),
implying weaker outflows here than for LBGs and SFGs at similar redshifts.

\item[$\bullet$]
If we assume that the detected outflows are different projections of
the same outflow from the \lya blob center, we can estimate the effects
of flow geometry on our measurements given that our large sample size
allows averaging over many lines-of-sight.  The absence of any strong
($\sim$1000\,\kms ) outflows among the eight galaxies tested is not a
projection effect: their covering fraction is $<$\,1/8 (13\%).  Likewise,
the lack of a blue-peak dominated \lya profile, at least in the direction
of the embedded galaxies (see point 6. below), implies that the covering
factor of any cold streams \citep{Keres05, Keres09, Dekel09} is less
than 13\%. The channeling of gravitational cooling radiation into \lya
may not be significant over the radii probed by our techniques here.

\end{enumerate} 

\item
Constraining the physical state of \lya-emitting gas in a \lya blob is a
critical step in understanding its emission mechanism and in comparing
to simulations.  For one \lya blob whose \lya profile and ISM metal
absorption lines suggest no significant bulk motion (CDFS-LAB14),
at least in its cool and neutral gas, we assume a simple RT model and
make the first column density measurement of gas in a embedded galaxy,
finding that it is consistent with a DLA.

\item
For one peculiar system (CDFS-LAB10), we discover blueshifted \lya
emission that is {\it not} directly associated with any embedded
galaxy. This \lya emitting gas is blueshifted relative to two embedded
galaxies, suggesting that it arose from a tidal interaction between
the galaxies or is actually flowing into the blob center.  The former
is expected in these overdense regions, where {\sl HST} images resolve
many galaxies.  The latter might signify the predicted but elusive cold
gas accretion along filaments.

\end{enumerate}

\acknowledgments
We thank the anonymous referee for her or his thorough reading of the
manuscript and helpful comments.
The authors thank Daniel Eisenstein for his contributions at the start
of this project.  
We thank Jason X.~Prochaska for helpful discussions.
YY thanks the MPIA ENIGMA group for the helpful discussions. YY
also thanks the Theoretical Astrophysics Center at the University of
California, Berkeley for the travel support, as well as Claude-Andr\'e
Faucher-Gigu{\`e}re, Du{\v s}an Kere{\v s} and Daniel Kasen for the
helpful discussions during that stay. YY also thanks Sangeeta Malhotra
for a helpful discussion regarding the LAE and LBG connection. 
YY acknowledges support from the BMBF/DLR grant Nr.\ 50 OR 1306.
A.I.Z.~thanks the Max-Planck-Institut f\"ur Astronomie and the Center
for Cosmology and Particle Physics at New York University for their
hospitality and support during her stays there.  A.I.Z.~acknowledges
support from the NSF Astronomy and Astrophysics Research Program through
grant AST-0908280 and from the NASA Astrophysics Data Analysis Program
through grant NNX10AD47G. She also thank the generosity of the 
John Simon Guggenheim Memorial Foundation.

\medskip
Facilities: \facility{VLT (X-shooter, SINFONI), Magellan (MagE)}

\appendix 
\section{Decomposition of \oiii profiles}

\begin{figure*}
\epsscale{1.15}
\ifpreprint\epsscale{0.95}\fi
\ifpreprint\vspace{-0.8cm}\fi
\begin{center}
\includegraphics[width=0.30\textwidth,trim=00 0 0 0,clip=true]{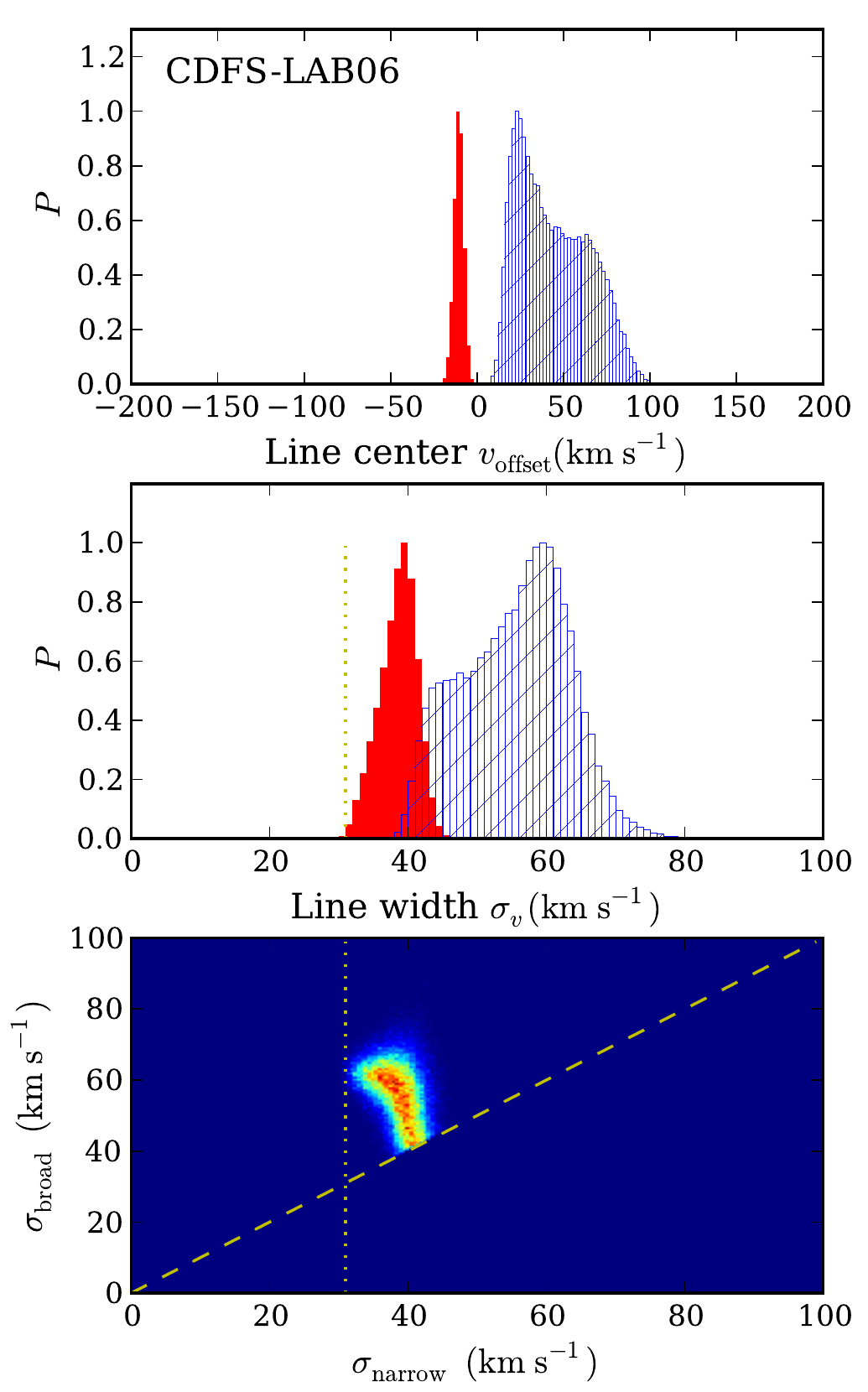}
\includegraphics[width=0.30\textwidth,trim=00 0 0 0,clip=true]{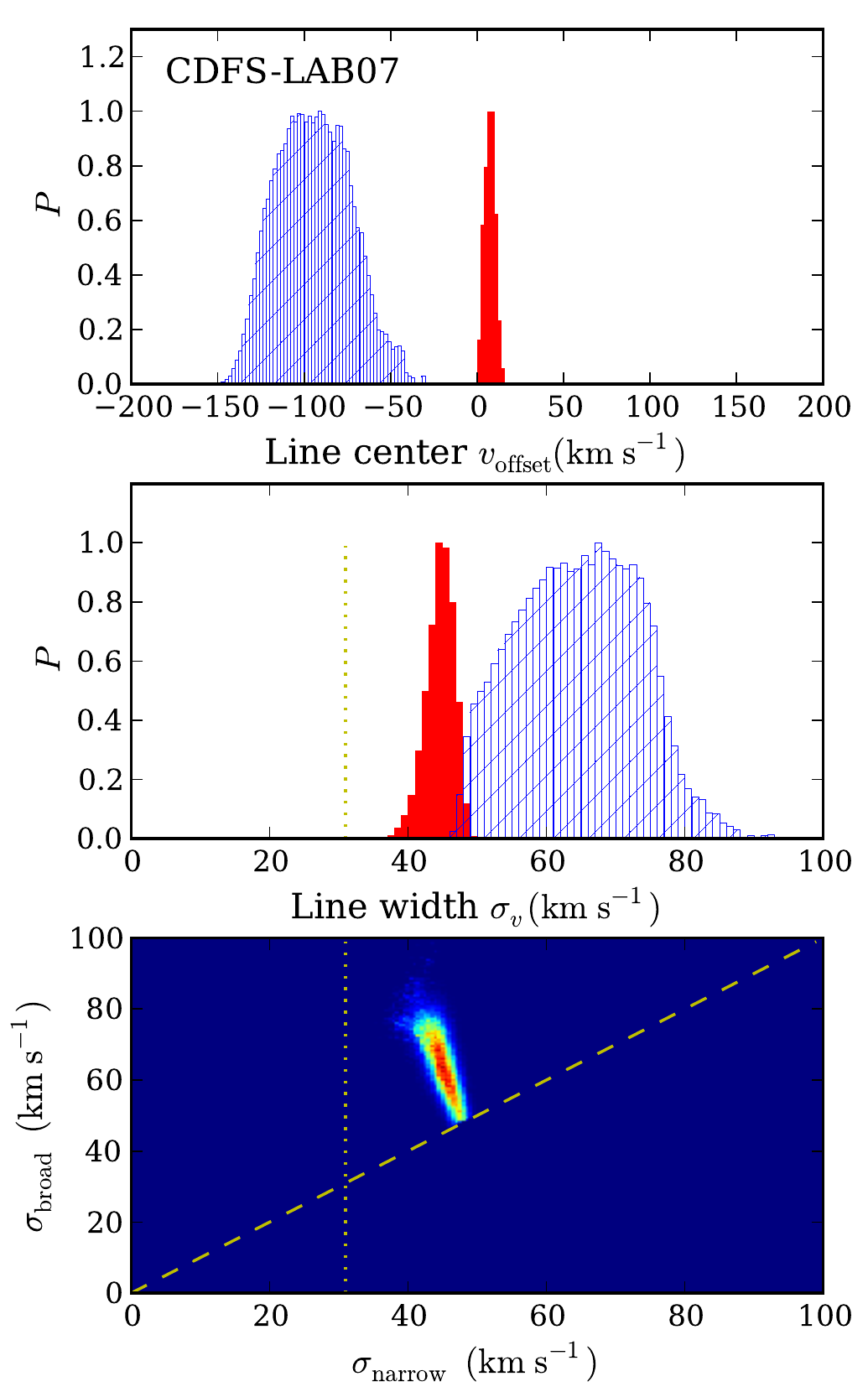}
\includegraphics[width=0.30\textwidth,trim=00 0 0 0,clip=true]{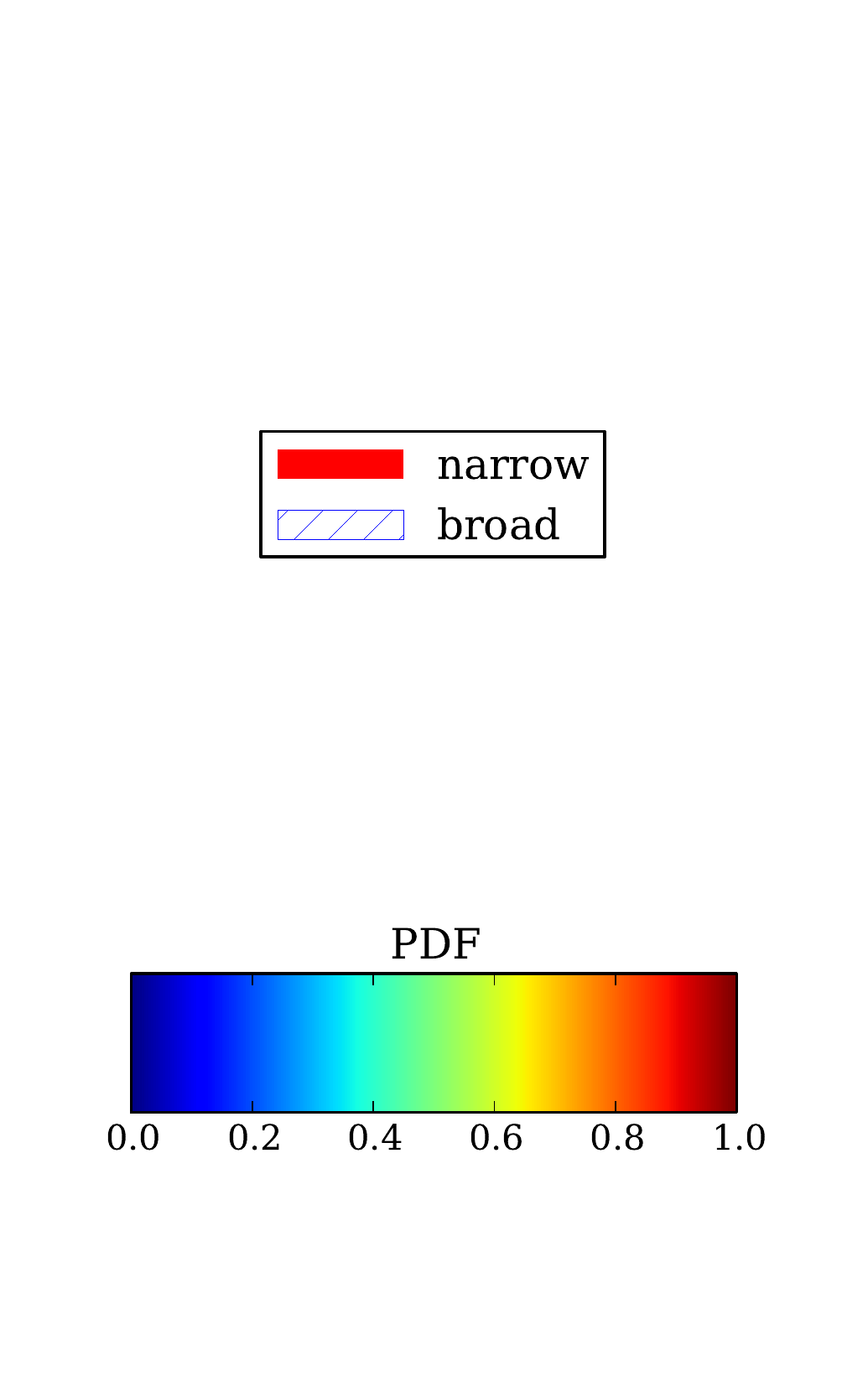}\\[2.5ex]
\includegraphics[width=0.30\textwidth,trim=00 0 0 0,clip=true]{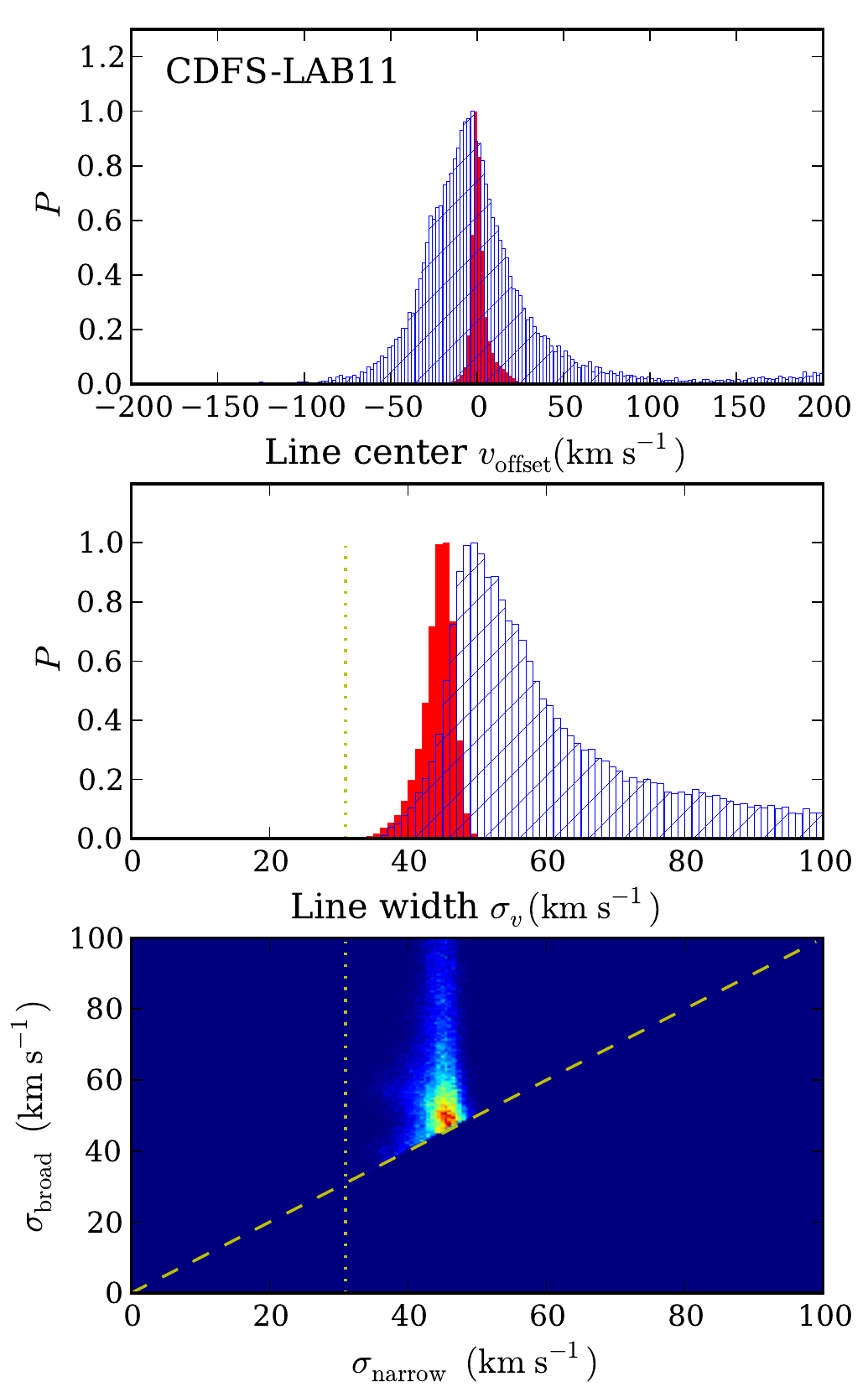}
\includegraphics[width=0.30\textwidth,trim=00 0 0 0,clip=true]{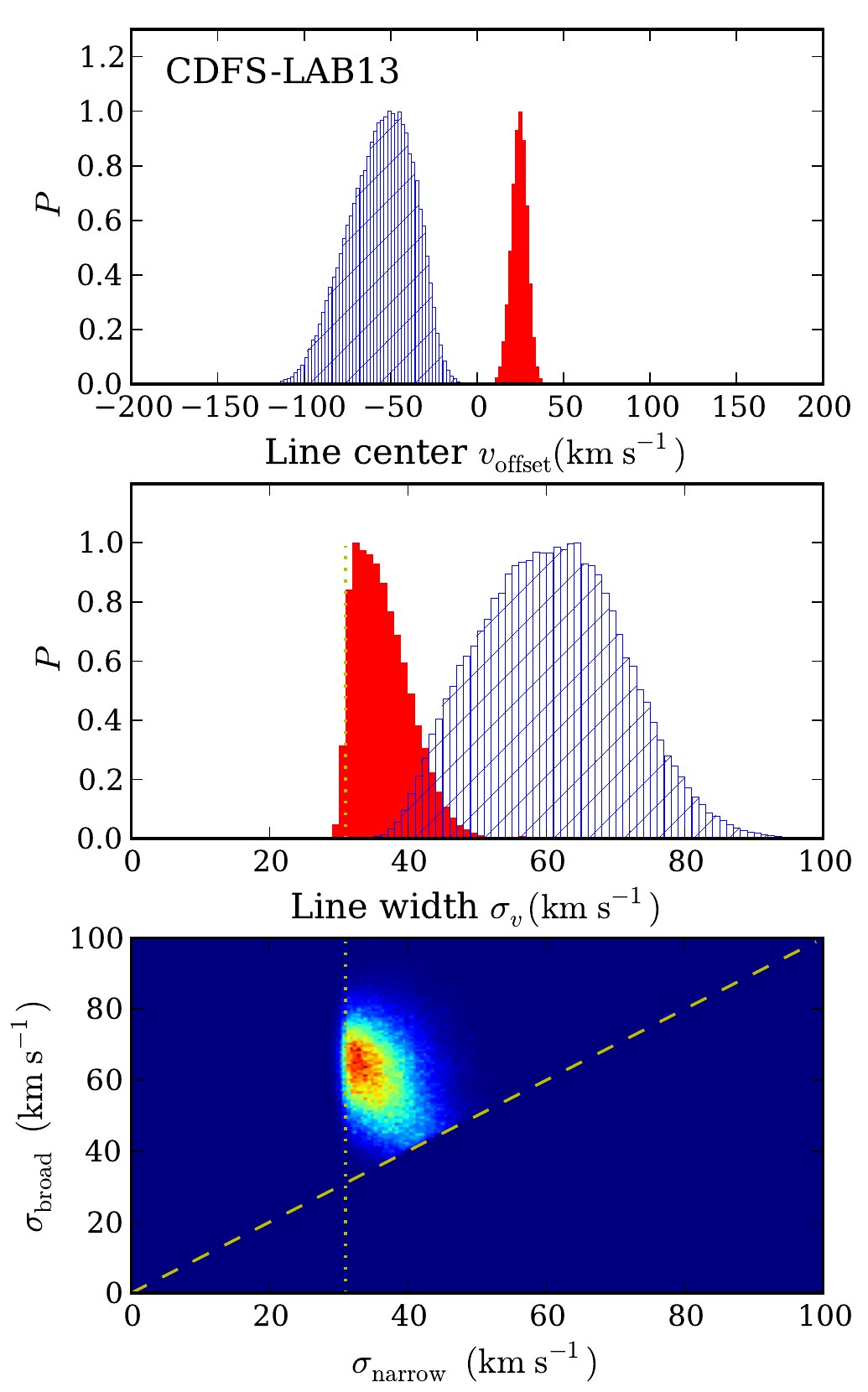}
\includegraphics[width=0.30\textwidth,trim=00 0 0 0,clip=true]{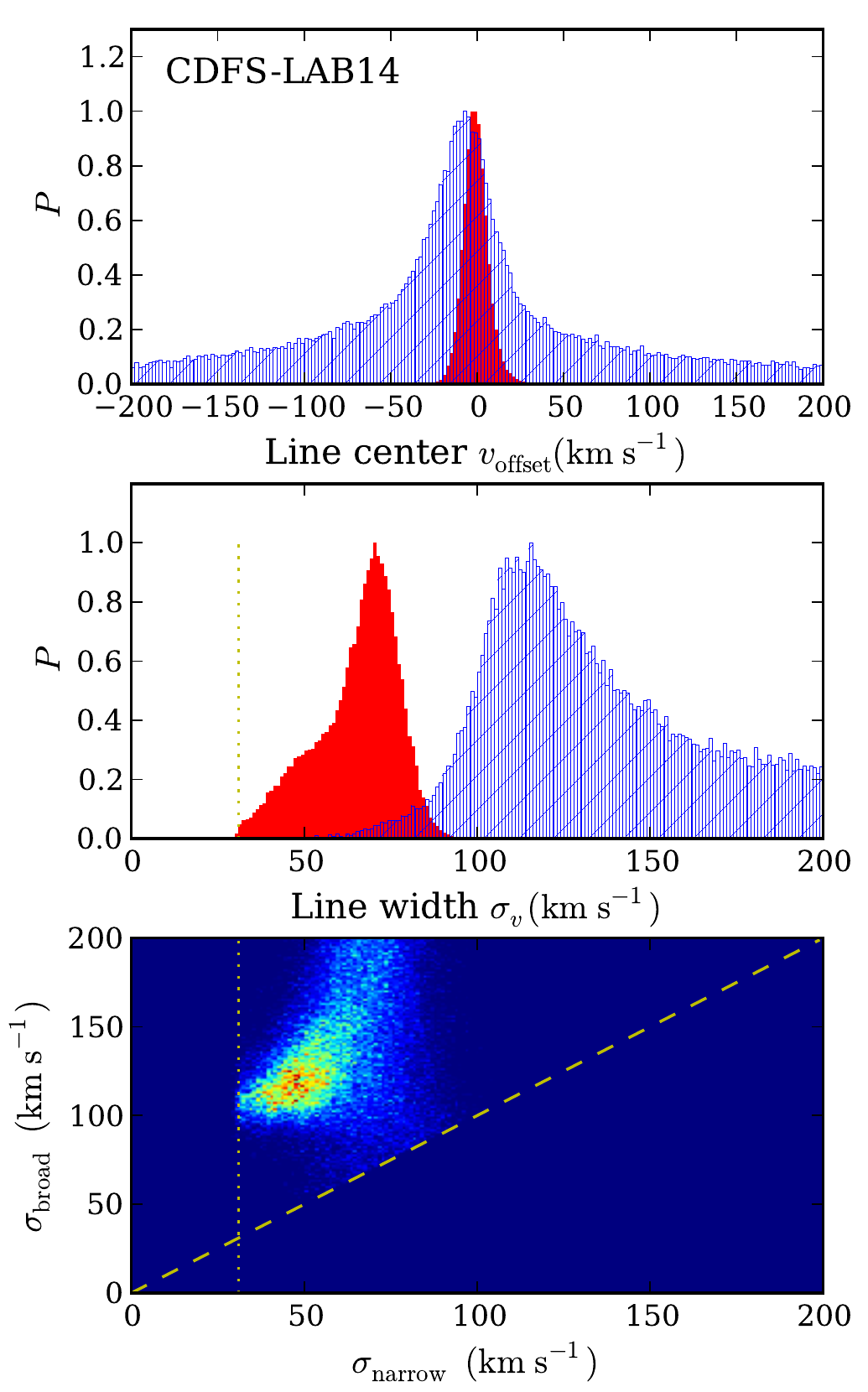}
\end{center}
\caption{
\ifpreprint\small\fi
({\it Top and middle}) Marginalized likelihood distributions of
the line centers and widths of the two Gaussian components.
({\it Bottom}) Joint 2--D likelihood distributions of the line widths
of the two Gaussian components.  The red and blue histograms represent
the narrow and broad component, respectively.  The vertical dotted
lines indicate the instrumental resolution ($\sigma_{\rm instr}$ =
31\,\kms).  The dash lines represent the locus for $\sigma_{\rm narrow}$
= $\sigma_{\rm broad}$.
}
\label{fig:O3MCMC}
\end{figure*}

To test if two velocity components are needed to fit our
spectrally-resolved \oiii profiles and to extract the line parameters,
we employ the \oiii profiles with two Gaussian functions using a Markov
Chain Monte Carlo (MCMC) technique. We use the {\tt emcee} software
\citep{emcee} to sample the distributions of six parameters: line center,
width, and fluxes for the two Gaussian profiles.
We require that (1) the line widths of both components are larger than
the instrumental line width (31\kms; the vertical dotted lines in Figure
\ref{fig:O3MCMC}), (2) the peak of each component should be at least 5\%
of the observed peak intensity, and (3) the peak of the narrow component
is higher than that of the broad component. The latter two priors are
imposed to prevent the MCMC chains from getting stuck in parameter spaces
with extremely broad lines but with negligible fluxes.

In Figure \ref{fig:O3MCMC}, we show the likelihood distributions of the
line widths of the two components: $\sigma_{\rm narrow}$ and $\sigma_{\rm
broad}$. Note that we exclude CDFS-LAB10 in this analysis because its
neighboring galaxy (CDFS-LAB10A) makes it difficult to reliably extract
its profile (see Sections \ref{sec:O3profile} and \ref{sec:cdfs-lab10}).
Except for CDFS-LAB11, the likelihood distributions of the two
line-widths do not overlap significantly, and the peaks of the joint
2--D distributions are not located near the $\sigma_{\rm narrow}$ =
$\sigma_{\rm broad}$ line (the dashed line).  Therefore, we conclude that
the remaining four systems (CDFS-LAB06, 07, 13, 14) are likely to consist
of two components.  The uncertainties of the parameters are determined
from the 68.2\% confidence interval of the marginalized distribution.

\def\arraystretch{1.0}
\begin{deluxetable}{c cc cc ccccc}
\tablewidth{0pt}
\tabletypesize{\scriptsize}
\tablecaption{Log of X-shooter Observations and Properties of Sample}
\tablehead{
\colhead{Name           }&
\colhead{$L$(\lya)\ff{a}}&
\colhead{EW(\lya)\ff{a} }&
\multicolumn{2}{c}{Coordinate}&
\colhead{Date   }&
\colhead{Airmass}&
\colhead{Seeing }&
\colhead{$T_{\rm exp}$ (UVB)}&
\colhead{$T_{\rm exp}$ (NIR)}\\
\colhead{          }&
\colhead{ ($10^{43}$ \unitcgslum) }&
\colhead{ (\AA)    }&
\colhead{ R.A.     }&
\colhead{ decl.    }&
\colhead{          }&
\colhead{          }&
\colhead{ (arcsec) }&
\colhead{ (hour)   }&
\colhead{ (hour)   }
}
\startdata
 CDFS-LAB06  &   1.57 $\pm$  0.07   &   56  $\pm$  3  &  03:32:19.78 & -27:47:31.5   &   2011-01-07  &     1.07   &    0.80    &    0.76   &       0.80     \\
 CDFS-LAB07  &   1.14 $\pm$  0.05   &   58  $\pm$  3  &  03:32:03.36 & -27:45:24.4   &   2011-01-06  &     1.05   &    0.77    &    0.76   &       0.80     \\
             &                      &                 &              &               &   2011-01-06  &     1.18   &    0.98    &    0.76   &       0.80     \\
             &                      &                 &              &               &   2011-01-28  &     1.12   &    0.81    &    0.76   &       0.80     \\
 CDFS-LAB10  &   0.71 $\pm$  0.04   &  318  $\pm$ 36  &  03:32:37.45 & -28:02:05.7   &   2010-11-06  &     1.00   &    0.71    &    0.76   &       0.80     \\
             &                      &                 &              &               &   2010-11-07  &     1.00   &    0.68    &    0.76   &       0.80     \\
             &                      &                 &              &               &   2010-11-07  &     1.08   &    1.13    &    0.76   &       0.80     \\
             &                      &                 &              &               &   2010-11-08  &     1.08   &    1.23    &    0.76   &       0.80     \\
 CDFS-LAB11  &   1.02 $\pm$  0.04   &  120  $\pm$ 10  &  03:32:43.25 & -27:42:58.3   &   2011-01-08  &     1.07   &    0.75    &    0.76   &       0.80     \\
             &                      &                 &              &               &   2011-01-08  &     1.23   &    0.85    &    0.76   &       0.80     \\
 CDFS-LAB13  &   0.94 $\pm$  0.04   &  184  $\pm$ 25  &  03:32:32.75 & -27:39:06.4   &   2011-01-07  &     1.22   &    0.79    &    0.76   &       0.80     \\
             &                      &                 &              &               &   2011-01-08  &     1.01   &    0.70    &    0.76   &       0.80     \\
 CDFS-LAB14  &   0.93 $\pm$  0.05   &   67  $\pm$  5  &  03:32:32.29 & -27:41:26.4   &   2011-01-09  &     1.07   &    0.80    &    0.76   &       0.80     \\
             &                      &                 &              &               &   2011-01-09  &     1.24   &    0.73    &    0.76   &       0.80     \\
\hline\\[-1.5ex]
 CDFS-LAB01  &   8.02 $\pm$  0.24   &  512  $\pm$ 50  &              &               &               &            &            &           &                \\
 CDFS-LAB02  &   2.88 $\pm$  0.12   &   43  $\pm$  2  &              &               &               &            &            &           &                
\enddata
\label{tab:obslog}
\tablecomments{Observations for CDFS-LAB01 and CDFS-LAB02 were presented in \citet{Yang11}.}
\tablenotetext{a}{\lya Luminosity and EW within an isophot of 5.5\E{-18} \unitcgssb.}
\end{deluxetable}

\renewcommand{\kms}        {\,km\,s$^{-1}$}
\renewcommand{\lya}        {Ly$\alpha$\xspace}
\renewcommand{\unitcgssb}  {erg\,s$^{-1}$\,cm$^{-2}$\,arcsec$^{-2}$}
\renewcommand{\unitcgsflux}{erg\,s$^{-1}$\,cm$^{-2}$}
\renewcommand{\unitcgslum} {erg\,s$^{-1}$}

\renewcommand\ff[1]{\tablenotemark{#1}}
\newcommand\B{\phn}
\def\arraystretch{1.0}

\begin{deluxetable}{c ccc r ccc l}
\tablewidth{0pt}
\tabletypesize{\scriptsize}
\ifpreprint\tabletypesize{\small}\fi
\ifpreprint\rotate\fi
\tablecaption{Properties of \lya Line}
\tablehead{
\multicolumn{1}{c}{}& 
\multicolumn{3}{c}{Red Peak}&
\multicolumn{1}{c}{}& 
\multicolumn{3}{c}{Blue Peak}\\
\cline{2-4}
\cline{6-8}\\[-1.5ex]
\colhead{Name}&  
\colhead{\dvlya}&  
\colhead{$\sigma_v$\ff{a}}&  
\colhead{Flux}&  
\colhead{}&
\colhead{$v_{\rm offset}$}&  
\colhead{$\sigma_v$\ff{a}}&  
\colhead{Flux}&
\colhead{Profile\ff{b}}\\[0.5ex]
\colhead{}&
\colhead{(\kms)}&
\colhead{(\kms)}&
\colhead{($10^{-17}$ \unitcgsflux)}&
\colhead{}&
\colhead{(\kms)}&
\colhead{(\kms)}&
\colhead{($10^{-17}$ \unitcgsflux)}&
\colhead{}
}
\startdata
  CDFS-LAB01     &   $-$65 $\pm$    20 &     228 $\pm$    14 &   45.2 $\pm$    1.6 &&             \nodata &             \nodata &             \nodata  &    (1) \ff{c} \\
  CDFS-LAB11     &      84 $\pm$     6 &      79 $\pm$     3 &   26.5 $\pm$    0.7 &&             \nodata &             \nodata &             \nodata  &    (1) \ff{c} \\
  CDFS-LAB10     &   \B247 $\pm$   147 &     425 $\pm$    97 &   18.7 $\pm$    2.4 &&             \nodata &             \nodata &             \nodata  &    (1) \ff{d} \\
  CDFS-LAB13     &     152 $\pm$    14 &     100 $\pm$   9\B &   13.7 $\pm$    0.7 &&  $-$277 $\pm$     8 &      43 $\pm$     8 &    4.0 $\pm$    0.5  &    (2)       \\
  CDFS-LAB06     &     123 $\pm$    15 &     150 $\pm$    13 &   25.7 $\pm$    1.3 &&             \nodata &             \nodata &             \nodata  &    (1)       \\
  CDFS-LAB02     &     211 $\pm$    43 &     192 $\pm$    35 &  \B8.2 $\pm$    0.9 &&  $-$342 $\pm$   117 &     273 $\pm$   137 &    3.3 $\pm$    1.3  &    (2)       \\
  CDFS-LAB07     &     181 $\pm$    15 &     168 $\pm$    10 &   22.2 $\pm$    0.8 &&             \nodata &             \nodata &             \nodata  &    (1)       \\
  CDFS-LAB14     &     371 $\pm$    34 &     202 $\pm$    24 &   18.0 $\pm$    1.2 &&  $-$404 $\pm$    17 &     176 $\pm$    18 &   12.4 $\pm$    1.1  &    (3)    
\enddata
\label{tab:Lyman}
\tablenotetext{a}{Corrected for the intrumental profile in the UVB ($\sigma_{\rm instr}$ $\simeq$ 39\,\kms).}
\tablenotetext{b}{Line morphology: (1) single (red) peak, (2) double peaked profile with a stronger red peak, (3) double peaked profile with similar intensity peaks.}
\tablenotetext{c}{The profiles of CDFS-LAB01 and 11 are not significantly broader than the \oiii lines.}
\tablenotetext{d}{The profile of CDFS-LAB10 is composed of multiple components (Figure \ref{fig:spec2d}). Here the total integrated profile is used for the fit.}
\end{deluxetable}

\renewcommand\ff[1]{\tablenotemark{#1}}
\renewcommand\B{\phn}
\def\arraystretch{1.0}

\begin{deluxetable}{c cc c cc c}
\tablewidth{0pt}
\tablecaption{Decomposition of \oiii Profile}
\tablehead{
\colhead{Name}&  
\colhead{$\sigma_{\rm narrow}$\ff{a}}&  
\colhead{$v_{\rm broad}$}&  
\colhead{$\sigma_{\rm broad}$}&  
\colhead{$F_{\rm broad}/F_{\rm narrow}$\ff{b}}&
\colhead{$F_{\rm broad}/F_{\rm total }$\ff{c}}&
\colhead{$\Delta v_{\rm max}$\ff{d}}\\[0.5ex]
\colhead{}&  
\colhead{(\kms)}&  
\colhead{(\kms)}&
\colhead{(\kms)}&
\colhead{}&
\colhead{}&
\colhead{(\kms)}
}
\startdata
 CDFS-LAB06 &     38  $\pm$      3\B  &   $+$39  $\pm$      27 &     55  $\pm$      10 &   0.51$^{    +0.58}_{    -0.29}$   &     0.35  $\pm$     0.17 &   152  $\pm$      15 \\[0.9ex] 
 CDFS-LAB07 &     44  $\pm$      2\B  &   $-$97  $\pm$      23 &     64  $\pm$      10 &   0.44$^{    +0.31}_{    -0.18}$   &     0.32  $\pm$     0.12 &   222  $\pm$      11 \\[0.9ex] 
 CDFS-LAB13 &     35  $\pm$      5\B  &   $-$56  $\pm$      21 &     60  $\pm$      11 &   0.84$^{    +0.55}_{    -0.38}$   &     0.46  $\pm$     0.13 &   174  $\pm$      18 \\[0.9ex] 
 CDFS-LAB14 &     67  $\pm$      16   &   $-$12  $\pm$      71 &    125  $\pm$      40 &   0.47$^{    +0.66}_{    -0.20}$   &     0.33  $\pm$     0.20 &   262  $\pm$      75 \\[0.9ex] 
 CDFS-LAB10 &     71  $\pm$      1\B  &   \nodata              &    \nodata            &   \nodata                          &     \nodata              &   \nodata           \\[0.9ex]
 CDFS-LAB11 &     47  $\pm$      1\B  &   \nodata              &    \nodata            &   \nodata                          &     \nodata              &   \nodata            
\enddata
\label{tab:O3profile}
\tablenotetext{a}{
  The line widths ($\sigma_{\rm narrow}$ and $\sigma_{\rm broad}$) are
  {\it not} corrected for the intrumental profile, because some of the
  line widths are comparable to the intrumental resolution 
  ($\sigma_{\rm instr}$ $\simeq$ 31\,\kms in the NIR).}
\tablenotetext{b}{
  $F_{\rm broad}$ and $F_{\rm narrow}$ are the fluxes in the broad and
  narrow components of the \oiii emision line, respectively.}
\tablenotetext{c}{
  $F_{\rm total}$ $=$ $F_{\rm narrow}$ + $F_{\rm broad}$.}
\tablenotetext{d}{
  The maximum blueshifted velocity $\Delta v_{\rm max}$ $\equiv$ $|v_{\rm broad}$ $-$ $2\,\sigma_{\rm broad} |$. 
  For CDFS-LAB06 where the broad component is redshifted, we list $|v_{\rm broad}$ $+$ $2\,\sigma_{\rm broad} |$.}
\end{deluxetable}

\end{document}